\documentclass[prb,twocolumn,amssymb,amsmath,floatfix,superscriptaddress,longbibliography]{revtex4-2}
\usepackage{amsmath}
\usepackage{hyperref}
\hypersetup{
	colorlinks=true,
	linkcolor=blue,
	filecolor=blue,
	citecolor = black,      
	urlcolor=cyan,
}

\usepackage{amssymb}
\usepackage{listings}
\usepackage{tabularx}
\usepackage{cancel}
\usepackage[normalem]{ulem}

\usepackage[a4paper, margin=0.5in]{geometry}

\usepackage[utf8]{inputenc}
\usepackage{color}
\usepackage{graphicx}
\usepackage{empheq}

\newcommand{\magn}[1]{\left|#1\right|}
\newcommand{\op}[1]{\mathcal{O}_{#1}}

\newcommand{\bra}[1]{\left\langle#1\right|}
\newcommand{\ket}[1]{\left|#1\right\rangle}
\newcommand{\bracket}[2]{\left\langle#1|#2\right\rangle}
\newcommand{\adj}[1]{#1^{\dagger}}

\newcommand{\tr}[1]{\mathrm{Tr}\left[ #1 \right]}
\newcommand{\trover}[2]{\mathrm{Tr}_{#1}\left[ #2 \right]}

\newcommand{\id}{\mathbb{I}}

\newcommand{\mvec}[1]{\mathbf{#1}}

\newcommand{\su}[1]{\mathrm{SU}(#1)}
\newcommand{\unrho}{\tilde{\rho}}
\newcommand{\singlet}[1]{(#1)}

\newcommand{\Ac}{\overline{A}}
\newcommand{\cer}{\zeta}    
\newcommand{\cek}{\mu}      

\newcommand{\inst}{I_*}
\newcommand{\Ibulk}{I_{\mathrm{bulk}}}

\usepackage[shortlabels]{enumitem}

\graphicspath{{Figure/}}

\begin{document}

\date{\today}
\title{Measurement-induced purification in large-$N$ hybrid Brownian circuits}
\author{Gregory S.~Bentsen}
\thanks{These authors contributed equally to this work.}
\affiliation{Martin A. Fisher School of Physics, Brandeis University, Waltham MA 02453, USA}
\author{Subhayan Sahu}
\thanks{These authors contributed equally to this work.}
\affiliation{Condensed Matter Theory Center and Department of Physics,
University of Maryland, College Park, MD 20742, USA}
\author{Brian Swingle}
\affiliation{Martin A. Fisher School of Physics, Brandeis University, Waltham MA, USA}
\affiliation{Condensed Matter Theory Center and Department of Physics,
University of Maryland, College Park, MD 20742, USA}
\affiliation{Maryland Center for Fundamental Physics and Joint
Center for Quantum Information and Computer Science,
University of Maryland, College Park, MD 20742, USA}

\begin{abstract}
Competition between unitary dynamics that scrambles quantum information non-locally and local measurements that probe and collapse the quantum state can result in a measurement-induced entanglement phase transition. Here we study this phenomenon in an analytically tractable all-to-all Brownian hybrid circuit model composed of qubits. The system is initially entangled with an equal sized reference, and the subsequent hybrid system dynamics either partially preserves or totally destroys this entanglement depending on the measurement rate. Our approach can access a variety of entropic observables which are distinguished by the averaging procedure, and for concreteness we focus on a particular purity quantity for which the averaging is particularly simple. We represent the purity as a path integral coupling four replicas with twisted boundary conditions. Saddle-point analysis reveals a second-order phase transition corresponding to replica permutation symmetry breaking below a critical measurement rate. The transition is mean-field-like and we characterize the critical properties near the transition in terms of a simple Ising field theory in $0+1$ dimensions. In addition to studying the purity of the entire system, we study subsystem purities and relate these results to manifestations of quantum error correction in the model. We also comment on the experimental feasibility for simulating this averaged purity, and corroborate our results with exact diagonalization for modest system sizes.
\end{abstract}




\maketitle


\section{Introduction}
\label{sec:intro}

As a quantum many-body system evolves in time, its state vector follows a trajectory in Hilbert space guided by unitary dynamics and measurements. Unitary evolution is generated by a system's Hamiltonian, while measurements are generated by coupling the system to a macroscopic apparatus that records the value of some observable and simultaneously collapses the state vector. If a quantum system is composed of many parts and if the interactions and measurements involve only a few of these parts at a time, then the operator which updates the quantum state has the general structure of a network composed of many elementary pieces glued together. When the number of elementary pieces is large and the time is long, the evaluation of such a network is akin to evaluating the partition function of a generalized statistical mechanics problem, analogous to an Ising model where one allows more local degrees of freedom and all kinds of few-body interactions with coupling parameters that may be complex or even random. This point of view has a long tradition in theoretical physics, with recently studied examples including~\cite{hayden2016holographic,nahum2018operator,zhou2019emergent,vasseur2019entanglement,hunterjones2019unitary,lopez2020mean}. 

From this point of view, computing the dynamics of quantum many-body observables is part of a very general class of problems that also includes evaluating partition functions of classical statistical models and studying imaginary time evolution of local quantum systems. Given this overarching framework, one goal is to classify and understand all possible distinct classes of behaviors (phases) and the transitions between them (phase transitions). Many of these problems directly relate to experimentally realizable observables; even in cases where direct experimental access is challenging, a general understanding of the space of possible behaviors can shed indirect light on measurable observables.

In this paper, we consider a recently discovered class of such phases and phase transitions which involve the interplay of unitary scrambling dynamics and single-body measurements~\cite{li2018quantum,skinner2019measurement,chan2019unitary}. The phenomena of interest arise from a competition between the unitary part, which tends to move the state away from a product form by generating entanglement, and the measurement part, which tends to move the state towards a product form by decreasing entanglement. We make progress towards an effective field theory description of this physics by defining and solving a mean-field-like model that exhibits a similar phase transition. 

In more detail, the competition between scrambling dynamics and measurements in hybrid local quantum circuits composed of 2-body unitary scrambling gates interspersed with local projective measurements leads to a measurement-induced phase transition (MIPT) from a volume-law entangled phase to an area-law entangled phase above a critical measurement rate~\cite{skinner2019measurement,li2018quantum,chan2019unitary,li2019measurement,szyniszewski2019entanglement}. These phenomena are also related to dramatic phase transitions that can occur in the entanglement structures of final states in noisy quantum computers \cite{aharonov2000quantum}. The  hybrid circuit discoveries were followed by a series of works studying related transitions in a variety of models, exploring the critical properties, and studying relations to quantum error correcting codes~ \cite{bao2020theory,jian2020measurement,fan2020self,li2020conformal,choi2020quantum,zabalo2020critical,tang2020measurement,turkeshi2020measurement,zhang2020nonuniversal,goto2020measurement,iaconis2020measurement,li2020statistical}. More recently, some papers have considered all-to-all models and again found analogous phases and phase transitions~\cite{nahum2020measurement,vijay2020measurement}. Similar transitions have also been observed recently in fermionic chains of coupled Sachdev-Ye-Kitaev (SYK) models with imaginary damping terms \cite{liu2020non}. Related dynamics in free-fermion systems have also been studied \cite{cao2018entanglement,chen2020emergent}.

Distinct from but related to the entanglement transition in local systems, it was found that measurements in quantum circuits can also drive a purification transition, where an initial mixed state is dynamically purified in constant time by repeated measurements occurring above a critical rate, while remaining mixed until exponential times for measurement rates below this critical rate~\cite{gullans2020dynamical,gullans2020scalable}. From a quantum information processing point of view, the volume-law phase or the mixed phase can be identified as a randomly-generated quantum error correcting code \cite{choi2020quantum,jian2020measurement,fan2020self,li2020statistical}, where the scrambling dynamics effectively `hides' the quantum information from local measurements.

\begin{figure}
    \centering
    \includegraphics[width=\columnwidth]{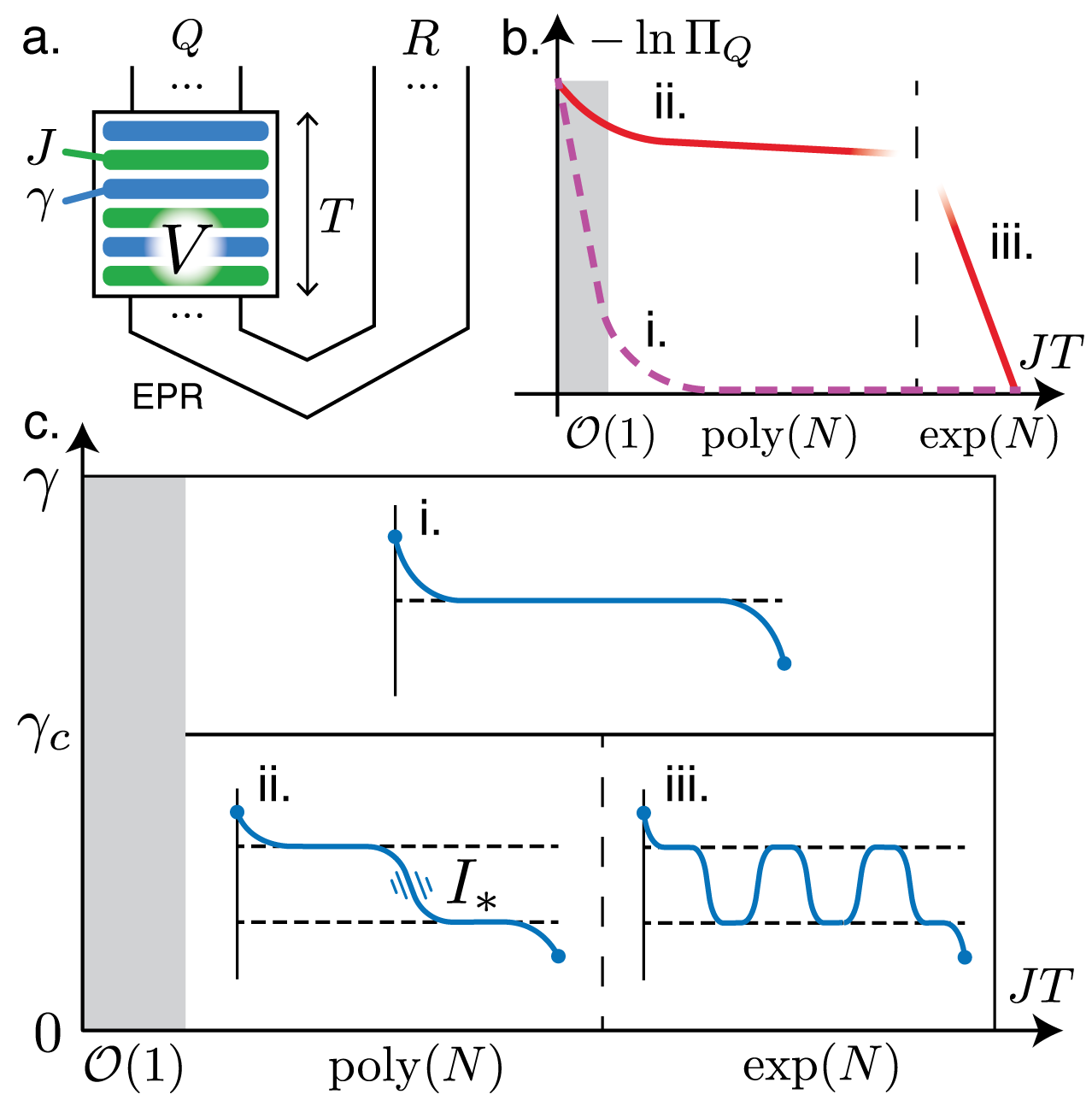}
    \caption{\textbf{Purification phase diagram for hybrid Brownian circuits.} (a) Hybrid Brownian circuits $V$ composed of alternating layers of unitary Brownian dynamics of strength $J$ (green) and non-unitary weak-measurement Brownian dynamics of strength $\gamma$ (blue) exhibit a measurement-induced purification transition diagnosed by the purity $\Pi_Q$ of qubits $Q$ that are initially maximally entangled with a reference system $R$. (b-c) Above the critical point $\gamma > \gamma_c$ (i) bulk fields (solid blue) traverse through a single saddle point (dotted black), leading to a pure phase with purity $\Pi_Q \sim 1$ (b, dotted purple). Below the critical point $\gamma < \gamma_c$ (ii) the bulk fields tunnel between two symmetry-broken saddle points (dotted black) via a single-instanton configuration, leading to a mixed phase with $\Pi_Q \sim T \exp{(-N I_*)} \ll 1$ for polynomially-long times $T$ (b, solid red). At exponentially long times the instantons proliferate and destroy the mixed phase (iii). Dynamics at early times (grey boxes) are also accessible in these models but are not the main focus of this work.}
    \label{fig:phasediagram}
\end{figure}

The interplay between unitary dynamics and measurement in quantum mechanics is an old and rich subject \cite{bohr1928quantum,aharonov2000quantum,wiseman2009quantum,wheeler2014quantum}. In particular, weak continuous measurements of the type studied here have a long history stemming largely from quantum optics and cold atoms \cite{wiseman1993aquantum,wiseman1993bquantum,wiseman1993interpretation,wiseman1994squeezing,lee2014heralded,barontini2015deterministic,kozlowski2016non,mazzucchi2016quantum,sorensen2018quantum,gross2018qubit,ivanov2020feedback,kroeger2020continuous}. In fact the monitored dynamics we consider here -- in which the quantum state evolution is conditioned upon obtaining particular measurement results -- is most naturally described in terms of the \emph{quantum trajectories} formalism, a standard tool in the analysis of open quantum systems \cite{wiseman2009quantum}. Continuous weak measurements of this kind have recently been used in experiments to engineer quantum Zeno dynamics capable of producing metrologically useful entanglement \cite{barontini2015deterministic,mazzucchi2016quantum} as well as to drive phase transitions in ensembles of cold atoms \cite{kroeger2020continuous}.

The novelty of our current interest is in the interplay of these weak measurements with the many-body scrambling dynamics of a strongly-interacting system, where the competition between these two forces drives a phase transition in the structure of the many-body entangled state \cite{bao2020theory,gullans2020dynamical,nahum2020measurement} that is underpinned by a dynamically-generated error-correcting code \cite{choi2020quantum,jian2020measurement,fan2020self}. Whereas existing theoretical and experimental work on weak measurements, quantum feedback, and quantum control has focused largely on single- or few-body dynamics, the hybrid dynamics we consider here is strongly chaotic and many-body. In these many-body systems the quantum state follows an increasingly complex trajectory that can rapidly explore arbitrary regions of the exponentially-large Hilbert space. For this reason, new techniques and approaches are required to characterize and understand hybrid entanglement dynamics in the many-body context.

Due to the inherently many-body nature of these entanglement transitions, the earliest work on measurement-induced phase transitions was driven by numerical simulations of Clifford circuits, in which a restricted subgroup of quantum operations can be simulated efficiently on a classical computer. Subsequently, Refs. \cite{bao2020theory,jian2020measurement} showed that Haar-random hybrid circuits in $1+1$d could be mapped exactly onto effective replica statistical mechanics problems in 2d. In these replica Potts models, the volume-law phase can be identified as the `low-temperature' replica permutation symmetry-broken phase of the Potts model, while the area-law phase is associated with replica-symmetric `paramagnetic' phase.
Moreover, as we emphasize below, there are actually families of observables and phase transitions related to different kinds of averaging procedures which translate to different kinds of replica limits.
These and other mappings are general, but analytical progress has been hampered by the presence of non-positive weights in the generalized partition function; in certain cases the problem can be ameliorated by considering the limit of large local Hilbert space dimension or restricting to special classes of circuits. One also has to take a delicate replica limit to access entanglement observables averaged with respect to Born probabilities.  

Throughout this body of work, a still outstanding goal is the construction of effective field theories which capture the universal physics of these various transitions. Here we make progress towards that goal by introducing new analytical tools for directly accessing measurement-induced entanglement transitions. In particular, we introduce a broad family of microscopic all-to-all hybrid Brownian circuit models exhibiting measurement-induced phase transition (MIPT) that also feature solvable large-$N$ (mean-field) solutions. We focus on a particularly simple member of this family featuring Brownian 2-body unitary dynamics \cite{lashkari2013towards} and Brownian 1-body weak measurements \cite{wiseman2009quantum,gross2018qubit} and characterize the MIPT in this model in terms of the purification dynamics of a system of spins in a cluster $Q$ that are initially maximally-entangled with a reference system $R$ as shown in Fig. \ref{fig:phasediagram}a. We write down and analyze a path integral representation for an average $\Pi_Q$ of the purity $\tr{\rho_Q^2}$ of the system which is controlled in the large-$N$ limit. The technology can be readily generalized to higher moments $\tr{\rho_Q^n}$ of the system density matrix as we show in Appendix \ref{app:deriv_pqpathintegral}. 

The path integral representation of the purity involves four replicas which are coupled by various collective fields and $N$ copies of an auxiliary few-spin path integral that depends on the collective fields. We analyze this path integral using a saddle-point approach. At large $N$ and time polynomial in $N$, the leading saddle point gives two distinct phases as a function of the measurement-to-scrambling ratio $\gamma / J$ as shown in Fig. \ref{fig:phasediagram}b-c. Sufficiently strong measurements $\gamma > \gamma_c$ yield a paramagnetic (replica symmetry unbroken) phase with a single dominant saddle point (Fig. \ref{fig:phasediagram}c.i) for all circuit depths $T \gg \mathcal{O}(1)$. As the measurement strength $\gamma$ decreases through the critical point $\gamma_c$, this saddle point continuously splits into a pair of degenerate saddle points, leading to a spontaneous breaking of the replica symmetry (Fig. \ref{fig:phasediagram}c.ii). Non-equal boundary conditions at times $t = 0,T$ promote the formation of \emph{instanton} configurations in the bulk with action $I_*$ that traverse between the two saddle points. At exponentially long times $T \sim \exp{(N I_*)}$, we must include subleading approximate saddles involving multi-instanton configurations. The instantons proliferate at long time and the replica symmetry is restored (Fig. \ref{fig:phasediagram}c.iii). 

These hybrid Brownian circuit models are motivated from several points of view. First, we wanted to consider continuous-time models for the more direct access they provide to path-integral and field-theory representations of the physics. Second, we wanted to consider models with all-to-all interactions which typically have simpler statistical properties, and we do observe that various distinct methods of averaging the purity are nearly identical for experimentally-accessible systems of modest size. Third, we wanted to analytically study the entanglement of subsystems to reveal how information about the reference is hidden from small subsystems, a key property of quantum codes. Fourth, we wanted to study models in which higher-body unitary dynamics and higher-body measurements could be seamlessly incorporated. Finally, while mean-field scrambling dynamics is also analytically accessible in random all-to-all fermion models, here we wanted to study models consisting of spins which are more directly related to potential near-term experimental realizations in cold-atom platforms. In the following we discuss each of these motivations in more detail and survey some of the main results.

First, the continuous-time path integral representation of our hybrid Brownian circuits makes it particularly easy to derive a field-theory representation of the purification phase transition. In particular, at low measurement rates close to but below the critical rate $\gamma_c$, we find in section \ref{sec:criticalexps} that the system entropy has critical exponent $3/2$, i.e., $-\ln \Pi_Q \sim N(\gamma_c-\gamma)^{\cer}$, with $\cer = 3/2$. The corresponding field theory is particularly simple, an effective $0+1$d Ising field theory, which hosts a phase transition in the limit of infinite $N$ for finite $T$. Moreover, by including subleading saddles at finite $N$, we show in section \ref{sec:latetime} that instantons destabilize the mixed (symmetry broken) phase leading to asymptotic purification of the system at exponentially long time.



Second, a crucial aspect of the MIPT phenomena is that they are 
visible only in entanglement-sensitive observables like the purity that are non-linear in the density matrix. Since measurement outcomes are fundamentally random in quantum physics, one must carry out a large number of experimental trials (exponential in the number of measurements) to generate even a few copies of a particular state associated to a fixed set of measurement outcomes, which would be necessary to estimate these observables.
These include so-called `forced' transitions where one post-selects on a particular measurement outcome and the quantum state evolves according to a fixed non-unitary transformation~\cite{nahum2020measurement,fuji2020measurement,biella2021many,jian2021yang,turkeshi2021measurementinduced} (our measurement setup is of this kind). It is also interesting to attempt to circumvent the experimental overheard by considering special hybrid circuits \cite{ippoliti2021postselection,ippoliti2021fractal}.

Given this exponential post-selection cost, we are justified in considering MIPTs in families of observables with comparable experimental accessibility, no harder than simulating the Born averaged observable. The above phase diagram applies to a simple kind of averaged purity in which we average the unnormalized purity and probability separately and then divide. This circumvents the theoretical difficulty associated with averaging the purity with respect to the Born probability. We show in section \ref{sec:orderparam} and in Appendix \ref{app:disorderavg} that this analytically-tractable averaging procedure actually corresponds to an experimentally-accessible observable which requires comparable experimental effort to the Born-averaged quantity. Moreover, we offer evidence from exact diagonalization that these two distinct averages are actually nearly identical in the mixed phase at modest system size.

Third, we can also directly access the purity $\Pi_A$ of subsystems $A \subset Q$ in our path integral representation by modifying the boundary conditions on the path integral. In particular, this allows us to make contact between our path integral representation and the dynamically-generated quantum error correcting codes generated in the mixed phase. We show in section \ref{sec:subsystems} that the subsystem purities are consistent with this code property and identify a critical subsystem fraction $k_c$ above which the subsystems of $Q$ with more than $k_c N$ qubits are entangled with the reference $R$ for any time $T$ polynomial in $N$. Within our model, we can also access the critical exponent of $k_c$ analytically, and we find $k_c-1/2 \sim (\gamma_c-\gamma)^{\cek}$, with $\cek = 1$. 

Fourth, while we focus most of our attention in this work on a single model, we emphasize that the tools we develop in sections \ref{sec:browniancircuit} and \ref{sec:saddlepointanalysis} are quite general. In particular, in Appendix \ref{app:deriv_pqpathintegral} we provide explicit path integral representations for a family of models indexed by $(p,q)$ featuring $p$-body unitary interactions and $q$-body weak measurements. Such models could be relevant for studying measurement-only transitions~\cite{lavasani2021measurement,sang2020measurement,ippoliti2021entanglement,lang2020entanglement,lavasani2021topological,bao2021symmetry} and we conjecture they will have an even higher degree of self-averaging for larger $p,q$. Furthermore, these all-to-all clusters can readily be placed on different geometries, such as chains and trees, and also with various kinds of experimentally-realizable long-range interactions. We reserve detailed study of these more general models to future work. 

Finally, we emphasize that hybrid Brownian dynamics similar to the type we study here could in principle be probed in experiments with optically-trapped cold neutral atoms coupled to a single-mode cavity. All-to-all interactions between atomic qubits mediated by photons in the optical cavity mode can be engineered to generate strong scrambling dynamics \cite{swingle2016measuring,bentsen2019treelike,davis2019photon}. Single-site measurements in principle could be performed using state-of-the-art single-site resolution imaging techniques \cite{bakr2009quantum,sherson2010single}. Alternatively, weak continuous measurements of collective spin observables could be performed by monitoring photons escaping from the rear port of the cavity \cite{gleyzes2007quantum,barontini2015deterministic} (see Appendix \ref{app:cavitymonitor}). To probe the purity $\Pi_Q$ one could prepare two identical copies of the state within a pair of atomic subensembles and interfere them \cite{islam2015measuring}. To guarantee identical measurement results in the two copies one would need to perform an exponentially large number of experimental trials and post-select on matching measurement records as noted above \cite{gullans2020scalable}.
While we acknowledge that these are daunting experimental challenges, in principle the analytical results we derive here provide precise predictions for the outcomes of real experiments that could be performed in the laboratory.


    
    
  

In the rest of the introduction we provide an outline for the rest of the paper. Section \ref{sec:browniancircuit} introduces the hybrid Brownian circuit models and describes how they may be converted into a large-$N$ path integral description. In section \ref{sec:model} we introduce the minimal $(2,1)$ hybrid Brownian circuit model, followed by a description the averaged purity $\Pi_Q$ that serves as our order parameter in section \ref{sec:orderparam}. We also explain the experimental interpretation of the averaged quantity in this section. In section \ref{sec:derivpathintegral} we derive the path integral representation for the purity $\Pi_Q$ and derive a simplification for spin-1/2 systems in section \ref{sec:spinpropagatorK}. We discuss the discrete replica permutation symmetry group and its representation in path integral language in section \ref{sec:replicasymm}.

In section \ref{sec:saddlepointanalysis} we study the purification transition in the $(2,1)$ hybrid Brownian model via large-$N$ methods and obtain the phase diagram shown in Fig. \ref{fig:phasediagram}. In section \ref{sec:bulksp} we consider time-independent saddle-point solutions of the path integral and show that the MIPT transition is driven by a spontaneously-broken $\mathbb{Z}_2$ symmetry in the bulk. In section \ref{sec:instantons} we consider the role of non-uniform boundary conditions and show that these drive instanton transitions between the two degenerate saddle points. In section \ref{sec:phasesofpathint} we show that these ingredients lead to the three phases shown in Fig. \ref{fig:phasediagram}b-c and compute analytical estimates for the purity $\Pi_Q$ in each of these phases. In section \ref{sec:criticalexps} we derive a field theory for the model near criticality and show that the critical exponent for the entropy $-\ln \Pi_Q$ is $\cer = 3/2$. We also study the path integral numerically using gradient descent methods and find a critical exponent $\cer = 1.44 \pm 0.07$ consistent with the analytical prediction. In section \ref{sec:timedependencepurity}, we show that the saddle-point approach can also capture the time-dependence of the purity at early times, and compare these predictions with results from numerical exact diagonalization in small systems. In section \ref{sec:latetime} we study the disintegration of the phase at exponentially long times due to the proliferation of instantons.

In section \ref{sec:subsystems} we consider the purity $\Pi_A$ of subsystems $A \subset Q$ as a function of subsystem fraction $k = \magn{A} / \magn{Q}$. In section \ref{sec:subsystemk} we show that measuring the purity of subsystems is equivalent to a straightforward modification of the boundary conditions in the path integral. In section \ref{sec:kcriticalexp} we use our field theory for the transition to find a critical exponent for subsystem fraction $\cek = 1$. In section \ref{sec:qecc} we interpret these results in terms of a dynamically-generated quantum error correcting code in the mixed phase. In section \ref{sec:discussion} we review our results and discuss directions for future work.




\section{Hybrid Brownian circuits}
\label{sec:browniancircuit}


In this section we define the microscopic model, a hybrid Brownian circuit combining time-dependent all-to-all $2$-spin interactions and post-selected local weak measurements, and show how to express the purity $\Pi_Q$ as a path integral expression with a large-$N$ limit. This model can be generalized to allow for $p$-spin interactions and post-selected weak measurements of $q$-spin operators as described in Appendix \ref{app:deriv_pqpathintegral}, but in the main text we focus on the simplest case with $p=2$, $q=1$. The system $Q$ consists of $N$ spins initialized in a maximally-entangled state with $N$ additional reference spins $R$. The system $Q$ is then evolved with the hybrid Brownian circuit $V$ while the reference $R$ is left untouched as shown in Fig. \ref{fig:phasediagram}a. 

In the large-$N$ limit, we expect this model to exhibit at least two hybrid dynamical phases and a measurement-induced phase transition between them as a function of the effective measurement rate. As discussed in the introduction, the physics of these phases is only visible in non-linear functions of the quantum state. We therefore construct path integral representations of the squared probability $P^2$ to obtain the post-selected state and of the unnormalized purity $Z_2$ of the system qubits. These are the simplest observables that both access the measurement-induced phase transition and are analytically calculable. Both integrals $P^2,Z_2$ can be analyzed in the large-$N$ limit by saddle-point analysis, yielding a mean-field description of the measurement-induced phase transition. In section \ref{sec:saddlepointanalysis} we analyze the physics of the $(p,q)=(2,1)$ model at large $N$ and demonstrate the phase structure illustrated in Fig. \ref{fig:phasediagram}.

\subsection{Model}
\label{sec:model}


Consider a system of $N$ spin-$S$ $\su{2}$ degrees of freedom $ S_i^{\alpha}$, $i = 1, \ldots, N$, $\alpha = x,y,z$ subject to an alternating sequence of unitary and non-unitary Brownian dynamics as illustrated in Fig. \ref{fig:phasediagram}a.  For the moment we leave the spin length $S$ unspecified, but we specialize to $S = 1/2$ in section \ref{sec:orderparam}.

On even timesteps $t = m \Delta t$, with $m$ an integer, the spins evolve under a Brownian unitary matrix $U(t) = \exp{[-i H(t) \Delta t / 2]}$ with Hamiltonian
\begin{equation}
    \label{eq:brham}
    H(t) = \sum_{\substack{i < j\\ \alpha, \beta}} J_{ij}^{\alpha \beta}(t) S_i^{\alpha} S_j^{\beta}
\end{equation}
with time-dependent all-to-all couplings $\mvec{J} = J_{ij}^{\alpha \beta}(t)$ \cite{lashkari2013towards}. These couplings are independent white-noise-correlated Gaussian random variables with zero mean and covariance
\begin{align}
    \label{eq:jrandvar}
    &\left\langle J_{ij}^{\alpha \beta}(t) J_{i'j'}^{\alpha' \beta'}(t') \right \rangle_{\mvec{J}} \nonumber \\
    &\quad \quad \quad = \frac{J}{N (S+1)^4} \frac{ \delta_{tt'}}{ \Delta t / 2} \delta_{ii'} \delta_{jj'} \delta^{\alpha \alpha'} \delta^{\beta \beta'}.
\end{align}
The scale of the fluctuations of the coupling is set by the coupling parameter $J \geq 0$. The normalization $1/N(S+1)^4$ ensures that the Hamiltonian \eqref{eq:brham} is extensive in $N$ and intensive in $S$, and the factor $\delta_{tt'} (\Delta t/2)^{-1}$ is a regularization of the Dirac delta function $\delta(t-t')$ for white-noise random variables.

On odd timesteps $t = (2m + 1)\Delta t/2$ the spins are subjected to single-site weak measurements along random spin directions. To perform each measurement we introduce an auxiliary qubit initialized in $\ket{\psi}_{\mathrm{aux}} = \ket{0}_{\mathrm{aux}}$ and couple it to the system via a unitary interaction,
\begin{equation}
    \label{eq:weakmeas}
    \exp{\left[- i \frac{\Delta t}{2} \op{}(t) \sigma^x_{\mathrm{aux}} \right] } \ket{\Psi}\ket{0}_{\mathrm{aux}},
\end{equation}
where $\ket{\Psi}$ is the state of the many-body system prior to the weak measurement, $\sigma^x_{\mathrm{aux}}$ is the Pauli-$x$ operator acting on the auxiliary qubit, and $\op{}(t) = \sum_{i,\alpha} n_i^{\alpha}(t) S_i^{\alpha}$ is the random spin operator to be measured. We then perform a projective measurement of the auxiliary qubit in the $\sigma^y_{\mathrm{aux}}$ eigenbasis and post-select for $+1$ results. The many-body state $\ket{\Psi}$ is thereby transformed to
\begin{align}
    \label{eq:nonunitarym}
    \ket{\Psi} &\to M(t) \ket{\Psi} \nonumber \\
    &= \left( 1 - \frac{1}{2} \op{} \Delta t - \frac{1}{8} \op{}^2 \Delta t^2 + \cdots \right) \ket{\Psi}
\end{align}
to lowest order in $\Delta t/2$ (note that $M(t) \neq \exp{[- \op{}(t) \Delta t / 2]}$). Note that under this measurement setup, the state evolves non-unitarily, and also deterministically, without any inherent measurement randomness. Similar to the unitary Brownian dynamics above, we take the coefficients $\mvec{n} = n_i^{\alpha}(t)$ to be independent white-noise-correlated Gaussian random variables with zero mean and covariance
\begin{equation}
    \label{eq:nrandvar}
    \left\langle n_i^{\alpha}(t) n_{i'}^{\alpha'}(t') \right \rangle_{\mvec{n}} = \frac{\gamma}{(S+1)^2} \frac{\delta_{tt'}}{\Delta t/2} \delta_{i i'} \delta^{\alpha \alpha'}.
\end{equation}
The fluctuations in $\mvec{n}$ are controlled by the parameter $\gamma \geq 0$. Due to the post-selection step the operator $M(t)$ does not conserve probabilities, and the resulting state $M(t) \ket{\Psi}$ is not normalized.

The full time evolution of the system is constructed by stacking alternating layers of $U(t)$ and $M(t)$ gates
\begin{equation}
    V \equiv \prod_{t = 0}^T M(t) U(t)
\end{equation}
as shown in Fig. \ref{fig:phasediagram}a. Given an initial state $\rho_0$ and a fixed disorder realization $\mvec{J},\mvec{n}$ this hybrid circuit produces the unnormalized output state $\unrho(V) = V \rho_0 \adj{V}$ with probability $P = \tr{\unrho(V)} \leq 1$. The relative strength of measurement and scrambling in this circuit is controlled by the dimensionless ratio $\gamma/J$.

\subsection{Phase structure and observables}
\label{sec:orderparam}

When $\gamma = 0$ the weak measurement layers have no effect and we recover a unitary Brownian circuit that strongly scrambles quantum information \cite{lashkari2013towards}. For a system $Q$ maximally entangled with a reference $R$ at time $t = 0$ (Fig. \ref{fig:phasediagram}a), the purely unitary dynamics obtained at $\gamma=0$ preserves the entanglement between $Q,R$ for all time. Specifically, if we measure the 2nd R\'enyi entropy of the system as a function of a time, it will remain at its maximal value $S_Q^{(2)}= N \ln 2$ for all time. This is analogous to a `volume-law' phase for the R\'enyi entropy of the system. We can also equivalently consider the purity $\Pi_Q = \exp{\left(-S_Q^{(2)}\right)}$. 

Once we consider a non-zero rate $\gamma$ of weak measurements, the purely unitary dynamics is modified to include processes that degrade entanglement. In particular, for sufficiently large $\gamma$ the measurements will dominate and all the entanglement between the system and the reference will be destroyed, thus \emph{purifying} the system. In this case, $S_Q^{(2)} = 0$ and $\Pi_{Q}=1$ (Fig. \ref{fig:phasediagram}b, dotted purple) \cite{gullans2020dynamical}. 

The R\'enyi entropy $S_Q^{(2)}$ or the purity $\Pi_{Q}$ therefore serve as order parameters for the \emph{purification transition}. Our goal in the remainder of this section is to derive a path integral expression for the purity $\Pi_Q$, computed for a particular analytically-tractable disorder average. We begin by specifying in more detail the quantity of interest.  

Consider a single realization of the circuit $V = V(\mvec{J},\mvec{n})$ which produces a pure unnormalized quantum state $\unrho(V)$ of the system and reference. To calculate the purity $\tr{\unrho^2_Q(V)}$ of the system's reduced density matrix $\unrho_Q (V)= \trover{R}{\unrho(V)}$ for this trajectory, we introduce a second identical copy $\unrho'(V) = \unrho(V)$  of the system and reference with the same post-selected measurement results and identical dynamics and compute the expectation value of the $\mathrm{SWAP}_{QQ'}$ operator \cite{ekert2002direct,horodecki2009quantum,daley2012measuring,islam2015measuring},
\begin{equation}
    \label{eq:z2}
    Z_2(V) \equiv \tr{\unrho_Q^2(V)} = \tr{(\unrho \otimes \unrho') \mathrm{SWAP}_{QQ'}},
\end{equation}
which gives the purity of the unnormalized state $\unrho_Q(V)$. The purity $\Pi_Q = \tr{\rho_Q^2} = Z_2 / P^2$ of the normalized state $\rho_Q = \unrho_Q / \tr{\unrho_Q}$ is obtained simply by dividing $Z_2$ by the squared probability for this trajectory,
\begin{equation}
    \label{eq:p2}
    P^2(V) \equiv \tr{\unrho_Q(V)}^2 = \tr{\unrho \otimes \unrho'}.
\end{equation}
From Eqs.~\eqref{eq:z2} and~\eqref{eq:p2} it is clear that the quantity $Z_2$ differs from $P^2$ only in the presence of the $\mathrm{SWAP}_{QQ'}$ operator. As we shall see, this SWAP operator modifies the initial and final boundary conditions of the system, leading to fundamentally different physics in $Z_2$ and $P^2$.

The normalized purity $\Pi_Q = Z_2 / P^2$ is in principle an experimentally-accessible observable and can be measured in the following way, which we discuss in more detail in Appendix \ref{app:disorderavg}. The experimentalist first fixes the parameters $\mvec{J},\mvec{n}$ and then applies the Brownian circuit dynamics $V(\mvec{J},\mvec{n})$, repeatedly performing the necessary projective measurements until the desired measurement record is obtained (i.e. $+1$ for all $\sigma^y_{\text{aux}}$ measurements). If the $\sigma^y_{\text{aux}}$ outcomes are close to equally likely, this will require a number of experimental runs scaling like $2^{N_{\text{aux}}}$, where $N_{\text{aux}}$ is the total number of auxiliary measurement qubits used over the whole circuit. Each successful run is stored as a quantum state $\unrho(V)$, and then once enough copies of the state have been obtained, the experimentalist can perform SWAP tests to estimate the value of the purity $\Pi_Q(V) = Z_2(V) / P^2(V)$ for this circuit realization $V$. If the purity is expected to be small, this estimate will require many samples to gather sufficient statistics. The total number of experiments required in this brute force approach is thus no more than $2^{a N  t + b N}$, where the $2^{a N t}$ piece represents the $2^{N_{\text{aux}}}$ experimental runs required for post-selection and the $2^{bN}$ piece represents extra copies needed to estimate the purity from SWAP tests.

Next, we can consider sampling the normalized purity $\Pi_Q(V)$ over different circuit realizations $V = V(\mvec{J},\mvec{n})$. The average of these samples then defines the circuit-averaged purity $\overline{\Pi_Q}$.
The circuit-averaged purity can be estimated experimentally by simply repeating the above procedure for each sample $\mvec{J},\mvec{n}$, yielding
\begin{equation}
    \label{eq:puritycircuitavg}
    \overline{\Pi_Q} = \sum_{V}\pi(V) \frac{Z_{2}(V)}{P^{2}(V)},
\end{equation}
where $\pi(V) = \pi(V(\mvec{J},\mvec{n}))$ is the probability for a particular circuit realization $\mvec{J},\mvec{n}$. While the experimental protocol for computing this disorder-averaged quantity is clear, this kind of average is difficult to calculate with, since the random variables $\mvec{J},\mvec{n}$ appear in numerator and denominator of Eq. \eqref{eq:puritycircuitavg}.

In this work we make analytical progress by sampling trajectories differently. A particularly convenient choice is to consider
\begin{equation}
    \label{eq:disavgpurity}
    \langle \Pi_{Q} \rangle = \frac{\langle Z_{2}\rangle}{\langle P^{2}\rangle},
\end{equation}
where both $\langle Z_2 \rangle = \sum_V \pi(V) Z_{2}(V)$ and $\langle P^2 \rangle = \sum_V \pi(V) P^{2}(V)$ are individually averaged over circuit realizations $V = V(\mvec{J},\mvec{n})$. While one might reasonably protest that the disorder-averaged quantity $\langle \Pi_{Q} \rangle$ is not as physical as the quantity $\overline{\Pi_Q}$, we show in Appendix \ref{app:disorderavg} that measuring $\langle \Pi_Q \rangle$ just corresponds to sampling the purity $\Pi_Q(V)$ over trajectories with a different probability distribution $\pi'(V)$ from the usual circuit probability distribution $\pi(V)$. Moreover, we demonstrate that the disorder-averaged quantity $\langle \Pi_Q \rangle$ requires only classical post-processing and no more quantum resources than simulating $\overline{\Pi_Q}$. 

In the rest of the paper, we consider the deterministic weak measurement setup for qubits $S = 1/2$, and suppress the $\langle \ldots \rangle$ notation for $\Pi_{Q}$, $Z_2$ and $P^2$, always referring to the particular averaged quantity whenever $\Pi_{Q}$, $Z_2$ and $P^2$ are considered. Also, we will make statements about the R\'enyi-2 entropy-like quantity $-\ln{\langle \Pi_{Q} \rangle}$ derived from the averaged purity $\langle \Pi_{Q} \rangle$. This is obviously not the same as the averaged R\'enyi-2 entropy of the system, since we are averaging the purity and then taking the logarithm. In the rest of the paper, when referring to the entropy of the system we always refer to the quantity $-\ln{\langle \Pi_{Q} \rangle}$, which is what we can access analytically.

\subsection{Path integral representation}
\label{sec:derivpathintegral}



\begin{figure*}
    \centering
    \includegraphics[width=\textwidth]{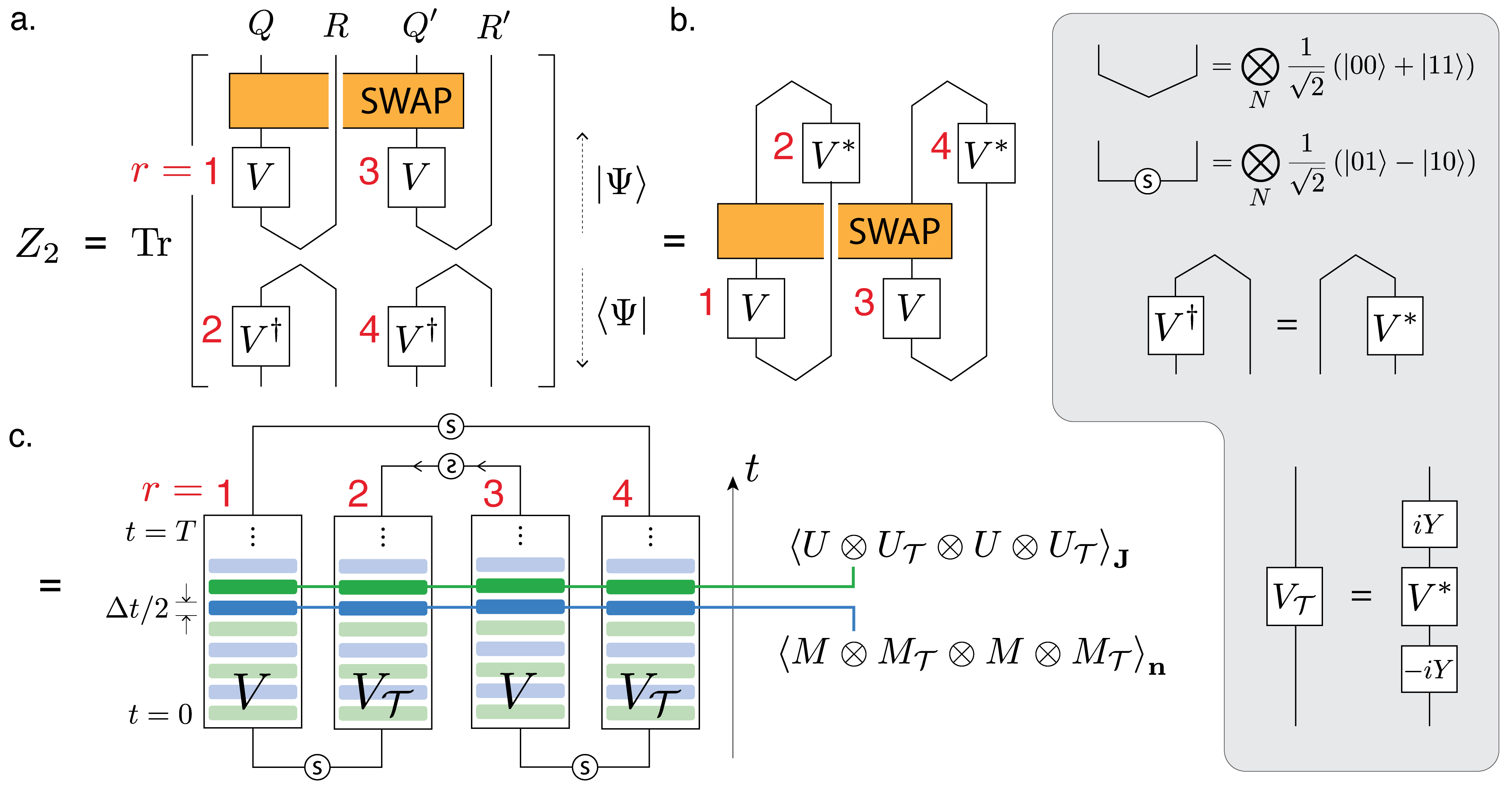}
    \caption{\textbf{Unnormalized purity for the hybrid Brownian circuit.} The purity $Z_2 = \tr{\unrho_Q^2}$ of the unnormalized state $\unrho_Q$ is equivalent to the expectation value of the $\mathrm{SWAP}_{QQ'}$ operator evaluated on two identical copies $\unrho,\unrho'$ of the system. Straightforward rearrangement of the circuit yields pure-state dynamics on four replicas $r = 1,2,3,4$ with nontrivial boundary conditions at $t = 0,T$. Because the Brownian coefficients $\mvec{J},\mvec{n}$ are uncorrelated in time, the disorder averages $\langle \ \cdot \ \rangle_{\mvec{J},\mvec{n}}$ at each circuit layer (solid green, solid blue) can be computed independently. Arrows on the $t = T$ boundary condition indicate a `reversed' singlet state $\ket{(32)} = - \ket{(23)}$. The corresponding circuit for the squared probability $P^2$ is identical except for the SWAP-ed boundary condition at $t = T$.}
    \label{fig:tensornetwork}
\end{figure*}

We have expressed the unnormalized purity $Z_2$ and squared probability $P^2$ in Eqs.~\eqref{eq:z2} and~\eqref{eq:p2} using two identical copies $\unrho' = \unrho$ of the unnormalized state. To express these quantities in path-integral language, we first use the Choi-Jamio\l{}kowski isomorphism to convert the mixed-state dynamics on two copies of the system into pure-state dynamics on four copies \cite{jamiolkowski1972linear,choi1975completely}. In its simplest form, this isomorphism just maps the two-copy density matrix $\ket{\psi} \bra{\psi} \otimes \ket{\psi'} \bra{\psi'}$ to the four-replica pure state $\ket{\psi} \ket{\psi} \ket{\psi'} \ket{\psi'}$. More generally, this isomorphism provides a mapping between quantum operators $\op{}$ acting on a Hilbert space $\mathcal{H}$ and pure quantum states $\ket{\op{}}$ living in a doubled Hilbert space $\mathcal{H} \otimes \mathcal{H}$.

In our calculation this conversion from mixed-state dynamics to doubled pure-state dynamics is easiest to see when the quantities $Z_2,P^2$ are represented graphically using a tensor-network representation as shown in Fig. \ref{fig:tensornetwork}a, where downward-facing external legs represent bras $\bra{\psi}$ and upward-facing legs represent kets $\ket{\psi}$. The two copies of the system $Q,Q'$ are initially maximally entangled with their respective reference systems $R,R'$ via EPR pairs $\ket{\mathrm{EPR}} = \bigotimes_N (\ket{00} + \ket{11}) / \sqrt{2}$. Hybrid Brownian dynamics $V$ are then applied to the system qubits $Q,Q'$, and the $\mathrm{SWAP}_{QQ'}$ operator (orange) exchanges qubits in the two systems to yield the purity $Z_2$.
Each instance of the time-evolution matrix $V$ has been labeled by a \emph{replica index} $r = 1,2,3,4$.

Evaluating the trace and using the identity
\begin{equation}
    \bra{\mathrm{EPR}}(\adj{V} \otimes \id) = \bra{\mathrm{EPR}} (\id \otimes V^*)
\end{equation}
we can bring the purity into the form shown in Fig. \ref{fig:tensornetwork}b. In this form the dynamics may be interpreted as pure-state dynamics on four replicas of the system, where replicas $r = 1,3$ are subject to time-evolution $V$ while replicas $r = 2,4$ are subject to complex-conjugated time-evolution $V^*$. The EPR pairs and $\mathrm{SWAP}_{QQ'}$ operator yield nontrivial boundary conditions on the initial and final quantum states as shown in Fig. \ref{fig:tensornetwork}c. These boundary conditions are analogous to those imposed at the input and output states of Haar-random tensor network models exhibiting a MIPT \cite{choi2020quantum}.


Complex conjugation $V^*$ is naturally related to time-reversal symmetry in quantum mechanics \cite{haake2010quantum}. For spin-1/2 degrees of freedom, the conventional definition of time reversal also includes conjugation by $i \sigma^y$, since complex conjugation alone only reverses the $y$ component of spin, 
\begin{eqnarray}
    (\sigma^x, \sigma^y, \sigma^z)^* = (\sigma^x, -\sigma^y, \sigma^z).
\end{eqnarray}
In replicas $r = 2,4$ we therefore reflect all spin components in the $\sigma^x,\sigma^z$ spin plane via the operator $i Y = i \prod_i \sigma_i^y$, such that
\begin{align}
    V^* &= (-i Y)(iY) V^* (-iY) (iY) \nonumber \\
    &= (-i Y) V_{\mathcal{T}} (iY)
\end{align}
where $V_{\mathcal{T}} = (i Y ) V^* (-i Y)$ is the properly time-reversed version of the time evolution operator $V$ \cite{haake2010quantum}. With this additional coordinate change the spin-1/2 Pauli matrices transform correctly as $\vec{\sigma} \rightarrow \vec{\sigma}_{\mathcal{T}} =  - \vec{\sigma}$ as required for angular momentum vectors under time-reversal $\mathcal{T}$. We regard this additional rotation as a convenient parameterization that makes the $\su{2}$ invariance of the problem manifest in the resulting path integral. 

The remaining factors of $\pm i Y$ introduced into the four-replica circuit by this change of coordinates serve to convert the initial and final EPR pairs into spin singlets:
\begin{align}
    \mp \frac{1}{\sqrt{2}} \left( \id \otimes i \sigma_y \right) \left( \ket{00} + \ket{11} \right)_{rs} &= \pm \frac{1}{\sqrt{2}} \left( \ket{01} - \ket{10} \right) \nonumber \\
    &= \pm \ket{\singlet{rs}}
\end{align}
where $\ket{\singlet{rs}}$ denotes a spin-singlet state between replicas $r,s$. The spin-singlet is antisymmetric under replica exchange $\ket{(rs)} = - \ket{(sr)}$, but because the four-replica circuit features pairs of identical singlet states at $t = 0,T$ these overall negative signs cancel such that $Z_2,P^2$ are always positive. The unnormalized purity $Z_2$ is initialized with spin-singlet pairs $\ket{(12)(34)}$ entangling replicas 1-2 and 3-4 as shown in Fig. \ref{fig:tensornetwork}c, while
the final state $\ket{(14)(32)}$ has spin singlets entangling replicas 1-4 and 3-2 due to the $\mathrm{SWAP}_{QQ'}$ operator. These non-equal boundary conditions in $Z_2$ lead to most of the interesting physical consequences explored in this work. By contrast, without the $\mathrm{SWAP}_{QQ'}$ operator the squared probability $P^2$ has identical singlet-pair states $\ket{(12)(34)}$ at both initial and final times.

With the unnormalized purity $Z_2$ and squared probability $P^2$ expressed in terms of pure-state dynamics on four replicas $r = 1,2,3,4$, we now perform the disorder average over the Brownian coefficients $\mvec{J},\mvec{n}$ and show that this leads to path-integral expressions for $Z_2,P^2$.
Because the disorder $\mvec{J},\mvec{n}$ is uncorrelated in time due to the delta function $\delta_{tt'}$, we may compute the disorder average for each circuit layer $U(t), M(t)$ separately as shown in Fig. \ref{fig:tensornetwork}c. Expanding each $U(t) = \exp{[-i H(t) \Delta t / 2]}$ to lowest order in $\Delta t$, the disorder average over a single unitary layer yields
\begin{widetext}
\begin{eqnarray}
    \left \langle U \otimes U_{\mathcal{T}} \otimes U \otimes U_{\mathcal{T}} \right \rangle_{\mvec{J}} 
    & =& 1 - \frac{\Delta t^2}{4} \sum_{r < s} (-1)^{r+s} \left \langle H^r H^s \right \rangle_{\mvec{J}} - \frac{1}{2} \frac{\Delta t^2}{4} \sum_r \left \langle H^r H^r \right \rangle_{\mvec{J}} + \mathcal{O}(\Delta t^4) \nonumber \\
    \left \langle H^r H^s \right \rangle_{\mvec{J}} &=& \frac{J}{\Delta t} \frac{N}{(S+1)^{4}} \left( \frac{1}{N} \sum_i \mvec{S}_i^r \cdot \mvec{S}_i^s \right)^2.
\end{eqnarray}
\end{widetext}
where terms linear in $\Delta t$ vanish because $\mvec{J}$ has zero mean and where $\mvec{S}_i^r \cdot \mvec{S}_i^s \equiv \sum_{\alpha} S^{\alpha,r}_i S^{\alpha,s}_i$ is the standard dot product.
The second line holds as an operator equation, where $H^{r,s}$ denote copies of the Hamiltonian \eqref{eq:brham} acting on replicas $r,s = 1,2,3,4$. The factor of $\Delta t$ in the denominator comes from the regularization of the white-noise random variables Eq. \eqref{eq:jrandvar}, while the overall factor of $N$ comes from the sum $\sum_i$ over spins and is ultimately responsible for large-$N$ control. The replica-dependent factor $(-1)^{r+s}$ is a consequence of the time-reversed dynamics in replicas $r = 2,4$, and is a crucial feature of the field theory governing the MIPT.

We can therefore express each disorder-averaged unitary circuit layer as a propagator $I_J[\mvec{S}_i^r]$ over $4N$ spins $\mvec{S}_i^r$:
\begin{align}
    \label{eq:disavgpropu}
    &\left \langle U \otimes U_{\mathcal{T}} \otimes U \otimes U_{\mathcal{T}} \right \rangle_{\mvec{J}} = e^{- N I_J[\mvec{S}_i^{r}] \Delta t}, \nonumber \\
    I_J[\mvec{S}_i^r] &\equiv \frac{J}{4 (S+1)^{4}} \sum_{r<s} (-1)^{r+s} \left(\frac{1}{N}\sum_{i}\mvec{S}_{i}^{r}\cdot \mvec{S}_{i}^{s}\right)^{2} \nonumber \\
    &+ \frac{J S^{2}}{2 (S+1)^{2}}
\end{align}
which holds as an operator equation to lowest order in $\Delta t$. Similar manipulations for the non-unitary circuit layers yield a propagator $I_{\gamma}[\mvec{S}_i^r]$
\begin{align}
    \label{eq:disavgpropm}
    &\left \langle M \otimes M_{\mathcal{T}} \otimes M \otimes M_{\mathcal{T}} \right \rangle_{\mvec{n}} = e^{- N I_{\gamma}[\mvec{S}_i^{r}] \Delta t}, \nonumber \\
    I_{\gamma}[\mvec{S}_i^r] &\equiv \frac{ -\gamma}{2 (S+1)^{2}} \sum_{r<s} (-1)^{r+s} \left(\frac{1}{N}\sum_{i}\mvec{S}_{i}^{r}\cdot \mvec{S}_{i}^{s}\right) \nonumber \\
    &+ \frac{\gamma S}{(S+1)}
\end{align}
where there is a relative minus sign in the first term compared to Eq. \eqref{eq:disavgpropu}. In terms of the propagators $I_J,I_{\gamma}$, the disorder-averaged unnormalized purity $Z_2$ (or squared probability $P^2$) is given by a stack of alternating unitary and non-unitary propagators with appropriate boundary conditions, i.e.
\begin{align}
    \label{eq:z2p2propagator}
    &\left \langle Z_2 \ \mathrm{or} \ P^2 \right \rangle_{\mvec{J},\mvec{n}} \nonumber \\
    &= \bra{\psi_T} \prod_t e^{-N I_{\gamma} \Delta t} e^{-N I_{J} \Delta t} \ket{\psi_0}
\end{align}
where the initial and final states
\begin{equation}
    \ket{\psi_0}, \ket{\psi_T} = \{ \ket{(12)(34)}, \ket{(14)(32)} \}
\end{equation}
are the singlet-pair states enforcing the non-uniform boundary conditions that distinguish $Z_2$ from $P^2$. Notice that both the unitary and non-unitary propagators in Eq. \eqref{eq:z2p2propagator} are preceded by a factor $N$ which allows for analytical control over fluctuations in the thermodynamic limit.

In performing the disorder average we have exchanged inter-site couplings $S^{\alpha}_i S^{\beta}_j$ in the original Hamiltonian \eqref{eq:brham} for inter-replica couplings $\mvec{S}_i^r \cdot \mvec{S}_i^s$ in the propagator \eqref{eq:z2p2propagator}. As a consequence,  the propagators $I_J,I_{\gamma}$ are functions only of the mean-field variables
\begin{equation}
    \label{eq:meanfields}
    G_{rs} = \left( \frac{1}{N} \sum_{i} \mvec{S}_i^r \cdot \mvec{S}_i^s \right) / (S+1)^2
\end{equation}
with $r < s$, which mediate all spin-spin interactions. Because these mean fields consist of a large number of independent and identical degrees of freedom, their dynamics is highly classical with fluctuations controlled by the system size $N$.
These simplifying features are typical of disorder-average calculations performed in the context of mean-field spin glass theory \cite{mezard1987spin,castellani2005spin} and the SYK model \cite{sachdev1993gapless,Kitaev2015,maldacena2016remarks}, where the high amount of connectivity between degrees of freedom naturally leads to mean-field behavior. Because our Brownian interactions are all-to-all, a similar phenomenon occurs in our hybrid model and the physics can be captured by the mean fields $G_{rs}$. 

Formally, we convert the propagator \eqref{eq:z2p2propagator} to a path integral by introducing an over-complete basis of coherent spin states $\ket{\mvec{\Omega}}_i^r$ parameterized by $\mathrm{SO}(3)$ unit vectors $\mvec{\Omega}$ and satisfying the eigenvalue equation
$\mvec{\Omega} \cdot \mvec{S}_i^r \ket{\mvec{\Omega}}_i^r = S \ket{\mvec{\Omega}}_i^r$
for each spin $\mvec{S}_i^r$ \cite{arecchi1972atomic,fradkin2013field}.
Using this basis we insert resolutions of the identity
\begin{equation}
    \id = \int \frac{2S+1}{4 \pi} d^2 \mvec{\Omega}_i^r \ket{\mvec{\Omega}_i^r} \bra{\mvec{\Omega}_i^r}
\end{equation}
at each timestep $\Delta t$ following the usual rules of path integration. This effectively converts the spin operators in the propagators $I_J,I_{\gamma}$ into classical $\mathrm{SO}(3)$ vectors $\mvec{S}_i^r \rightarrow (S+1) \mvec{\Omega}_i^r$. The standard path integral derivation also generates `kinetic' or Berry phase terms $\sim \Omega \partial_t \Omega$ in the path integral coming from overlaps $\bracket{\mvec{\Omega}_i^r(t)}{\mvec{\Omega}_i^r(t+\Delta t)}$ of the coherent spin states at consecutive time steps \cite{fradkin2013field} (see Appendix \ref{app:coherentstatepathint} for more details).

Next, to enforce the identification \eqref{eq:meanfields} we introduce six time-dependent mean fields $G_{rs}(t)$ and six Lagrange multiplier fields $F_{rs}(t)$ into the path integral via the identity
\begin{align}
    \label{eq:deltafunctiongrs}
    1 = & \int \prod_{r<s} \mathcal{D} F_{rs} \mathcal{D} G_{rs} \\
    & \exp \left[ \int dt \sum_{r<s} i F_{rs} \left( G_{rs} - \frac{1}{N} \sum_i \mvec{\Omega}_i^r \cdot \mvec{\Omega}_i^s \right) \right] \nonumber
\end{align}
With this delta-function constraint now explicit in the path integral, we may simply substitute the mean fields $G_{rs}$ for any mean-field Heisenberg terms $\sum_i \mvec{\Omega}_i^r \cdot \mvec{\Omega}_i^s / N$ that appear in the propagators $I_J, I_{\gamma}$. In particular, the unitary part $I_J$ of the path integral Eq. \eqref{eq:disavgpropu} contributes quadratic terms $\propto J G_{rs}^2$ while the non-unitary part $I_{\gamma}$ Eq. \eqref{eq:disavgpropm} contributes linear terms $\propto \gamma G_{rs}$. After making this replacement the only spins $\mvec{\Omega}_i^r$ explicitly remaining in the path integral are those coupled to the Lagrange multiplier fields $iF_{rs}$ coming from the delta-function constraint Eq. \eqref{eq:deltafunctiongrs}.

Thus in the limit $\Delta t \rightarrow 0$ at fixed $T$ we finally arrive at
\begin{widetext}
\begin{eqnarray}
    \label{eq:pathintegral}
    Z_{2} \ \mathrm{or} \ P^2 &=& \int \left(\prod_{r<s}\mathcal{D}F_{rs}\mathcal{D}G_{rs}\right)\exp{\left[-N I\left[F_{rs},G_{rs}\right]\right]}\nonumber\\
    I[F_{rs},G_{rs}] &=& \int_0^T dt\left[ \frac{J}{4} \sum_{r<s} (-1)^{r+s} G_{rs}^{2} - \frac{\gamma}{2}\sum_{r<s}(-1)^{r+s}G_{rs}-i\sum_{r<s}F_{rs}G_{rs}\right]-\ln K[F_{rs},\psi_0,\psi_T] \nonumber\\
    K[F_{rs},\psi_0,\psi_T] &=& \bra{\psi_T} \exp{\left[- \int_0^T dt \sum_{r < s} i F_{rs} \frac{\mvec{S}^r \cdot \mvec{S}^s}{(S+1)^{2}} \right]} \ket{\psi_0}.
\end{eqnarray}
\end{widetext}
where we have expressed the path integral over spins $\mvec{S}^r$ as a time-ordered exponential propagator $K[F_{rs},\psi_0,\psi_T]$ and the choice of boundary states $\ket{\psi_0},\ket{\psi_T}$ determines whether the expression corresponds to $Z_2$ or $P^2$. The path integral expression \eqref{eq:pathintegral} is the main technical result of this section.
Here the quadratic terms in the action $\propto J G_{rs}^2$ correspond to the unitary part of the dynamics while the linear terms $\propto \gamma G_{rs}$ correspond to the non-Hermitian weak measurement part. The competition between these two terms as a function of $\gamma / J$ is what drives the measurement-induced phase transition in this model. 


Due to the overall factor of $N$ preceding the action $I[F_{rs},G_{rs}]$ in Eq. \eqref{eq:pathintegral}, we can evaluate the path integral via steepest-descent when $N$ is large. In this limit, the action $I$ may be viewed as a classical Lagrangian where the first two terms describe effective potential energies for the mean fields $G_{rs}$ and the third term is a coupling between the $G_{rs}$ and the Lagrange multipliers $F_{rs}$. The propagator $K$ evolves an initial state $\ket{\psi_0}$ of four spin-1/2 degrees of freedom $\mvec{S}^r$ to a final state $\ket{\psi_T}$ under the influence of time-dependent external fields $i F_{rs}(t)$. We perform this steepest-descent analysis for time-independent fields $F_{rs},G_{rs}$ in section \ref{sec:bulksp}, and consider time-dependent fields leading to instanton transitions in section \ref{sec:instantons}. This large-$N$ analysis yields an analytically tractable description of the system over a large portion of the phase diagram shown in Fig. \ref{fig:phasediagram}.

Although our focus in this work is on the simplest mean-field model \eqref{eq:pathintegral} featuring bilinear spin interactions ($p = 2$), single-spin weak measurements ($q = 1$), and probed by the purity $\tr{\unrho_Q^n}$ with $n = 2$, we show in Appendix \ref{app:deriv_pqpathintegral} how this $(2,1)$ model can be easily generalized to arbitrary $(p,q)$ hybrid Brownian models featuring higher-order interaction and measurement terms and probed by $n$th-order moments $\tr{\unrho_Q^n}$ of the density matrix. From this more general derivation we find that the choice of $p,q$ only affects the $G_{rs}$-dependent part of the action $I$, while the propagator $K$ is unaffected by the order of the unitary interactions or weak measurements. By contrast, the parameter $n$ changes the number of replicas $r,s$ but otherwise leaves the path integral \eqref{eq:pathintegral} unchanged. We reserve the study of these more general models for future work.


\subsection{Simplification of spin-1/2 propagator $K$}
\label{sec:spinpropagatorK}

\begin{figure}
    \centering
    \includegraphics[width=0.35\textwidth]{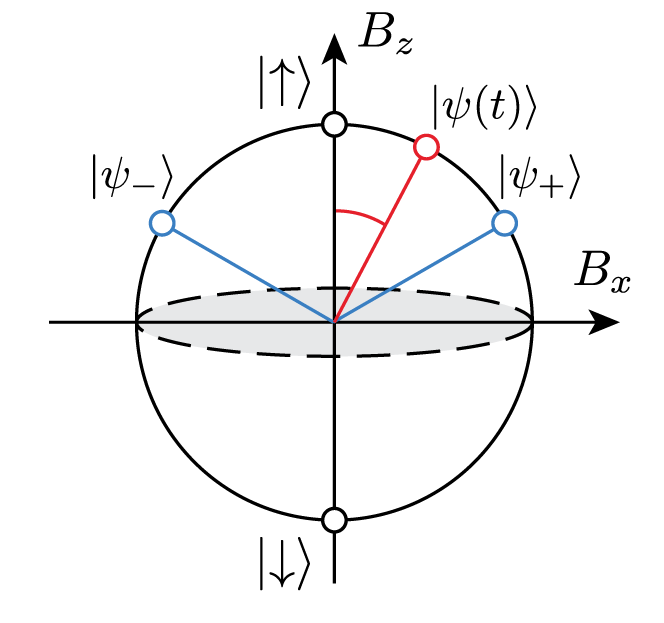}
    \caption{\textbf{Bulk two-level r-bit subspace $\ket{\uparrow},\ket{\downarrow}$}. For $S=1/2$, the $\su{2}$ symmetry of the problem kinematically constrains the dynamics to a single effective qubit or \emph{r-bit} $\ket{\psi(t)}$ (red) living in the $\mvec{S}^{\mathrm{Tot}} = 0$ subspace spanned by $\ket{\uparrow},\ket{\downarrow}$. The r-bit's trajectory $\ket{\psi(t)}$ must begin on the singlet-pair state $\ket{\psi_+} = \ket{(12)(34)}$, and end on the same singlet-pair state for $P^2$ or on the SWAP-ed singlet-pair state $\ket{\psi_-} = \ket{(14)(32)}$ for $Z_2$.}
    \label{fig:blochsphere}
\end{figure}

Before performing a saddle-point analysis of the action \eqref{eq:pathintegral} at large-$N$, however, it is convenient to first simplify the four-replica propagator $K[F_{rs},\psi_0,\psi_T]$ for the special case $S = 1/2$.
This case is particularly easy to calculate because the Heisenberg coupling terms $\mvec{S}^r \cdot \mvec{S}^s$ in the propagator as well as the initial and final singlet-pair states are all manifestly $\su{2}$ invariant. In fact, because the initial and final states $\ket{(12)(34)}$, $\ket{(14)(32)}$ are pairs of $\su{2}$ spin singlets, the four-replica system is constrained at all times to live in the subspace of total spin $\mvec{S}^{\mathrm{Tot}} = 0$, in which the total spin operators $\mvec{S}^{\mathrm{Tot}} = \sum_r \mvec{S}^r$, summed over all four replicas, act trivially and the Casimir operator $\mvec{S}^{\mathrm{Tot}} \cdot \mvec{S}^{\mathrm{Tot}} = \mathcal{S} (\mathcal{S} + 1)$ has eigenvalue $\mathcal{S} = 0$. For four spin-1/2 degrees of freedom $\mvec{S}^r$ with $r = 1,2,3,4$ this subspace is two-dimensional and is spanned by the basis vectors
\begin{align}\label{eq:basis2spins}
    \ket{\uparrow} &= \frac{1}{2 \sqrt{3}} \left(2\ket{1010} + 2\ket{0101} - \ket{0011} - \ket{1100} \right. \nonumber \\
    & \left. \quad \quad \quad \quad \quad \quad \quad \quad - \ket{1001} - \ket{0110} \right) \nonumber \\
    \ket{\downarrow} &= \frac{1}{2} \left( \ket{0011} + \ket{1100} - \ket{1001} - \ket{0110} \right)
\end{align}
which transform trivially under global $\su{2}$ rotations generated by the total spin operators $\mvec{S}^{\mathrm{Tot}}$. We view this two-dimensional subspace of the replica space as encoding an effective two-level system $\ket{\psi(t)} = \psi_{\uparrow}(t) \ket{\uparrow} + \psi_{\downarrow}(t) \ket{\downarrow}$ which we refer to as a \emph{replica-bit} or \emph{r-bit}. The instantaneous state of the r-bit may be drawn on a Bloch sphere as shown in Fig. \ref{fig:blochsphere}.

In the two-dimensional basis $\ket{\uparrow}, \ket{\downarrow}$ the initial and final boundary conditions may be simply written in terms of the states
\begin{equation}\label{eq:pm_states}
    \ket{\psi_{\pm}} = \frac{\sqrt{3}}{2} \ket{\uparrow} \pm \frac{1}{2} \ket{\downarrow},
\end{equation}
where the initial and final states $\ket{\psi_0},\ket{\psi_T}$ take the values:
\begin{equation}
\label{eq:boundaryconditionsz2p2}
\begin{tabular}{ |m{0.5cm}||m{1cm}|m{1cm}|  }
 \hline
  & \ $t = 0$ & \ $t = T$ \\
 \hline
 $Z_2$ & $\ \ket{\psi_+}$ & $\ \ket{\psi_-}$ \\
 $P^2$ & $\ \ket{\psi_+}$ & $\ \ket{\psi_+}$ \\
 \hline
 \end{tabular}
 \end{equation}
such that $Z_2, P^2$ differ only in the final boundary conditions at $t = T$.
The difference between the non-uniform boundary conditions for $Z_2$ compared to the uniform boundary conditions for $P^2$ will be crucial in distinguishing between the mixed and purified phases.

The propagator in the two-dimensional r-bit subspace simplifies to
\begin{align}
    &K[\vec{B},\psi_0,\psi_T] = \nonumber \\ 
    & \bra{\psi_T} \exp{\left[ \frac{1}{2} \int_0^T dt \ \vec{B}(t) \cdot \vec{\sigma} \right]} \ket{\psi_0} e^{B_0 T / 2}
\end{align}
where $\vec{\sigma}$ are the $2 \times 2$ Pauli matrices acting on the r-bit $\ket{\uparrow},\ket{\downarrow}$ subspace and $\vec{B}(t)$ is a time-dependent `magnetic field' with components
\begin{align}
    B_x &= \frac{2}{3\sqrt{3}} \left(i F_{12} + i F_{34} - i F_{14} - i F_{23} \right) \nonumber \\
    B_y &= 0 \nonumber \\
    B_z &= \frac{2}{9} \sum_{r<s} i F_{rs} - \frac{2}{3} \left( i F_{13} + i F_{24} \right)
\end{align}
and where terms proportional to the identity within the r-bit subspace have been collected into the term
\begin{equation}
    B_0  = \frac{2}{9} \sum_{r < s} i F_{rs}.
\end{equation}
The time-dependent bulk fields $B_x(t),B_z(t)$ encode the relevant mean-field dynamics of the r-bit $\psi(t)$, and in general must execute nontrivial motions in the bulk in order to satisfy the non-equal boundary conditions $\ket{\psi_0},\ket{\psi_T}$. By contrast, the remaining fields in the action $I$ appear simply as quadratic Gaussian fields and may therefore be trivially integrated out of the path integral, leading to the effective action
\begin{widetext}
\begin{align} \label{eq:2fieldPathIntegralmain}
     Z_{2} \ \mathrm{or} \ P^2 &= \int \mathcal{D}B_x\mathcal{D}B_z \exp{\left[-N I[\vec{B}]\right]}\nonumber\\
     I[\vec{B}] &= \int_{0}^{T} dt \left[\frac{27 B_{x}^{2}}{4 J} - \frac{81 B_{z}^{2}}{4 J} + B_{z}(1+18\gamma)-\frac{J}{72}-\frac{4\gamma^{2}}{J}-\frac{\gamma}{2}\right]-\ln K[\vec{B},\psi_0,\psi_T]\nonumber\\
     K[\vec{B},\psi_0,\psi_T] & = \bra{\psi_{T}}\exp\left[\frac{1}{2}\int_{0}^{T} dt \left(B_x \sigma_{x}+B_z \sigma_{z}\right)\right]\ket{\psi_{0}}.
\end{align}
\end{widetext}
which is a simplification of the general path integral \eqref{eq:pathintegral} for the special case $S = 1/2$.

In this form, one can view the magnetic field variables $\vec{B}(t)$ as `guiding fields' for the bulk r-bit, in the sense that the propagator $- \ln K$ in \eqref{eq:2fieldPathIntegralmain} is minimized when the r-bit $\ket{\psi(t)}$ is in the instantaneous ground state of the effective `magnetic-field' Hamiltonian $H({\vec{B}}) = -\vec{B}(t) \cdot \vec{\sigma} / 2$ appearing in the propagator $K$. As a result of this coupling between the magnetic field variables $\vec{B}(t)$ and the bulk r-bit $\ket{\psi(t)}$, we expect the fields $\vec{B}(t)$ to be strongly time-dependent near $t = 0,T$ in order to guide the r-bit $\ket{\psi(t)}$ to its appropriate boundary conditions $\ket{\psi_{0,T}}$. We see these expectations borne out in gradient-descent numerics in section \ref{sec:saddlepointanalysis}.

A crucial ingredient in the path integral representation of $Z_2$ or $P^2$
in Eq. \eqref{eq:2fieldPathIntegralmain} is the contour of integration of the $B_x$ and $B_z$ fields. Due to the minus sign preceding the $B_z^2$ term in Eq. \eqref{eq:2fieldPathIntegralmain} we conclude that for the path integral to be well defined, $B_x$ must be integrated along the real axis, while $B_z$ must be integrated along the imaginary axis. We shall discuss this issue of contour integration more fully in section \ref{sec:bulksp}.
Note also that until now we have not made any assumptions about the time-dependence of the fields $\vec{B}(t)$. The simplified expression \eqref{eq:2fieldPathIntegralmain} for the path integral over the fields $B_x,B_z$ follows solely from the $\su{2}$ symmetry of the four-replica propagator $K$ and the fact that the boundary states $\ket{\psi_0},\ket{\psi_T}$ belong to a particular spin sector.

For the rest of the paper we will focus on the properties of the path integral \eqref{eq:2fieldPathIntegralmain}, which we shall discuss in terms of classical field configurations $\vec{B}(t)$. But it should always be borne in mind that this is only a particular example of the more general path integral expression \eqref{eq:pathintegral} and further higher-body generalizations.


\subsection{Replica symmetry}
\label{sec:replicasymm}

The pure-state dynamics on four replicas shown in Fig. \ref{fig:tensornetwork}c possesses a number of discrete symmetries. The microscopic bulk dynamics $\mathbb{V} = V^{(1)}\otimes V^{(2)}_{\mathcal{T}}\otimes V^{(3)}\otimes V^{(4)}_{\mathcal{T}}$ on replicas $r = 1,2,3,4$ is manifestly invariant under the replica symmetry group
\begin{equation}
    \label{eq:replicasymmgroup}
    G = (S_2\times S_2)\rtimes \mathbb{Z}_{2},
\end{equation}
where the inner $S_2 \cong \mathbb{Z}_2$ denote symmetric groups permuting the time-reversed or non-time-reversed replicas amongst themselves with generators $\sigma: 1 \leftrightarrow 3$ and $\sigma': 2 \leftrightarrow 4$ \cite{nahum2020measurement}.
The outer $\mathbb{Z}_2$ in the semidirect product is generated by an operation $\tau$ corresponding to time-reversal $\mathcal{T}$ on all four replicas followed by exchange of even and odd replicas $1 \leftrightarrow 2$, $3 \leftrightarrow 4$, where $\sigma' = \tau \sigma \tau$. 
Crucially, the generator $\tau$ is antiunitary; as we discuss in Appendix \ref{app:replicasymm}, this fact constrains the spectrum of $\mathbb{V}$ to be real or for for its eigenvalues to come in complex-conjugate pairs. This is the same mechanism that guarantees the reality of the spectrum in non-Hermitian PT-symmetric quantum mechanics \cite{bender1998real,bender2005introduction,gopalakrishnan2020entanglement,garcia2021replica}. 

To be explicit we can express the symmetry generators $\sigma,\tau$ directly in terms of their effects on the path integral expressions \eqref{eq:pathintegral} and \eqref{eq:2fieldPathIntegralmain}. In Eq. \eqref{eq:pathintegral} the generator $\sigma$ simply exchanges spins $\mvec{S}_i^1 \leftrightarrow \mvec{S}_i^3$ in the propagator $K$, while the generator $\tau$ exchanges even and odd replicas and flips the sign of all spins $\mvec{S}_i^1 \leftrightarrow -\mvec{S}_i^2$, $\mvec{S}_i^3 \leftrightarrow -\mvec{S}_i^4$. 
If we ignore the boundary conditions $\ket{\psi_{0,T}}$, each of these transformations can be undone by an appropriate redefinition of the fields $F_{rs},G_{rs}$, leaving the bulk action $I$ invariant. In the reduced spin-1/2 action Eq. \eqref{eq:2fieldPathIntegralmain}, $\sigma$ generates a reflection about $\sigma^z$ in the r-bit subspace:
\begin{equation}
    \sigma: \vec{\sigma} \rightarrow \sigma^z \vec{\sigma} \sigma^z
\end{equation}
Ignoring the boundary conditions, this transformation can be undone by a redefinition of the $x$-component of the magnetic field $B_x \rightarrow -B_x$, which leaves \eqref{eq:2fieldPathIntegralmain} invariant. The action is trivially invariant under $\tau$ as this generator acts trivially in the r-bit space $\ket{\uparrow},\ket{\downarrow}$.

The boundary conditions $\ket{\psi_{\pm}} = \ket{(12)(34)},\ket{(14)(32)}$ break the replica symmetry group down to a subgroup $H \subset G$ generated by the mutually-commuting generators $\tau,c$, where
\begin{equation}
    c \equiv \sigma \tau \sigma
\end{equation}
corresponds to performing time-reversal $\mathcal{T}$ on all four replicas followed by a `reflection' in replica space $1234 \leftrightarrow 4321$. The generators $\tau,c$ leave the boundary states $\ket{(12)(34)},\ket{(14)(32)}$ invariant, while the generator $\sigma$ transforms these two states into one another:
\begin{equation}
    \sigma \ket{(12)(34)} = \ket{(14)(32)}
\end{equation}
The generator $\sigma$ therefore represents a $\mathbb{Z}_2$ symmetry generator of the bulk symmetry group $G$ that is explicitly broken by the boundary states.

We shall find in the next section that this same $\mathbb{Z}_2$ symmetry is also spontaneously broken in the bulk of the four-replica system below the critical point $\gamma < \gamma_c$, leading to two ordered phases that transform into one another via the generator $\sigma$. By imposing non-equal boundary conditions $\ket{(12)(34)},\ket{(13)(24)}$ at times $t = 0,T$ that explicitly break the $\sigma$-symmetry, we force the system in the ordered phase to transition somewhere in the bulk between the two symmetry-broken ordered phases via a domain wall or `kink'. This is entirely analogous to imposing non-equal boundary conditions on either end of a conventional Ising chain (in dimension $d > 1$) in thermal equilibrium \cite{bao2020theory}. In both cases, the bulk systems undergo spontaneous $\mathbb{Z}_2$ symmetry breaking transitions as a function of measurement rate or temperature, respectively, with an ordered phase below the critical point. In the ordered phase, the non-equal boundary conditions force the creation of domain walls or `kinks' in the bulk where the system rapidly transitions from one symmetry-broken phase to the other in order to satisfy the boundary conditions. We shall see this picture emerge explicitly for the hybrid Brownian model in the next section, where we study the path integral Eq. \eqref{eq:2fieldPathIntegralmain} in the large-$N$ limit.

\section{Measurement-induced purification at large $N$ in the $(2,1)$ hybrid Brownian circuit}

\label{sec:saddlepointanalysis}

The path integral \eqref{eq:pathintegral}, along with its simplification \eqref{eq:2fieldPathIntegralmain} for spin-1/2 degrees of freedom, expresses the disorder-averaged purity $Z_2$ for hybrid Brownian dynamics in terms of mean-field variables $F_{rs},G_{rs}$ (or $\vec{B}$) whose fluctuations are controlled by the large parameter $N$. In the thermodynamic limit $N\rightarrow \infty$, the factor of $N$ preceding the action $I$ in the path integral \eqref{eq:pathintegral},\eqref{eq:2fieldPathIntegralmain} allows for analysis via steepest-descent methods (also known as saddle-point or stationary phase methods \cite{zinn2002quantum}). At infinite $N$, the leading contribution to the path integral comes from the dominant saddle point (or saddle point manifold). This section primarily focuses on this regime of the path integral, meaning large $N$ at fixed $T$. We will also focus on the regime where $T > J^{-1},\gamma^{-1}$ is larger than the various microscopic time-scales in the problem, which is a quasi-steady-state regime in which early time transients have died away.

At finite $N$, there are two kinds of corrections to the leading saddle point answer, fluctuations around the saddle point, which are perturbative in $1/N$, and additional subleading saddles, which are non-perturbative in $1/N$, e.g. $e^{-N} = e^{-1/(1/N)}$. We leave the study of perturbative corrections from fluctuations to future work. We do, however, consider contributions from subleading saddles in section \ref{sec:latetime}. These give rise to important new effects in the long-time limit at fixed $N$, leading to the disintegration of the mixed phase.

Our main purpose in this section is to analyze the path integrals for $Z_2$ and $P^2$ in the saddle-point approximation using a combination  of analytical and numerical tools. We will show that the model exhibits two phases separated by a continuous phase transition as a function of the measurement strength $\gamma$. The boundary conditions at $t = 0,T$ deserve special attention as they are a departure from typical large-$N$ or saddle-point analyses. Typically, such calculations are concerned with the equilibrium physics of a Hamiltonian $H$ at inverse temperature $\beta$ with the path integral constructed to compute the partition function $Z = \tr{e^{-\beta H} \ldots}$. Because of the trace, such a path integral naturally has time-translation symmetry, and the relevant saddle points may usually be taken to be time independent. In our case, however, the propagator $K[\vec{B},\psi_0,\psi_T]$ explicitly breaks time translation invariance, and we are forced to consider time-dependent saddle points in the analysis. Nevertheless, when $T$ is large compared to microscopic scales, the relevant saddle-point configurations will be approximately time-independent for a majority of the time domain.


The plan for the remainder of this section is as follows. In sections \ref{sec:bulksp} and \ref{sec:instantons} we outline the different components that are used to construct saddle-point solutions. These components include `bulk' configurations which are time-independent, instanton-like configurations that are localized in time, and boundary effects which are concentrated near the boundaries $t = 0,T$. In section \ref{sec:phasesofpathint} we use these components to show the existence of and analyze two distinct phases in the purity. In section \ref{sec:criticalexps} we consider the critical point between these two phases and analyze the resulting effective field theory of the transition. In section \ref{sec:timedependencepurity} we consider the dynamics of the purity at early times from the perspective of numerical gradient descent and exact diagonalization. Finally, in section \ref{sec:latetime} we comment on late times at fixed $N$, which we analyze by summing over subleading saddles with multi-instanton configurations.

\subsection{Components of saddle-point configurations}

\subsubsection{Bulk (time-independent) configurations}
\label{sec:bulksp}

While we expect general field configurations $\vec{B}(t)$ to be time-dependent, especially near the boundaries $t = 0,T$, we first consider the physics of the path integral \eqref{eq:2fieldPathIntegralmain} deep in the bulk, i.e. dynamics occurring at times $1/J \ll t \ll T$ very far from either the initial or final boundary. Because the propagator $K$ is local in time, fields $\vec{B}(t)$ deep in the bulk are largely unaffected by the faraway $t = 0,T$ boundary conditions. Moreover, time-dependent variations $|\partial_t \vec{B}| > 0$ are penalized in the action $\eqref{eq:2fieldPathIntegralmain}$ via kinetic-energy terms in the propagator $K$. We therefore expect fields deep within the bulk to be time-independent.

For a time-independent magnetic field $\vec{B}(t) = \vec{B}$, the propagator \eqref{eq:2fieldPathIntegralmain} is easy to evaluate and one obtains
\begin{align}
    K \approx \exp\left(\frac{TB}{2}\right)+\exp\left(-\frac{TB}{2}\right),
\end{align}
where $B \equiv \sqrt{B_x^2+B_z^2}$ and where we have dropped the contributions from the boundary states $\ket{\psi_0},\ket{\psi_T}$, which are subdominant in the limit of large $T$ (and have not been treated properly by our assumption of time-independence anyway). Assuming real and positive $B$ and large $T$, we may simply replace $\ln K \to BT/2$ in the action $I[\vec{B}]$ \eqref{eq:2fieldPathIntegralmain}.

\begin{figure}
    \centering
    \includegraphics[width=\columnwidth]{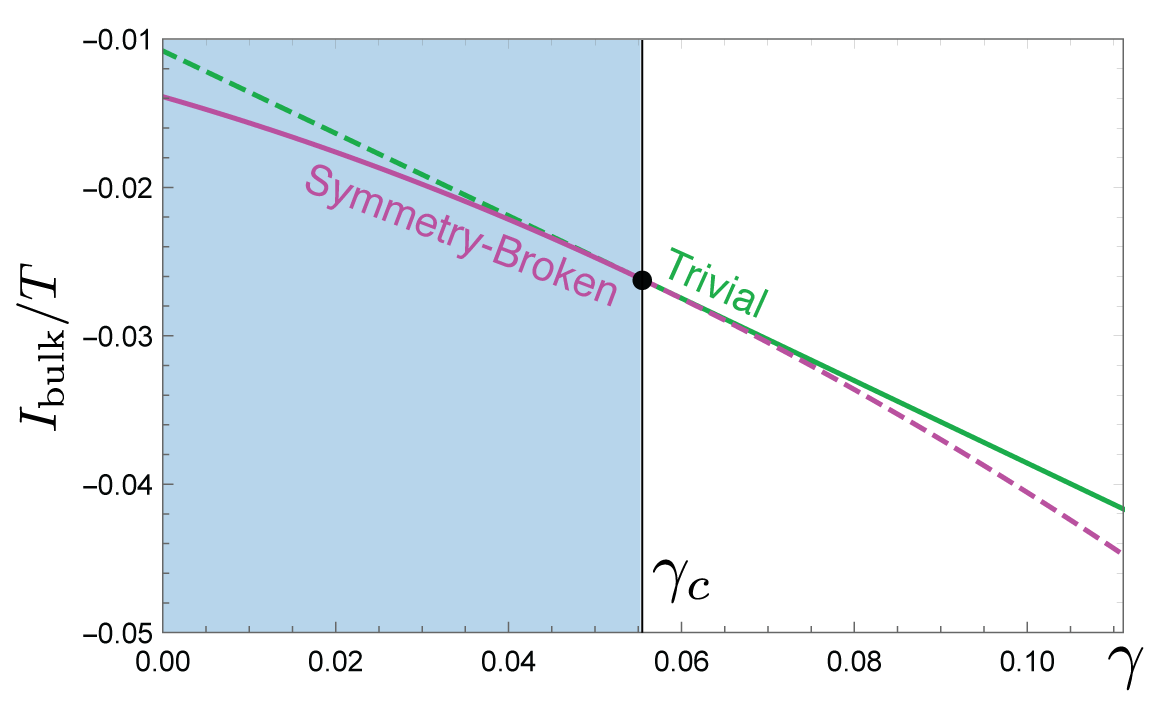}
    \caption{\textbf{Classical bulk action for the trivial and symmetry-broken saddles.} For all values of $\gamma$ the bulk action $I_{\mathrm{bulk}}$ for the symmetry-broken saddles (purple) is always smaller than the trivial saddle (green), but above the critical point $\gamma>\gamma_c$ the symmetry-broken saddle points (dotted purple) are imaginary and do not contribute to the path integral (see Fig. \ref{fig:sbulk_contour}). The symmetry-broken saddle points (solid purple) therefore dominate below the critical point $\gamma < \gamma_c$ while the trivial saddle point (solid green) dominates above $\gamma > \gamma_c$. This smooth exchange of saddle-point dominance at $\gamma = \gamma_c$ is responsible for the second-order measurement-induced purification transition in the $(2,1)$ model.}
    \label{fig:sbulk_comp}
\end{figure}

\begin{figure*}
    \centering
    \includegraphics[width=\textwidth]{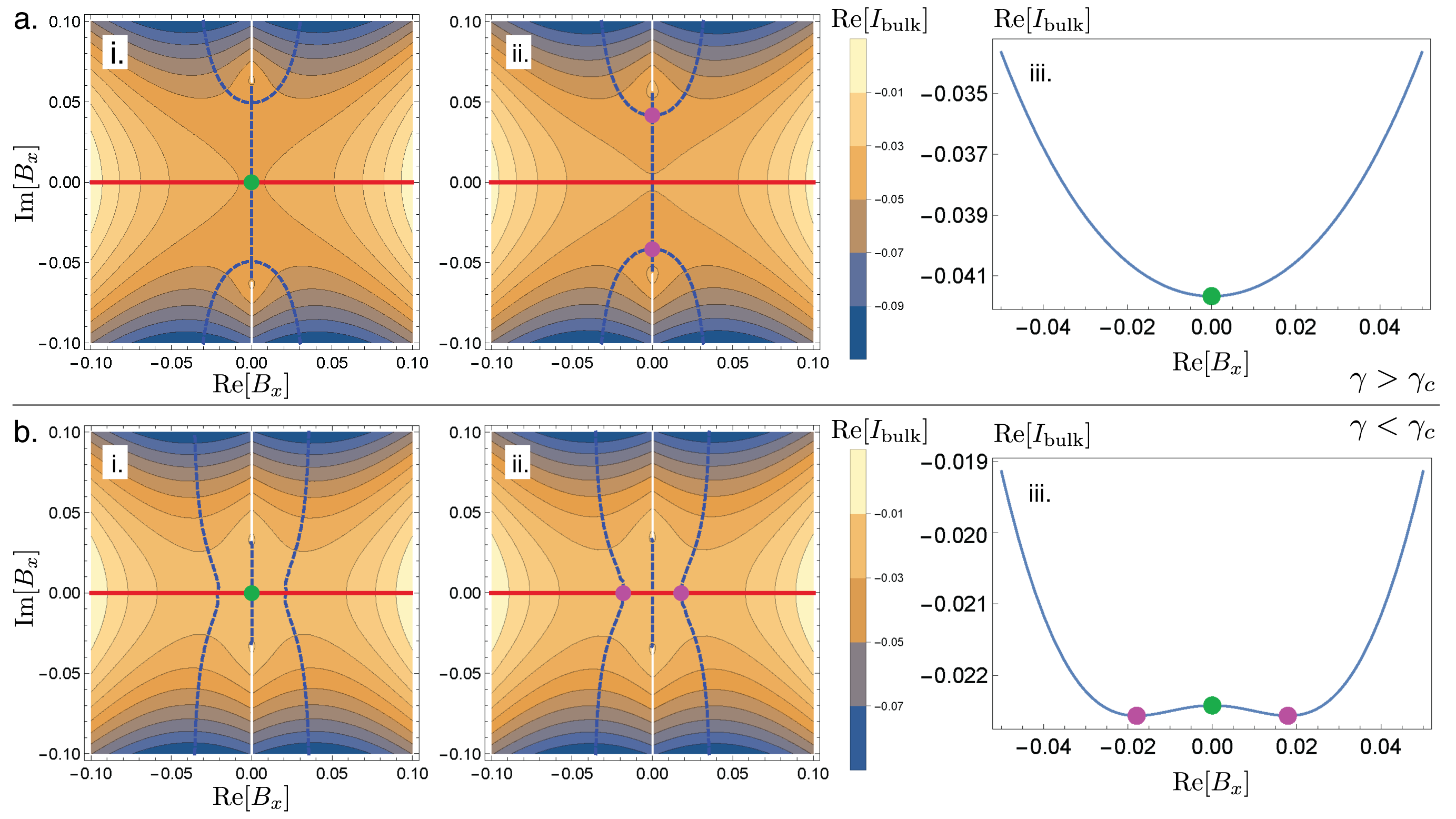}
    \caption{\textbf{Time-independent saddle-point analysis.} With $B_z$ fixed to its saddle-point value, plots of $\mathrm{Re}[\Ibulk]$ in the complex $B_x$ plane reveal the $\mathbb{Z}_2$ symmetry breaking in the bulk responsible for the purification transition. Dotted blue and solid red lines show contours of steepest descent. (a) Above the critical point $\gamma > \gamma_c$, the trivial (i., green) and symmetry-broken (ii., purple) saddle-points lie on the imaginary-$B_x$ axis. Because the integration contour for $B_x$ in the path integral lies along the real axis (solid red), only the trivial saddle point contributes to the effective bulk action $\Ibulk$ (iii). (b) Below the critical point $\gamma < \gamma_c$, all three saddle points lie on the real-$B_x$ axis (i,ii) and therefore all three contribute to the bulk action (iii), where the symmetry-broken saddle-points (purple) minimize the effective bulk action $\Ibulk$.}
    \label{fig:sbulk_contour}
\end{figure*}

With this replacement, one can now easily determine the time-independent saddle points of the action $I[\vec{B}]$ \eqref{eq:2fieldPathIntegralmain} in the large-$N$ limit by solving the Euler-Lagrange equations $\partial_{B_x} I = \partial_{B_z} I = 0$. We find one symmetric saddle point,
\begin{align}\label{eq:bulksaddlestrivial}
B_x &= 0 \nonumber \\
B_z &= \frac{4}{9} \left(\gamma + \frac{1}{2} \gamma_c \right),
\end{align}
and a pair of degenerate symmetry-broken saddle points 
\begin{align}\label{eq:bulksaddles}
    B_x &= \pm\frac{1}{3} \sqrt{\left(\gamma_c - \gamma \right) (\gamma + 3 \gamma_c)} \nonumber \\ 
    B_z &= \frac{1}{3} (\gamma + \gamma_c),
\end{align}
where $\gamma_c \equiv J/18$ is the critical point where all three saddle-point solutions coincide. The symmetric saddle is invariant under the replica permutation symmetry $B_x \rightarrow - B_x$, while the second pair of saddle points explicitly break this replica symmetry. 

The bulk action $I_{\mathrm{bulk}}$ for these saddle points are plotted in Fig. \ref{fig:sbulk_comp} as a function of the measurement rate $\gamma$. Note that the symmetry-broken saddle points (purple) always have lower action than the trivial saddle point (green); the dominant saddle point, however, depends on the direction of the integration contour. In the region $\gamma < \gamma_c$, the symmetry-broken saddles are indeed dominant, but for $\gamma > \gamma_c$, the trivial saddle controls the path integral. We now explain these points.

In Fig. \ref{fig:sbulk_contour}, we plot the contours of the real part of the action in Eq. \eqref{eq:2fieldPathIntegralmain} in the complex plane of $B_x$ with $B_z$ set to its saddle-point value. The original contour of integration is along the real $B_x$ axis. For $\gamma < \gamma_c$ the replica symmetry broken saddles (Fig. \ref{fig:sbulk_contour}b, purple dots) lie along the contour of integration, and are the minimum-action saddles as confirmed in Fig. \ref{fig:sbulk_contour}b(iii). However, for $\gamma>\gamma_c$, the corresponding symmetry broken saddles lie along the imaginary axis (Fig. \ref{fig:sbulk_contour}a). The contour of integration cannot be deformed to pass through these saddles while maintaining a valid estimate of the integral using only the saddle point value. 

In order for the saddle-point value to make the only important contribution to the integral, the contour of integration must pass though the saddle point in such a way that the saddle is a local minimum. However, for $\gamma > \gamma_c$, the desired integration contour hits a branch cut that leads off to infinity in a direction for which the value of the integrand diverges. This means that the symmetric saddle (Fig. \ref{fig:sbulk_contour}a(iii), green dot) is actually the relevant saddle to estimate the integral when $\gamma > \gamma_c$. Hence, there is a bulk phase transition at $\gamma = \gamma_c$ across which the replica permutation symmetry is spontaneously broken. 

This symmetry breaking appears explicitly in the problem if we plot the real part of the action $I$ as a function of $B_x$ above and below the transition point (Fig. \ref{fig:sbulk_contour}a,b(iii)), which reveals a straightforward double-well potential with spontaneously-broken $\mathbb{Z}_2$ symmetry $B_x \leftrightarrow -B_x$ below the transition $\gamma < \gamma_c$.

\subsubsection{Time-dependent configurations}

\label{sec:instantons}

The time-independent analysis of the previous section revealed the key $\mathbb{Z}_2$ symmetry-breaking physics that is responsible for the purification transition in the $(2,1)$ hybrid model. But the time-independent bulk solutions alone cannot be the whole story: indeed, as the bulk action $I$ is identical for $Z_2$ and $P^2$, the time-independent bulk solutions alone appear to predict a purity $Z_2/P^2 = 1$ for all $\gamma$, which is clearly incorrect. Neglected in this time-independent analysis are the non-equal boundary conditions at times $t = 0,T$ which explicitly break time translation symmetry in the problem.

To correctly evaluate the path integral expressions \eqref{eq:2fieldPathIntegralmain} in the large-$N$ limit (or \eqref{eq:pathintegral} and its generalizations in Appendix \ref{app:deriv_pqpathintegral}), we must expand the action $I$ around `classical' time-dependent configurations of the fields $\vec{B}(t)$ ($G_{rs}(t),F_{rs}(t)$) that properly account for boundary effects at $t = 0,T$. Quantum fluctuations around these classical configurations are controlled by the factor of $N$ preceding the action $I$, analogous to the role played by $1/\hbar$ in semiclassical (WKB) expansions \cite{zinn2002quantum}. By definition the classical field configurations $\vec{B}(t)$ obey the time-dependent Euler-Lagrange equations that extremize the action $I$, including time-derivatives terms $\partial_t B_{x,z}$ coming from the path-integral expansion of the propagator $K$ as well as boundary terms associated with the boundary states $t = 0,T$. 

These full time-dependent Euler-Lagrange equations can be solved numerically in the general case, and analytically in some special cases, including with time-independent configurations and near the critical point. In the discussion below, we focus on the regime where $T$ is much larger than any microscopic scale, so we are not considering transients associated with times of order $\gamma^{-1}$ or $J^{-1}$. We will return to consider dynamics on these short timescales in section \ref{sec:timedependencepurity}.

There are two kinds of time-dependent configurations that will be important. The first are edge configurations in which, under the influence of the boundary states $\ket{\psi_0},\ket{\psi_T}$ in the propagator $K$, the $\vec{B}(t)$ fields near $t = 0,T$ are deformed away from their time-independent values deep in the bulk. These boundary contributions are relevant for any value of $\gamma$. The second are instanton-like configurations in which the $\vec{B}(t)$ fields traverse from one symmetry-broken saddle to another. These are only relevant for $\gamma < \gamma_c$.

\begin{figure*}
    \centering
    \includegraphics[width=\textwidth]{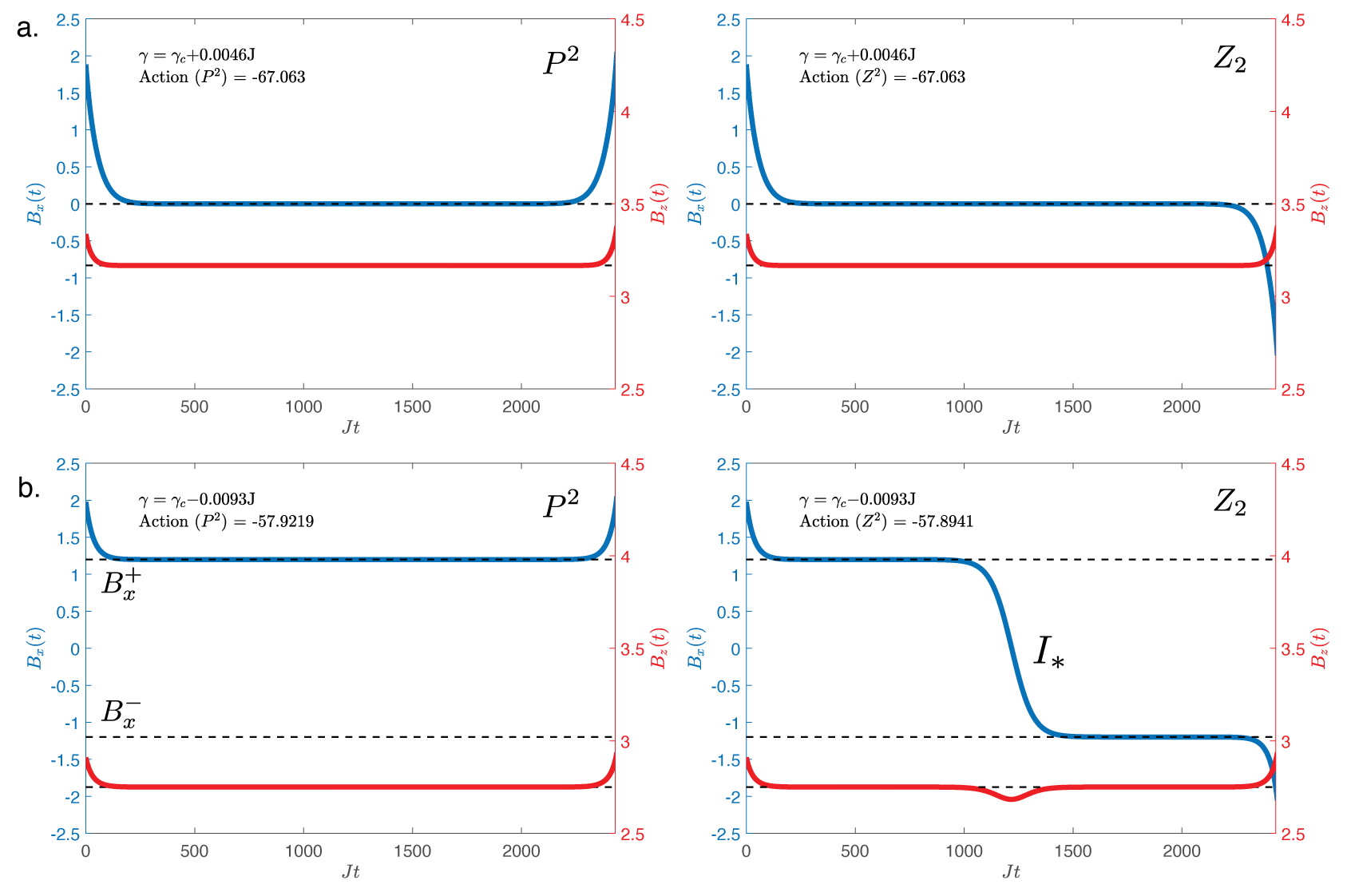}
    \caption{\textbf{Time-dependent classical field configurations $\vec{B}(t)$ from numerical gradient descent.} Optimal classical field configurations $B_x(t)$ (blue), and $B_z(t)$ (red) as obtained by numerical gradient descent over $\mathrm{Re} \ I[\vec{B}]$ of the `magnetic-field' action Eq. \eqref{eq:2fieldPathIntegralmain}. Gradient descent is performed by taking $J\delta t = 0.05$, until the threshold $\delta I = 10^{-7}$ is reached for the action difference, requiring $\sim 10^4$ iterations for the parameters considered here. (a) Above the critical point $\gamma > \gamma_c$ the configurations are dominated by a single trivial time-independent saddle point (dotted black), where the different boundary conditions in $Z_2$, $P^2$ lead to nontrivial boundary dynamics in the field $B_x(t)$ near $t = 0,T$. (b) Below the critical point $\gamma < \gamma_c$, the optimal configurations are dominated by a pair of symmetry-broken saddle points (dotted black). The non-uniform boundary conditions in $Z_2$ promote the formation of an instanton with action $I_*$ somewhere in the bulk that traverses between the two saddle points.}
    \label{fig:numgraddesc}
\end{figure*}

We first discuss the edge configurations, focusing on the regime $\gamma > \gamma_c$ as illustrated in Fig.~\ref{fig:numgraddesc}a. These configurations are found using a numerical gradient descent algorithm which takes the action \eqref{eq:2fieldPathIntegralmain}, discretizes the time direction to approximate the kernel $K$, and then searches through classical configuration space $\vec{B}(t)$ to find the time-dependent fields $B_x(t)$ and $B_z(t)$ that extremize the action; for further details of the gradient-descent numerics, see Appendix \ref{app:gradientdesc}. Deep in the bulk the fields $B_x(t), B_x(t)$ take their trivial saddle-point values $B_x = 0, B_z = 2 \gamma_c / 3$ as expected (Fig. \ref{fig:numgraddesc}a). Near the boundaries, however, the fields $B_x(t),B_z(t)$ differ considerably from their bulk value due to the influence of the boundary states in the definition of $K$. Crucially, since the initial and final states $\ket{\psi_{0}},\ket{\psi_T}$ are symmetric about the $B_x = 0$ saddle-point, the action evaluated on the time-dependent configurations are identical for $P^2$ and $Z_2$ up to corrections that are exponentially small in $T$. As a result, these contributions cancel in the ratio $\Pi_Q = Z_2 / P^2$, yielding $\Pi_Q \approx 1$. 

Similar edge configurations are also relevant for $\gamma < \gamma_c$, but instanton configurations now also play a role. For $\gamma < \gamma_c$, there are two symmetry-broken bulk saddle configurations $B_x^{\pm}$ (dotted black in Fig.~\ref{fig:numgraddesc}b). An important new ingredient is that the bulk saddles are distinguished by the boundary conditions. The boundary state at $t=0$ in the definition of $P^2$ and $Z^2$ favors the $B_{x}^{+}$ saddle (upper dashed black line), while the boundary state at $t=T$ in the definition of $Z_2$ favors the $B_{x}^{-}$ saddle (lower dashed black line). The identical boundary conditions in $P^2$ favor a configuration shown in the left panel of Fig. \ref{fig:numgraddesc}b in which the bulk saddle is $B_x^+$ and there are identical edge configurations near $t=0$ and $t=T$. By contrast the non-equal boundary conditions in $Z_2$ favor a configuration that traverses from $B_{x}^{+}$ to $B_{x}^{-}$ over a localized time window. This single-instanton configuration is shown in the right panel of Fig. \ref{fig:numgraddesc}b. Due to the reflection symmetry $B_x \leftrightarrow -B_x$ of the propagator $K$ under the generator $\sigma$, the overlap between $\ket{\psi_0}$ and the ground state of the $B_{x}^{+}$ saddle Hamiltonian is equal to the overlap between $\ket{\psi_T}$ and the ground state of the $B_{x}^{-}$ saddle Hamiltonian. As a result, the edge contributions to the action are approximately the same in the left and right panels of Fig.~\ref{fig:numgraddesc}b, and these contributions therefore cancel from the ratio $\Pi_Q = Z_2 / P^2$.

The time-translation symmetry of the bulk implies that the instanton is approximately free to move in time, giving rise to a zero mode in the path integral as is typical for instanton physics. This means that for $\gamma < \gamma_c$, we do not have just an isolated saddle point but a continuous family of nearly-degenerate saddle points. For this reason, the instanton contributes an additional `entropic' factor $\propto (T - T_0)$ to the purity $Z_2$ with $T_0$ some short-time regulator that arises because the instanton cannot get too close to the $t = 0,T$ boundaries without changing its action.

We note that there are other possible time-dependent configurations that could be relevant for $Z_2$. In particular, one might try to avoid the action cost of the instanton by considering a configuration that adheres closely to the $B_x^+$ saddle until times of order $T$. Then around $t=T$, the $\vec{B}(t)$ fields could bend towards the $B_x^-$ saddle to some degree. Such a configuration might be viewed as a partial instanton `bound' to the $t=T$ boundary. One can find approximate solutions of roughly this form, but at least when $T$ is large, the unbound instanton configuration always has lower action than such a bound configuration in every calculation we have done.

\subsection{Phases of the path integral}
\label{sec:phasesofpathint}

With the above ingredients in hand, we can now obtain the structure of the purity $\Pi_Q$ as a function of $\gamma$. The bulk phase transition at $\gamma = \gamma_c$ drives the transition in $\Pi_Q$, but to correctly compute this quantity we must include edge and instanton effects as discussed above. 

For $\gamma > \gamma_c$, there is only a single symmetric bulk saddle and the edge contributions to the action are identical for $Z_2$ and $P^2$ (Fig. \ref{fig:numgraddesc}a). Again, this follows from the important fact that the boundary states at $t=0$ and $t=T$ as well as the trivial bulk saddle are symmetric with respect to the reflection symmetry $B_x \leftrightarrow -B_x$. For this reason, the classical actions for $Z_2$ and $P^2$ are identical, up to corrections that decay exponentially with $T$ and we therefore expect $Z_2 / P^2 \approx 1$. Moreover, although we do not explicitly consider $1/N$ fluctuations in this work, we note that it seems plausible that the $1/N$ corrections are also equal order-by-order in $Z_2$ and $P^2$ up to corrections that decay exponentially with $T$.

For $\gamma < \gamma_c$, there are two symmetry-broken bulk saddles (Fig. \ref{fig:numgraddesc}b), and the instanton configuration with action $I_*$ and `entropy' $\propto (T-T_0)$ controls the $Z_2$ path integral. Once again, due to the symmetry under $B_x \leftrightarrow - B_x$ we expect the edge contributions to the action for $P^2$ and $Z_2$ to be identical and the only difference arises from the extra instanton in $Z_2$. 

Combining these results together, we find that at large $N$ and fixed $T > J^{-1},\gamma^{-1}$ (with $T$ large compared to microscopic scales), the purity exhibits two phases, 
\begin{equation} \label{eq:purity_exp_k1}
    \Pi_{Q} = \frac{Z_2}{P^2} = \begin{cases} \frac{T-T_0}{a(T)} \exp\left(-N \inst(\gamma)\right) & \ \gamma<\gamma_c\\
    1 & \ \gamma \geq \gamma_c,
    \end{cases}
\end{equation}
where $(T-T_0)/a(T)$ is the ratio of the functional determinants entering $Z_2$ and $P^2$ (and we are only really interested in the explicit $T$ dependence, although $a(T)$ may have some weak $T$ dependece as well). Note that for $\gamma > \gamma_c$ we expect the $Z_2$ and $P^2$ functional determinants to be approximately equal (it would be good to check this expectation explicitly). Finally, we remind the reader that this result concerns large $N$ and fixed $T$; we discuss late-time dynamics at fixed $N$ in section \ref{sec:latetime}.

\subsection{Phase transition}
\label{sec:criticalexps}

From the above analysis, which yields the large-$N$ estimate \eqref{eq:purity_exp_k1} for the purity $\Pi_Q$, we find that the value of the purity in the mixed phase $\gamma < \gamma_c$ is governed almost entirely by the instanton action $I_*$. For $\gamma \ll \gamma_c$, this depends on the details of the spin propagator $K$. However, in the vicinity of the critical point $\gamma = \gamma_c$, it is possible to analytically determine the instanton action as a function of $\gamma$. In this section, we outline the effective field theory of the transition and compute the instanton action. This allows us to determine various critical exponents which have an expected mean-field character arising from the large-$N$ limit.

Near the critical point, the symmetry breaking field $B_x$ has magnitude $B_x \propto \sqrt{J (\gamma_c-\gamma)}$ which vanishes at the critical point, while the field $B_z$ remains finite and of the order of $J$ by Eqs. \eqref{eq:bulksaddlestrivial} and \eqref{eq:bulksaddles}. Thus, near the critical point, one can find the time-dependence of the instanton configuration analytically by expanding the action in terms of the small field $B_x$. For the instanton configurations, the $B_z$ field in the action can be set to be a constant value set by the bulk saddle point, in agreement with our observations from gradient descent numerics (Fig. \ref{fig:numgraddesc}). Keeping in mind that the $\sigma_x,\sigma_z$ terms in the `magnetic-field' Hamiltonian $H({\vec{B}}) = -\vec{B}(t) \cdot \vec{\sigma} / 2$ in the propagator $K[\vec{B},\psi_0,\psi_T]$ (Eq. \eqref{eq:2fieldPathIntegralmain}) do not commute at different times $t$, one can expand the term $-\ln K$ in orders of $B_x \propto \sqrt{J(\gamma_c-\gamma)}$ which is the small parameter, while keeping the time-dependence explicit. Keeping terms up to second order in $B_x(t)$ (and dropping constant terms) the action in Eq. \eqref{eq:2fieldPathIntegralmain} can be rewritten as
\begin{widetext}
\begin{align}\label{eq:time-dependent-action1}
    I[\vec{B}] &\approx \int dt \frac{27}{4 J} B_{x}^{2}(t)-\frac{1}{8 J}\int dt_{1} dt_{2}B_x(t_1)B_x(t_2)f(t_1,t_2) + \mathcal{O}(B_x^{4}) \nonumber \\
    & \text{ with the kernel } f(t_1,t_2) = \frac{\cosh \alpha (T-2|t_1-t_2|)}{\cosh \alpha T}, \ \alpha = \frac{\gamma+\gamma_c}{6J}
\end{align}
\end{widetext}
where we have fixed the field $B_z = (\gamma + \gamma_c) / 3$ to its time-independent saddle-point value.

One can take $T\to\infty$ in the integration kernel safely, since most of the instantons occur far from the boundary, thereby simplifying the kernel to $f(t_1,t_2) = e^{-2\alpha |t_1-t_2|}$. Since the kernel is tightly-peaked near $t_2 = t_1$, one can expand in $t_2$ near $t_1$ and obtain the time-dependence of the field $B_x(t)$. The lowest order of time dependence occurs at quadratic order $B_x^2$, and all higher orders of time-dependence are suppressed at least to quartic order $B_x^{4}$. Thus, keeping only the lowest order of time dependence, we can just consider the time-independent part of the $B_x^4$ term in the action. This can be easily obtained from the bulk saddle solutions Eq. \eqref{eq:bulksaddles} and expanding the time-independent action to order $B_x^4$, which gives a contribution $\int dt B_x^4/(128 \alpha^3J^3)$. Combining these results, and also extracting the time-independent part of the kernel in Eq. \eqref{eq:time-dependent-action1} for simplicity, the action $I[\vec{B}]$ can be approximated as,
\begin{align}
    \label{eq:nearcriticalaction}
    I[\vec{B}] \approx &\int dt \ B_x(t)\left( \frac{27}{4J}B_x(t) + \frac{1}{128\alpha^3J^3} B_x^3(t) \right. \nonumber \\
    &\left. -\frac{1}{8J}\int ds B_x(s)e^{-2\alpha |t-s|}\right),
\end{align}
where the first two terms correspond to the time-independent contributions and the final term captures the time dependence of $B_x(t)$.

\begin{figure*}
    \centering
    \includegraphics[width=\textwidth]{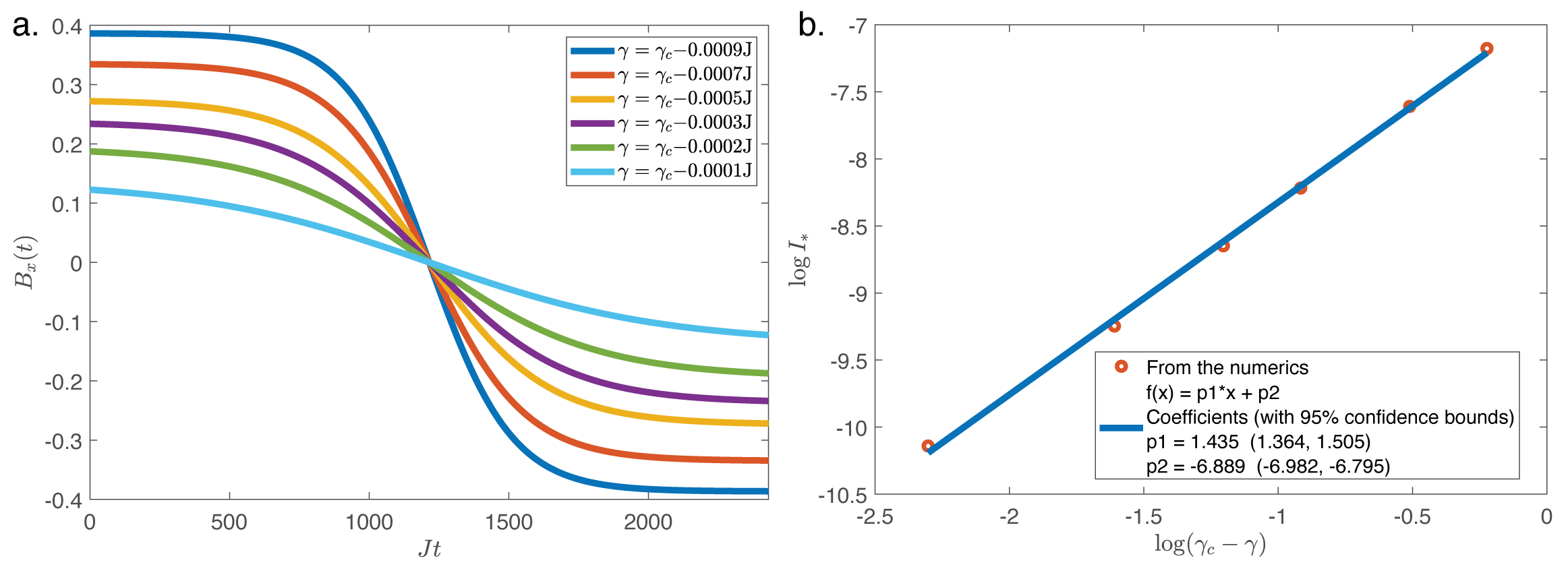}
    \caption{\textbf{Instanton configurations near criticality and critical exponent from gradient descent numerics.} (a) Bulk instanton configurations $\vec{B}(t)$ obtained from numerical gradient descent for measurement rates $\gamma = \gamma_c - \Delta \gamma$ just below the critical point. Gradient descent is performed by taking $J\delta t = 0.05$, until the threshold $\delta I = 10^{-7}$ is reached for the action difference. For the close-to-critical $\gamma$ considered, the analytically obtained instanton configuration in Eq. \eqref{eq:exact_instanton} are fixed points of the gradient descent algorithm. (b) Critical scaling of the instanton action $I_*$ shows a critical exponent $\cer = 1.44\pm0.07$, which is consistent with the theoretical expectation, $\cer = 3/2$.
    }
    \label{fig:crit_exp}
\end{figure*}

One can easily check that there exist static solutions $B_x(t) = B_x$ satisfying the time-independent equations of motion for the action \eqref{eq:nearcriticalaction}. These time-independent equations of motion are
\begin{equation}
    \frac{27B_{x}}{2J} + \frac{B_x^3}{32\alpha^3J^3} -\frac{B_x}{4J}\int_{-\infty}^{+\infty} ds e^{-2\alpha |t-s|}=0.
\end{equation}
Evaluating the integral this simplifies to
\begin{equation}
    B_x^3 = 32 \alpha^3J^3\left(\frac{1}{4J\alpha}-\frac{27}{2J}\right)B_x = \delta B_x
\end{equation}
which has static solutions
\begin{equation}
    B_x = 0, \pm\sqrt{\delta}
\end{equation}
where we have introduced the parameter $\delta \equiv 2(\gamma+\gamma_c)^2(\gamma_c-\gamma)/J$. Close to criticality, this static solution for $B_x$ is consistent with the earlier time-independent results, approximately the same as the Eq. \eqref{eq:bulksaddles}, differing only at order $\mathcal{O}(\gamma_c-\gamma)^{3/2}$ for $\gamma \lesssim \gamma_c$.


One can also easily find time-dependent solutions $\vec{B}(t)$ to the action \eqref{eq:nearcriticalaction}. Since the kernel $e^{-2 \alpha \magn{t-s}}$ is tightly peaked near $s = t$, we can Taylor-expand the field as $B_x(s) = B_x(t) + (s-t)B^{\prime}_x(t)+ (s-t)^2B^{\prime\prime}_x(t)/2 + ... $ in the equation of motion. After some algebra we obtain the time-dependent equation of motion for the field $B_x(t)$ with a second-order time derivative,
\begin{equation} \label{eq:eom_instanton}
    B_x^{\prime \prime}(t) = -\delta B_x(t) +B_x^{3}(t).
\end{equation}
Eq. \eqref{eq:eom_instanton} is exactly the equation of motion of a scalar field in a $\phi^4$ potential with a mass set by $\delta$, which vanishes at criticality, $\delta \to 0$. This is the correct theory near criticality, as any higher order time derivatives are suppressed either by factors of $1/\alpha$ or $\delta$. The bulk field theory model close to criticality $\delta \to 0$ is thus given by,
\begin{align}\label{eq:bulk_field_theory}
    &I[\vec{B}] = \int dt \left(\frac{1}{2} ( \partial_t B_x)^2 + V(B_x) \right),
    \nonumber\\
    &V(B_x) = - \delta \frac{B_x^2}{2} + \frac{B_x^4}{4}
\end{align}
which is just the action for a scalar $\phi^4$ theory, where $B_x$ is the scalar field and $\delta$ is the mass. The purification transition in the $(2,1)$ hybrid Brownian circuit model is therefore captured by the same universal physics as a 0+1d Ising model.


In the mixed phase $\delta > 0$, we expect time-dependent instanton transitions between the static solutions $B_x = \pm \sqrt{\delta}$ just as in section \ref{sec:phasesofpathint}. The instanton configuration has a field profile that asymptotes from $B_x = \sqrt{\delta}$ in the far past to $B_x = -\sqrt{\delta}$ in the far future. The equation of motion in Eq.~\eqref{eq:eom_instanton} has instanton solutions of the required form, with
\begin{align}\label{eq:exact_instanton}
    B_x^{*}(t) &= - \sqrt{\delta} f_{*}(t), \nonumber \\
    f_{*}(t) &= \tanh t \sqrt{\delta / 2}.
\end{align}
We can plug this solution back into the action \eqref{eq:nearcriticalaction} to compute the action cost of the instanton $I_*$, relative to a background that stays in one saddle for all time. This calculation yields
\begin{equation}\label{eq:instanton_scaling}
    \inst =  \delta^{3/2} \int \frac{dy}{\sqrt{2}} \left(  (f'_*)^2 - \frac{f_*^2}{2 } +  \frac{f_*^4}{4 } + \frac{1}{4 } \right).
\end{equation}
The integral is just a numerical constant independent of $\delta$, so the instanton action contribution goes like $\inst \sim \delta^{3/2}$. 

We confirm that this instanton configuration is correct for the full action in Eq. \eqref{eq:2fieldPathIntegralmain} by feeding the instanton solution \eqref{eq:exact_instanton} into the action and checking if there are nearby configurations with smaller action. In Fig. \ref{fig:crit_exp}, we find time-dependent configurations of the field $B_x(t)$ from numerical gradient descent that are indistinguishable from the analytically-obtained solution in Eq. \eqref{eq:exact_instanton} within the threshold for gradient descent. Furthermore, by numerically computing the action cost of these optimal configurations as a function of measurement rate $\gamma$ we find a critical exponent $\zeta = 1.44 \pm 0.07$ consistent with the analytically-obtained $\zeta = 3/2$. Hence, as the purity undergoes a transition at $\gamma = \gamma_c$, the entropy $-\ln \Pi_Q$ has a scaling form
\begin{equation}
    -\ln \Pi_Q  \sim N(\gamma_c-\gamma)^{\cer}.
\end{equation}
with critical exponent $\cer = 3/2$.

\subsection{Time-dependence of purity}
\label{sec:timedependencepurity}

\begin{figure*}
    \centering
    \includegraphics[width=\textwidth]{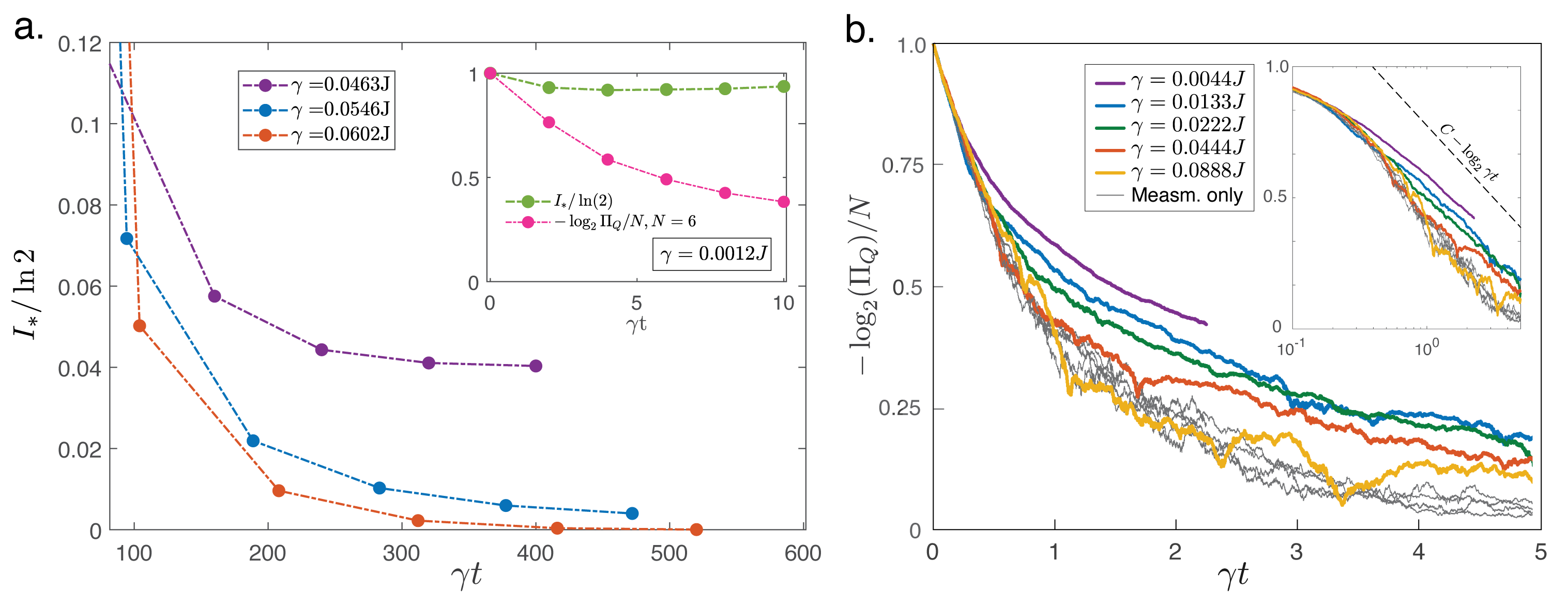}
    \caption{\textbf{Time dependence of R\'enyi-2 entropy, from saddle-point calculation and exact diagonalization.} (a) 
    We plot the instanton action $I_{*}$ as a function of time at different $\gamma$, obtained by performing gradient descent of the action in Eq. \eqref{eq:2fieldPathIntegralmain} for field configurations at different time intervals. Note, $\gamma_c = J/18 = 0.0556J$. For $\gamma > \gamma_c$ (red), $I_{*}$ goes to zero, while for $\gamma < \gamma_c$ (blue, purple), it approaches a finite non-zero plateau at late times. Close to criticality (blue), this plateau value is small, approaching zero, $I_{*}\to 0$ as $\gamma \to \gamma_c$. This result is true for $N = \infty$, where the saddle-point solution is exact. Inset shows estimated R\'enyi-2 entropy of the system for $\gamma < \gamma_c$ deep in the mixed phase accounting only for the instanton action (green), and including the $-\ln T/N$ term for $N = 6$ (pink) to show the logarithmic decay in entropy at late times. Gradient descent is performed by taking $J\delta t = 0.1$, until the threshold $\delta I = 10^{-6}$ is reached for the action difference. (b) We probe the time dependence for finite N, for system size $|Q| = |R| = 6$, via exact diagonalization. We note that for $\gamma > \gamma_c$ (yellow), and for measurement-only dynamics (J = 0) (gray), the entropy largely follows an exponential decay to zero.  However, for $\gamma <\gamma_c$, the time plots deviate from the exponential decay at later times. In the inset, we find at the latest times, there is a logarithmic decay in the entropy, $-\log_{2}\Pi_{Q}\propto -\log T$. For exact diagonalization via Krylov method, averaging is done over 50 disorder realizations, with $J\delta t = 0.01$, $Jt = 200$ and $N_K = 8$ Krylov subspace dimension.}
    \label{fig:timedependencepurity}
\end{figure*}

The statements made so far have been for the purity of the system at long times $T > J^{-1},\gamma^{-1}$ after some initial early-time transients controlled by the microscopic parameters. In this section we study these early-time dynamics for times $T \sim J^{-1},\gamma^{-1}$, 
accessing the purity on $O(1)$ time-scales using the saddle-point approach as well as exact diagonalization numerics. In Fig. \ref{fig:timedependencepurity}a, we plot the R\'enyi-2 entropy $-\log_{2}\Pi_{Q}$ or $S^{(2)}_{Q}(t)$, as a function of time for different $\gamma$, by finding minimal action configurations of the fields at different time intervals. To access this numerically, we perform gradient descent with the action in Eq. \eqref{eq:2fieldPathIntegralmain}, and interpret the results using the formula, $S^{(2)}_{Q}/N\sim I_{*} - \ln(T)/N$, from Eq. \eqref{eq:purity_exp_k1}. For $\gamma > \gamma_c$, we find that the instanton action $I_{*}$ goes to zero (equivalently, the system is purified) at $O(1)$ timescales, preceded by an exponential decay. For $\gamma < \gamma_c$, $I_{*}$ exponentially decays to a finite non-zero values (this is most clearly evident in the numerics for low $\gamma$, deep in the mixed phase). For the entropy, this plateau region ultimately gives in to a logarithmic decay with time (see Eq.\eqref{eq:purity_exp_k1}), which is not captured using gradient descent to estimate $I_{*}$. Directly using $S^{(2)}_{Q}/N\sim I_{*} - \ln(T)/N$, we can visualize the actual time dependence of the entropy, even for a modest choice $N = 6$ for $\gamma$ deep in the mixed phase, in the inset of Fig. \ref{fig:timedependencepurity}a. The $\ln T$ term becomes more and more important when we fix a finite $N$ and increase $T$. 


To access the dynamics at finite $N$, we resort to exact diagonalization using Krylov subspace methods \cite{lanczos1950iteration,park1986unitary,liesen2013krylov}. The unitary layers \eqref{eq:brham} of the Brownian circuit are computed via the conventional Krylov subspace technique with subspace dimension $N_K = 8$ and timestep $\delta t = 0.01$ in dimensionless units where $J = 1$. The non-unitary measurement layers \eqref{eq:nonunitarym} are computed by using the identity
\begin{align}
    M(t) = &\left(\frac{1}{2} - \frac{i}{2} \right) \exp{\left[-i \op{}(t) \delta t / 2 \right]} \nonumber \\
    + &\left(\frac{1}{2} + \frac{i}{2} \right) \exp{ \left[ i \op{}(t) \delta t / 2 \right]}.
\end{align}
In our numerical simulations we compute the exponentiations $\exp{[\pm i \op{}(t) \delta t / 2]} \ket{\Psi}$ separately using the conventional Krylov subspace technique and sum the results with appropriate complex coefficients to give $M(t) \ket{\Psi}$. At each timestep $t = m \delta t$ we independently sample coefficients $J_{ij}^{\alpha \beta}(t),n_i^{\alpha}(t)$ from normal distributions with zero mean and variance given by Eqs (2),(5) respectively. Using this disorder realization we then construct the Brownian generators $H(t),\op{}(t)$, and compute the time-evolved unnormalized state $\ket{\Psi (t+\delta t)} = U(t) M(t) \ket{\Psi(t)}$. For each disorder realization we compute both the purity $\tr{\unrho_Q^2(t)}$ and squared probability $\tr{\unrho_Q(t)}^2$ of the reduced unnormalized state $\unrho_Q(t) = \trover{R}{\ket{\Psi(t)} \bra{\Psi(t)}}$ as a function of time. 

These methods allow us to simulate hybrid Brownian dynamics for modest system sizes, $N = |Q| = |R| = 6$ and for times as long as $Jt = 200$. In Fig. \ref{fig:timedependencepurity}b we plot the resulting R\'enyi entropy $-\log_2 \Pi_Q$ at different rates of measurement $\gamma$. We find that for $\gamma > \gamma_c$, the behavior is qualitatively similar to measurement only dynamics (without the unitary part of the circuit), for which the entropy exponentially decays to 0 (although very small values of the entropy are inaccessible in the exact diagonalization numerics). For $\gamma < \gamma_c$, it is difficult to distinguish the plateau region (since here $N\sim 1$), and the eventual decay due to the entropic factor $\ln T$. However, the plots here already show that the late time behavior qualitatively deviates from the exponential decay, instead showing a much slower decay at late times. We identify in the inset of Fig. \ref{fig:timedependencepurity}b that this late time behavior is consistent with the factor of $\ln T$ in $-\ln\Pi_Q$, which comes from the entropic freedom of the instanton in Eq. \eqref{eq:purity_exp_k1}. In the next section, we discuss the role of the entropic enhancement of the purity in the eventual late time disintegration of the mixed phase.

\subsection{Phase disintegration at late times}
\label{sec:latetime}


Although the mixed phase $\gamma < \gamma_c$ is robust to repeated single-qubit measurements over extensive timescales $T \sim \mathrm{poly}(N)$, at very long times $T$ exponential in the system size, the measurements ultimately destroy entanglement between $R,Q$ and the mixed phase disintegrates \cite{gullans2020dynamical,li2020statistical}. In the path integral representation \eqref{eq:2fieldPathIntegralmain} this disintegration occurs due to the proliferation of instantons, which are heavily favored in the path integral at long times due to the `entropic' factor $(T-T_0)/a(T)$ found in section \ref{sec:phasesofpathint}. At large $N$ these multi-instanton configurations are subleading saddle-point configurations in the path integral and therefore do not contribute to the result at strictly $N \to \infty$. For very large but finite $N$, however, we must sum over these additional subleading saddles in the path integral.


To see the breakdown of the mixed phase explicitly, first consider the contribution $z_{\ell}$ to the path integral \eqref{eq:2fieldPathIntegralmain} coming from a configuration $\vec{B}_{\ell}(t)$ consisting of $\ell$ instantons. Due to the boundary conditions, the full instanton contribution is a sum over all odd $\ell$ for $Z_2$ or over all even $\ell$ for $P^2$. For sufficiently large $T$, we may apply the dilute-gas approximation in which the $\ell$ instantons are assumed to be widely separated in time and non-interacting \cite{sakita1985quantum}. In this limit, each instanton independently contributes an action penalty $e^{- N \inst}$ and an `entropic' factor $\propto (T-T_0)/a(T)$ coming from integration over the zero-mode \cite{sakita1985quantum,muller2006introduction}. In this approximation, and ignoring the contribution of the boundary conditions at $t = 0,T$ the $\ell$-instanton configuration has amplitude
\begin{align}
    z_{\ell} &\equiv \int \mathcal{D} \eta \ \exp{\left[ - N \int_0^T dt \ I[\vec{B}_{\ell}(t) + \eta(t)] \right]} \nonumber \\
    &\approx \frac{1}{\ell!} e^{- \ell N \inst} \left( T / a \right)^{\ell} z_0 = \frac{1}{\ell!} \mathcal{R}^{\ell} z_0 
\end{align}
where $\mathcal{R} \equiv z_1 / z_0 = e^{- N \inst} (T -T_0) / a(T)$, and the `entropic' term $(T-T_0) / a(T)$ comes from the functional determinant capturing the quantum fluctuations $\eta(t)$ around the classical configuration $\vec{B}_{\ell}(t)$ as discussed in section \ref{sec:instantons} \cite{sakita1985quantum}. We have to divide by $\ell!$ because instantons must always precede anti-instantons. The numerical value of the $\mathcal{O}(1)$ constant $a(T)$ depends on the details of the action $I[\vec{B}]$ and can be found by computing a functional determinant in the path integral after removing the zero-mode fluctuations of the instanton \cite{sakita1985quantum,muller2006introduction}.

Summing over even (or odd) $\ell$ we find
\begin{align}
    \sum_{\ell \ \mathrm{even}} z_{\ell} &= z_0 \cosh{\mathcal{R}} \nonumber \\
    \sum_{\ell \ \mathrm{odd}} z_{\ell} &= z_0 \sinh{\mathcal{R}} 
\end{align}
which gives a disorder-averaged purity
\begin{equation}
    Z_2 / P^2 = \tanh \mathcal{R}.
\end{equation}
At intermediate times $T \sim \mathrm{poly}(N)$ the exponential penalty $e^{-N \inst}$ dominates and instantons are disfavored $\mathcal{R} \ll 1$ such that $ Z_2  /  P^2  \approx \mathcal{R}$ as found in section \ref{sec:phasesofpathint}. At exponentially long times $T \sim \exp{(\inst N)}$, however, instantons become much more attractive $\mathcal{R} \rightarrow 1$ due to the `entropic' factor $(T-T_0)/a(T)$. We therefore find that instantons proliferate at late times with $ Z_2 /  P^2  \rightarrow 1$ as $T \rightarrow \infty$, corresponding to purification of the state and the disappearance of the mixed phase. We note that the instanton action $\inst$, along with the functional determinant $(T-T_0)/a(T)$, determines the crossover point $T \sim \exp{(\inst N)}$. This late-time destruction of the phase is similar to that found in phenomonological studies of MIPTs in 1+1d using capillary-wave theory \cite{li2020statistical}.

\section{Purification dynamics for subsystems}
\label{sec:subsystems}

In the previous section we considered the purity $\Pi_Q$ of the full system consisting of all $\magn{Q} = N$ qubits as an order parameter for the transition in the $(2,1)$ model. It is also interesting to ask whether the transition can be probed using only a fraction $k = \magn{A} / \magn{Q}$ of the system's qubits, with $A \subset Q$ \cite{gullans2020scalable}. In section \ref{sec:subsystemk}, we study the disorder-averaged purity $\Pi_A$ for variable-size subsystems $A$ and show that the purification transition is only visible for sufficiently large $k > k_c(\gamma) \geq 1/2$, leading to the phase diagram shown in Fig. \ref{fig:kphasediag}. Using a modified version of the bulk field theory \eqref{eq:bulk_field_theory} we identify the critical point $k = k_c$ in this diagram as a second-order phase transition in section \ref{sec:kcriticalexp} and compute its critical exponent $\cek = 1$ using analytical and numerical methods. Finally, we show in section \ref{sec:qecc} that these results can be interpreted in the language of quantum error correcting codes.



\subsection{Subsystem purity}
\label{sec:subsystemk}

\begin{figure}
    \centering
    \includegraphics[width=\columnwidth]{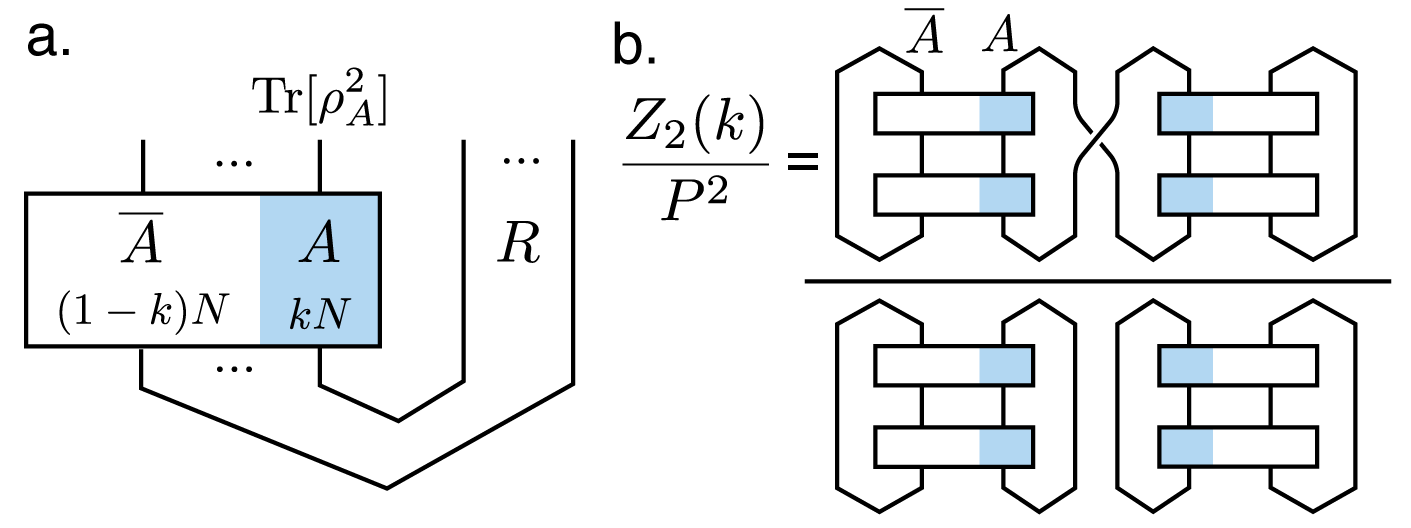}
    \caption{\textbf{Purification dynamics for subsystems $A \subset Q$.} (a) Subsets $A,\Ac$ of the system $Q$ are both initially maximally entangled with the reference $R$, but only the purity $\Pi_A$ of the subsystem $A$ is computed, while the remaining qubits $\overline{A}$ are traced over. (b) The disorder-averaged purity $Z_2(k)/P^2$ represented as a quantum circuit, where in the numerator the $\mathrm{SWAP}_{AA'}$ operator has been applied only to qubits in subsystem $A$.}
    \label{fig:kdependence}
\end{figure}



Consider a modified circuit setup shown in Fig. \ref{fig:kdependence}a where we compute the purity $\Pi_A = \tr{\rho_A^2}$ of a portion $\magn{A} = k N$ of the system qubits using a $\mathrm{SWAP}_{AA'}$ operator while the remaining $\magn{\Ac} = (1-k)N$ qubits are traced over.
Similar to section \ref{sec:browniancircuit} we compute the disorder-averaged purity $Z_2(k)$ and probability $P^2$ of the unnormalized state $\unrho_A = \trover{R,\overline{A}}{\unrho(V)}$ as shown in Fig. \ref{fig:kdependence}b, where the $\mathrm{SWAP}_{AA'}$ operator in this case leads to nontrivial boundary conditions only between the $A,A'$ subsystems. Converting this to a path integral expression leads to an action identical to \eqref{eq:2fieldPathIntegralmain} except for the replacement
\begin{align}
    \label{eq:pathintegral_k}
    \ln K \rightarrow k \ln K_A + (1-k) \ln K_{\Ac} \nonumber \\
    K_A = K[\vec{B},\psi_+,\psi_-] \nonumber \\
    K_{\Ac} = K[\vec{B}, \psi_+, \psi_+]
\end{align}
in the unnormalized purity $Z_2(k)$; the probability $P^2$ is left unchanged by the $k$-dependence. Here $K[\vec{B},\psi_0,\psi_T]$ is the propagator from Eq. \eqref{eq:2fieldPathIntegralmain} and the boundary states $\ket{\psi_{\pm}}$ have been defined in Eq. \eqref{eq:pm_states}. An analogous replacement can be made to compute subsystem purities in the general path integral \eqref{eq:pathintegral}. For $k = 1$ these path-integral expressions reduce to their original forms \eqref{eq:2fieldPathIntegralmain},\eqref{eq:pathintegral} as required. We therefore find that the bulk physics remains entirely unchanged by varying $k$ and inherits the same set of time-independent saddle points as discussed in section \ref{sec:bulksp}. Dependence on the subsystem fraction $k$ enters only through the boundary effects in the propagators $K_{A,\Ac}$.

\begin{figure*}
    \centering
    \includegraphics[width=\textwidth]{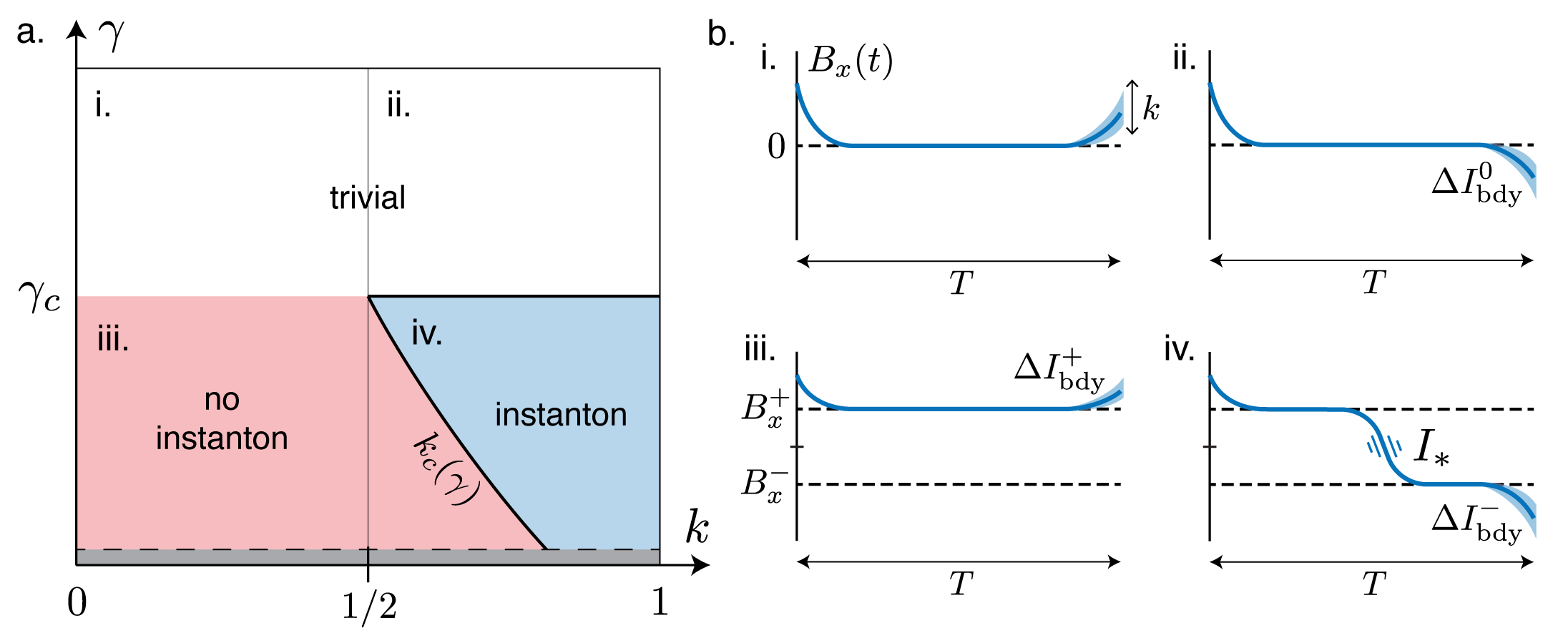}
    \caption{\textbf{Subsystem purity phase diagram.} (a) At times $T \sim \mathrm{poly}(N)$, the subsystem purity $\Pi_A$ exhibits three distinct phases as a function of $\gamma,k$ which are governed by the corresponding classical field configurations $\vec{B}(t)$ (b). Above the critical point $\gamma > \gamma_c$ the bulk fields (solid blue) primarily occupy the trivial saddle point $B_x = 0$ (dotted black), leading to a trivial (purified) phase for all $k$ (i-ii). Below the critical point $\gamma < \gamma_c$, the zero-instanton configuration (iii) dominates for small $k < k_c(\gamma)$ while the single-instanton configuration (iv) with action $I_*$ is dominant for large $k > k_c(\gamma)$.}
    \label{fig:kphasediag}
\end{figure*}

The interplay of bulk physics and boundary effects in $Z_2(k), P^2$ leads to a nontrivial phase diagram as a function of $\gamma,k$ as shown in Fig. \ref{fig:kphasediag}a. We can argue through the major features of this phase diagram by comparing the action costs $I[\vec{B}(t)]$ of various time-dependent classical field configurations $\vec{B}(t)$, which dominate the path integral at large $N$. Eqs. \eqref{eq:2fieldPathIntegralmain} and \eqref{eq:pathintegral_k} indicate that this action cost near the $t = 0$ boundary will be the same for $Z_2(k)$ and $P^2$ regardless of the value of $k$, and hence will cancel out in the purity $Z_2(k) / P^2$.
It is the future boundary condition at time $t = T$, generated by the $\mathrm{SWAP}_{AA'}$ operator acting on subsystems $A,A'$, that distinguishes between the different phases.

Let us first discuss the relevant time-dependent configurations of the $\vec{B}(t)$ fields. In Fig. \ref{fig:kphasediag}b, we consider classical configurations of the field $B_x(t)$ at different values of $k$ and $\gamma$ at intermediate times $T \sim \mathrm{poly}(N)$. As shown in Fig. \ref{fig:kphasediag}b(i-ii), for $\gamma>\gamma_c$ the classical path (solid blue) begins in a configuration $B_x > 0, B_z < 0$ that is bent towards the $\ket{\psi_+}$ state, traverses through the single trivial saddle point $B_x = 0$ (dotted black) and either returns to $B_x > 0$ for $k < 1/2$ (Fig. \ref{fig:kphasediag}b(i)) or continues on to $B_x < 0$ for $k > 1/2$ (Fig. \ref{fig:kphasediag}b(ii)). The action cost associated with the future boundary deflection is identical for $k$ and $1-k$ because of the symmetry $B_x \rightarrow - B_x$ and $k \rightarrow 1-k$.

For $\gamma<\gamma_c$, the situation is more complicated as shown in Fig. \ref{fig:kphasediag}b(iii-iv) due to the presence of the two symmetry-broken saddle points $B_x^{\pm}$ (dotted black), which can host instanton transitions between them. For small $k < 1/2$ (Fig. \ref{fig:kphasediag}b(iii)) the purity $Z_2(k)$ is dominated by the $\ket{\psi_+}$ future boundary condition, so $B_x$ spends most of its time on the nearest bulk saddle point $B_x^+$ with the deflection at the future boundary similar to (but not identical to) that of the past boundary. As the fraction $k$ increases, however, the $\ket{\psi_-}$ contribution begins to significantly affect the future boundary condition and $B_x$ field is pulled towards $B_x < 0$ in order for the r-bit $\ket{\psi(t)}$ to have higher overlap with $\ket{\psi_-}$ at $t = T$. For sufficiently large $k > k_c$, the future boundary condition forces an instanton to appear somewhere in the bulk (Fig. \ref{fig:kphasediag}b(iv)).

Because of the extra action cost $\inst(\gamma)$ of the instanton, the transition point between the zero- and single-instanton configurations Fig. \ref{fig:kphasediag}b(iii-iv) in the mixed phase $\gamma < \gamma_c$ always occurs at a critical fraction $k = k_c(\gamma) > 1/2$ larger than half the system size. By the same reasoning, we also expect $k_c\to1/2$ as $\gamma \to \gamma_c$ due to vanishing instanton cost $\inst \rightarrow 0$ as the symmetry-broken saddle-points $B_x^{\pm}$ rejoin at the critical point. Together, these arguments allow us to map out the major features of the $k,\gamma$ phase diagram Fig. \ref{fig:kphasediag} for subsystem purity at times $T\sim \mathrm{poly}(N)$.

Similar to section \ref{sec:phasesofpathint}, we can estimate the purity $Z_2(k) / P^2$ in each of these phases by computing the action cost $I[\vec{B}(t)]$ of the classical time-dependent field configurations $\vec{B}(t)$ discussed above. Above the critical point $\gamma > \gamma_c$, the action gets contributions from the trivial bulk saddle point $B_x = 0$ as well as the boundary contributions near $t=0,T$ as illustrated in Fig \ref{fig:kphasediag}b(i-ii). Similar to what we found in section \ref{sec:phasesofpathint}, the bulk contribution and the $t=0$ boundary contributions cancel in the ratio $Z_2(k)/P^2$, so the purity in this phase is controlled entirely by the difference of future boundary contributions $\Delta I_{\mathrm{bdy}}^0(k,\gamma)$. Note that $\Delta I_{\mathrm{bdy}}^0(1-k,\gamma) = \Delta I_{\mathrm{bdy}}^0(k,\gamma)$ from the $k \leftrightarrow 1-k$ symmetry present for $\gamma > \gamma_c$.

Below the critical point $\gamma < \gamma_c$ and for small subsystems $k < k_c$, the zero-instanton configuration Fig. \ref{fig:kphasediag}b(iii) dominates and we obtain nontrivial contributions $I_{0,T}$ to the action from the boundary dynamics near $t = 0,T$ and from the bulk saddle point $B_x^+$. As with $\gamma > \gamma_c$, the bulk contribution and the $t = 0$ boundary contribution $I_0$ are common to both $Z_2(k)$ and $P^2$, so the purity is controlled by the difference of the $t=T$ boundary contributions denoted  $\Delta I_{\mathrm{bdy}}^+(k,\gamma)$. For larger subsystems $k \geq k_c$ the single-instanton configuration Fig. \ref{fig:kphasediag}b(iv) dominates and we obtain nontrivial contributions in the action from the boundary dynamics $I_{0,T}$, from the bulk saddle value, and from the bulk instanton with action $\inst$. Again, the bulk saddle contribution and $t=0$ boundary contribution are common to $Z_2(k)$ and $P^2$, so the ratio is controlled by $\inst + \Delta I_{\mathrm{bdy}}^-$ where $\Delta I_{\mathrm{bdy}}^-$ denotes the difference of future boundary contributions in the presence of an instanton.

Combining these results, we find estimates for the subsystem purity $Z_2(k) / P^2$ in all three regions of the phase diagram Fig. \ref{fig:kphasediag}:
\begin{widetext}
\begin{equation}
    \label{eq:Z2kestimates}
    \frac{Z_2(k)}{P^2} = \begin{cases}
    \frac{T-T_0}{a'(T)} \exp\left[-N\left( \inst(\gamma)+ \Delta I_{\mathrm{bdy}}^-(k,\gamma)\right)\right] \ & \gamma<\gamma_c, \ k\geq k_c\\
    \exp\left[-N \Delta I_{\mathrm{bdy}}^+(k,\gamma) \right] & \gamma<\gamma_c, \ k<k_c\\
     \exp\left[-N \Delta I_{\mathrm{bdy}}^0(k,\gamma) \right] & \gamma \geq \gamma_c
    \end{cases}
\end{equation}
\end{widetext}
where we have included the `entropic' term $(T-T_0)/a'(T)$ (with a possibly different prefactor $a'(T)$) in the single-instanton configuration coming from the zero-mode motion of the instanton. These $k$-dependent purity estimates are a generalization of the $k=1$ purity estimates in Eq. \eqref{eq:purity_exp_k1}.

The expressions \eqref{eq:Z2kestimates} are similar to those obtained using capillary-wave theory as a phenomenological description of measurement-induced transitions in $1+1$d systems \cite{li2020statistical}. In this picture, the bulk of the circuit is viewed as a two-dimensional statistical mechanics system supporting a collection of domains separated by domain walls. In the mixed phase, small subsystems $A$ with nontrivial boundary conditions at the late-time boundary are unable to force most of the bulk to transition between phases and the system therefore has a domain wall pinned near the subsystem $A$. This is analogous to the case $\gamma < \gamma_c, k < k_c$ in Eq. \eqref{eq:Z2kestimates}, where the boundary action $\Delta I_{\mathrm{bdy}}^+$ is analogous to the energy cost of the pinned domain wall. 

Sufficiently large subsystems $A$ at the late-time boundary, by contrast, can force the entire bulk to transition, leading to a `domain-wall decoupling' effect where it is entropically favorable for the system to support two decoupled domain walls: one in the bulk that is free to move in time, and another that is pinned near the small subsystem $\overline{A}$. This situation is analogous to the case $\gamma < \gamma_c, k \geq k_c$ in Eq. \eqref{eq:Z2kestimates} where the instanton action $I_*$ corresponds to the energy cost of the bulk domain wall and the boundary action $\Delta I_{\mathrm{bdy}}^-$ corresponds to the energy cost of the pinned domain wall. In particular, the entropic prefactor $(T-T_0)/a'(T)$ coming from the zero-mode motion of the instanton corresponds to the entropy of the decoupled domain wall in the bulk; in both cases, this additional entropy is the reason why the single-instanton (or decoupled domain-wall) configuration is favorable despite the additional cost $I_*$ of creating the instanton (or domain wall) \cite{li2020statistical}. While the capillary-wave theory is phenomenological and specifically tailored for $1+1$d systems, in Eq. \eqref{eq:Z2kestimates} we have obtained similar expressions for the same physical phenomena starting from an exactly-solvable all-to-all microscopic model.

\subsection{Critical scaling of $k_c$ near $\gamma_c$}
\label{sec:kcriticalexp}

\begin{figure}
    \centering
    \includegraphics[width=\columnwidth]{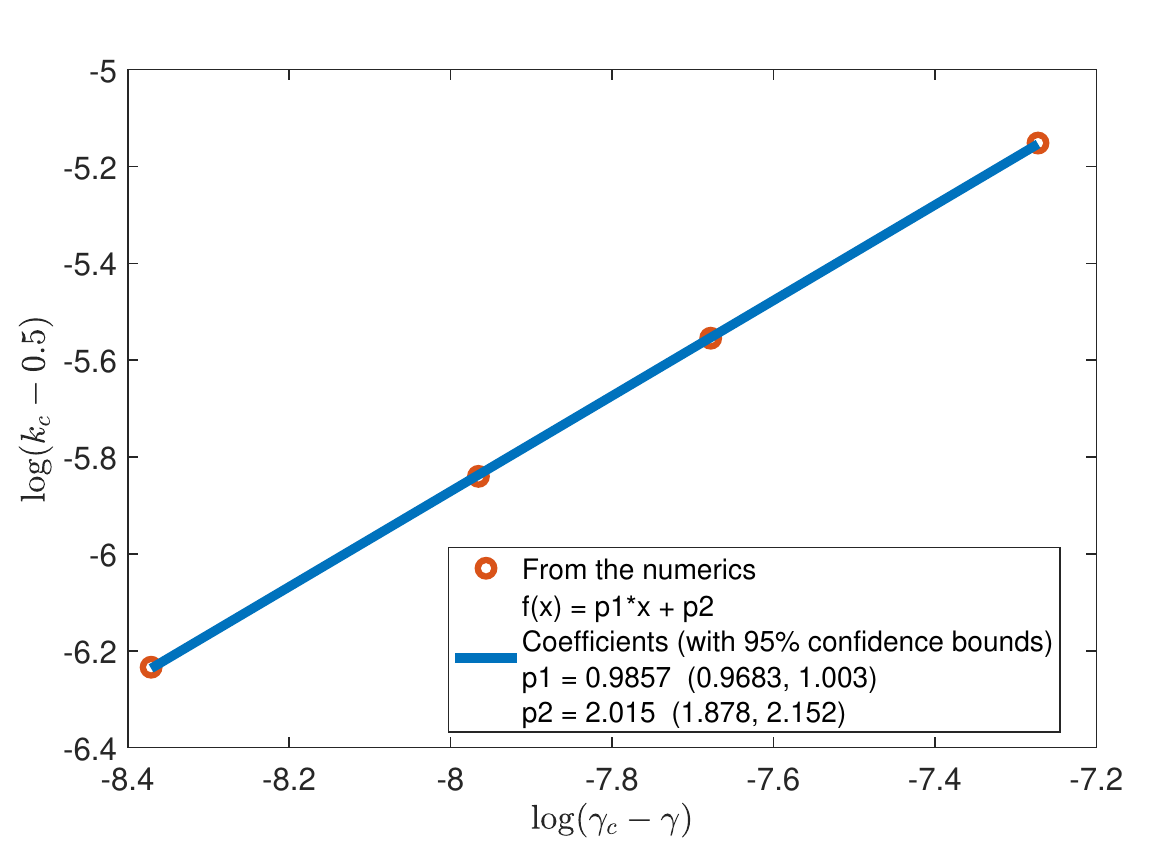}
    \caption{\textbf{Subsystem purity critical exponent $\mu$ from gradient descent numerics.} The critical subsystem fraction $k_c(\gamma)$ is identified for measurement rates $\gamma < \gamma_c$ just below the critical point by finding points in the $k,\gamma$ plane (red) where the boundary action $\Delta I_{\mathrm{bdy}}$ is equal to the single-instanton action $I_*$ (see Fig. \ref{fig:k_crit_appendix} of Appendix \ref{app:gradientdesc}). A linear fit (blue) gives an estimate $\mu = 0.99 \pm 0.01$, consistent with $\mu = 1$ from analytical arguments.}
    \label{fig:k_crit_exp}
\end{figure}

We can also study the behavior of $k_c(\gamma)$ close to the bulk phase transition $\gamma = \gamma_c$, in the symmetry broken phase. For this, we can numerically compare the action penalty for the boundary $\Delta I_{\text{bdy}} \equiv \Delta I_{\text{bdy}}^{+}-\Delta I_{\text{bdy}}^{-}$ with the instanton action $\inst$. We perform an optimization of the fields with the initial configurations corresponding to the two cases in Fig. \ref{fig:kphasediag}b(iii-iv), to locally minimize the action in Eq. \eqref{eq:pathintegral_k} for $\gamma\lesssim\gamma_c $. We identify the critical fraction $k_c \gtrsim 1/2$ by interpolating to find the value of $k$ above which $\Delta I_{\text{bdy}}>\inst$, such that the single-instanton configuration is dominant (see Appendix \ref{app:gradientdesc} for more details). We find numerically in Fig. \ref{fig:k_crit_exp} that $k_c$ scales with $\gamma_c - \gamma$ as 
\begin{equation}
    k_c - \frac{1}{2} \sim (\gamma_c-\gamma)^{\mu}, \ \text{for $\gamma$ close to $\gamma_c$.}
\end{equation}
A linear fit yields an estimate $\mu  = 0.99 \pm 0.01$ for the critical exponent.

Close to the bulk critical point, we can also adopt the previously described critical field theory model, now embellished with a boundary term, to analytically argue that $\mu = 1$.
We showed earlier that the bulk field theory close to criticality is given by the action \eqref{eq:bulk_field_theory}. We now model the boundary effect at $t= T$ with a delta function pinning field with action
\begin{equation}\label{eq:field_theory_bulk_bdy}
   I[\vec{B}] \rightarrow I[\vec{B}] +  \int dt \, h B_x \delta(t-T).
\end{equation}
where $h$ is an additional field controlling the strength of the pinning effect.

The delta function is regulated by setting $\delta(t-T) \rightarrow \delta(t-T+\epsilon)$ and taking $B_x(t) = B_{x,\mathrm{bdy}}$ (a constant) for $t\in[T-\epsilon,T]$. The equation of motion then implies that $\partial_t B_x$ jumps across $t=T-\epsilon$,
$- \partial_t B_x(T) + \partial_t B_x(T-\epsilon)  + h =0$. Since $ \partial_t B_x(T) =0$, we find that $ \partial_t B_x(T-\epsilon) = - h$. In other words, $-h$ is the slope of the $B_x(t)$ configuration at $t = T^{-}$. Close to criticality, we expect the following scaling,
\begin{equation} \label{eq:h_scaling}
    h\propto (k-1/2).
\end{equation}
This is because, for $k>1/2$, the SWAP-ed boundary condition should dominate, and the $B_x$ field at the future boundary should go lower than the $B_x\approx 0$ bulk saddle point, leading to $-h<0$ i.e. $h>0$. On the other hand, for $k<1/2$, the trivial boundary condition should dominate, and the $B_x$ field at the future boundary should go higher than the $B_x\approx 0$ bulk saddle point, leading to $h<0$. Close to $k\sim 1/2$, the linear scaling of $h\propto (k-1/2)$ can thus be justified and we expect our simplified model of the boundary condition to capture the universal physics.

The value of $B_{x,\mathrm{bdy}}$ is determined by appealing to a conservation law. For $t<T-\epsilon$, the quantity
\begin{equation}
    \mathcal{H}[\vec{B}] = \frac{1}{2} (\partial_t B_x)^2 - V(B_x)
\end{equation}
is conserved -- in the language of classical mechanics this is the statement that the classical Hamiltonian $\mathcal{H}[\vec{B}]$ corresponding to the Lagrangian $I[\vec{B}]$ is conserved. If we consider solutions that asymptote to a saddle point in the far past, then we know that
$\mathcal{H} = - V(\sqrt{\delta}) =  \delta^2/4$. Now, we can consider the two different cases as before -- firstly where the $B_x$ field configuration asymptotes to $B_x^{+} = \sqrt{\delta}$, and secondly to $B_x^{-} = - \sqrt{\delta}$, in the far past. Consider $B_{x,\mathrm{bdy}} = B_x(-\infty)+a$, where for the two cases, $B_x(-\infty) = \pm \sqrt{\delta}$. Because of the conservation law, we can solve for $a$ close to criticality where $|a/\sqrt{\delta}|\ll 1$ and we obtain  $a \propto - h/\sqrt{2\delta}$. Note, we have selected the correct sign of $a$ consistent with the fact that higher $h$ should lower the boundary field compared to the bulk saddle. 

Finally, we read off the excess boundary actions $\Delta I^{\pm}_{\text{bdy}}$ for both cases from Eq. \eqref{eq:field_theory_bulk_bdy}. To leading order in $a$ we find
\begin{align}
    &\Delta I^{\pm}_{\text{bdy}} = \pm h\sqrt{\delta} + h \ \mathcal{O}(a) \nonumber \\ 
    &\implies \Delta I_{\text{bdy}} = 2h\sqrt{\delta}.
\end{align}
Recall that in section \ref{sec:criticalexps} we found that $\inst \propto \delta^{3/2}$. Thus the condition for the zero-instanton and single-instanton configurations exchanging dominance $\Delta I_{\text{bdy}}>\inst$ occurs when $h\propto \delta$. Combined with Eq. \eqref{eq:h_scaling}, we find that $k_c$ scales as $(k_c-1/2) \propto \delta$, and
thus $\mu = 1$.

\subsection{Mutual information and error-correction}
\label{sec:qecc}

We can also understand the mixed phase at low measurement rate in this model through the lens of quantum error correction. Consider $\gamma\lesssim\gamma_c$, and $k\gtrsim k_c(\gamma)$. For this case, the dominant saddle-point configuration for the field $B_x(t)$ will be the single-instanton configuration shown in Fig.~\ref{fig:kphasediag}b(iv). On the other hand, for a subsystem fraction $k' = (1-k)$, the dominant saddle point will be the zero-instanton configuration shown in Fig.~\ref{fig:kphasediag}b(iii). Since the boundary conditions are symmetric with respect to the saddle points (i.e. the overlap between $\ket{\psi_{+}}$ and the  r-bit state favored by $B_x^{+}$ is equal to the overlap between $\ket{\psi_{-}}$ and the $B_x^{-}$ state), we can deduce a strong relation between the purities for $k$ and $(1-k)$.

In particular, this symmetry dictates that $\Delta I_{\text{bdy}}^{+}(1-k) = \Delta I_{\text{bdy}}^{-}(k)$. For the R\'enyi-2 entropies at $\gamma\lesssim\gamma_c$ and $k\gtrsim k_c(\gamma)$, we therefore have
\begin{align}
    S^{(2)}_{k} & = \Delta I_{\text{bdy}}^{-}(k) + I_{*}, \nonumber\\
    S^{(2)}_{1-k} & = \Delta I_{\text{bdy}}^{+}(1-k), \nonumber\\
    S^{(2)}_{1} & =  I_{*}.
\end{align}
From these relations, and the identity relating the two boundary effects, we may deduce
\begin{align}\label{eq:entropy-eqk}
    S_{k}^{(2)} = S_{1-k}^{(2)}+S_{1}^{(2)}
\end{align}
at $\gamma\lesssim\gamma_c$ and $k\gtrsim k_c(\gamma)$.

Because we may interpret the SWAP$_{AA'}$ operator shown in Fig. \ref{fig:kdependence} as acting either at the $t = 0$ boundary or at the $t = T$ boundary (this is equivalent to cyclically permuting the SWAP operator in the trace), the entropy $S^{(2)}_{k}$ can be identified as the R\'enyi-2 entropy of either a subsystem of fraction $k$ of the system $Q$ or of the reference $R$. As a result, Eq.~\eqref{eq:entropy-eqk} can be understood as the statement that for sufficiently large $k > k_c$, the mutual information between the $(1-k)N$ qubits in the subsystem $\overline{A}$ and the $N$ qubits in the reference $R$ vanishes identically, when measured by the disorder-averaged R\'enyi-2 entropy. That is,
\begin{align}
    \label{eq:mutualinfo}
    I^{(2)}&\left(\overline{A}:R\right)  \nonumber \\& = S^{(2)}\left(\overline{A}\right) +S^{(2)}\left(R\right)-S^{(2)}\left(\overline{A},R\right)\nonumber \\&
   = S^{(2)}\left(\overline{A}\right) +S^{(2)}\left(R\right)-S^{(2)}\left(A\right) \nonumber \\&
    =S^{(2)}_{1-k} +S^{(2)}_1-S^{(2)}_k = 0.
\end{align}
where in the second line we have used the fact that $S^{(2)}(\overline{A},R) = S^{(2)}(A)$ because the state $\unrho(V)$ is pure.

This result is consistent with the system forming a quantum error correcting code. The physical interpretation of the result \eqref{eq:mutualinfo} is that small parts of the system $\overline{A}$ contain no information about the reference $R$, after some of the system-reference entanglement is destroyed by the measurements. To be explicit, consider using the entire remaining system purity to encode information. We have $S^{(2)}_{1}$ logical qubits encoded within $N$ qubits and a reference entangled with those $S^{(2)}_1$ logical qubits. Eq. \eqref{eq:mutualinfo} states that any subsystem $\overline{A}$ of size less than $1-k_c < 1/2$ has zero R\'enyi mutual information with the reference $R$. If these statements also held for the mutual information $I(\overline{A}: R)$ computed from the von Neumann entropies, then the existence of a recovery channel would be guaranteed which could undo the erasure of the subsystem $\overline{A}$. 

Of course, these statements are only established here for the 2nd R\'enyi mutual information defined via a certain averaging procedure. More work is therefore needed to establish the existence of a recovery map since the mutual information can depend on the R\'enyi index $n$, and also on the averaging procedure. In order to make the connection to quantum error correction rigorous, we would also in principle need to include $1/N$ corrections and carefully work with approximate recovery maps. These are interesting topics to pursue in future work; here we content ourselves with describing the analogous phenomenon at the level of the 2nd R\'enyi entropy.

\section{Discussion}
\label{sec:discussion}

In this work we introduced new tools for analyzing measurement-induced purification transitions in the large-$N$ limit. Specifically, the $(2,1)$ hybrid Brownian circuit introduced in section \ref{sec:browniancircuit} exhibits a purification transition described by a relatively simple mean field theory that is analytically tractable at large $N$. We represented a particular disorder average over the purity as a path integral coupling four replicas, and derived the critical properties of the replica permutation breaking in the system, which is manifested as the purification transition for the system. Since the model is all-to-all, and the saddle-point point analysis depends on taking the large $N$ limit, the resulting field theory can be viewed as a minimal mean field description for the purification transition. Furthermore, since the resulting theory is a simple Ising field theory in $0+1$ dimensions, the critical exponents can be analytically understood, and also sheds light on the late time purification in the mixed phase through the mechanism of instanton proliferation. We also derived an entropic relation between subsystem and the reference, which allows us to identify the mixed phase as being a dynamically generated quantum error correcting code.  

This work adds to the growing paradigm of interpreting entanglement dynamics in quantum circuits through statistical mechanical models in the replica space in the context of hybrid circuits \cite{bao2020theory,jian2020measurement,fan2020self,liu2020non,lopez2020mean,garcia2021replica,choi2020quantum} and more broadly in random circuits \cite{hayden2016holographic,nahum2018operator,zhou2019emergent,vasseur2019entanglement,hunterjones2019unitary}. However, this model differs from the earlier works in considering the large-$N$ limit that allows us to make progress in interpreting entropy-like quantities as path integrals dominated by their saddle points. This is distinct from the large local Hilbert space dimension which is often necessary to make analytical progress in the random and hybrid circuits, and these two limits can lead to distinct physics. Our analysis here focused on contributions to lowest order in $1/N$. However, one can also study the subleading $1/N$ effects, which we reserve for future studies.




We emphasize that the $(p,q) = (2,1)$ model studied here using the purity ($n = 2$) is only one example of a large family of hybrid Brownian circuit models with measurement-induced transitions. Straightforward generalizations of the Brownian circuit layers introduced in section \ref{sec:browniancircuit} can generate $p$-body unitary interaction terms and $q$-body non-unitary measurement terms. Further, by introducing additional copies of the state $\unrho$ one can probe the phase transition using higher moments of the density matrix $n > 2$. We show in Appendix \ref{app:deriv_pqpathintegral} that each of these models leads to a distinct $(p,q)_n$ path integral representation with a large-$N$ limit. In particular, for reasonably small $n = 3,4,5,\ldots$ we expect that the combination of $\su{2}$ and replica symmetry will kinematically constrain the system to subspaces of small dimension similar to what we found in section \ref{sec:spinpropagatorK} for $n = 2$, allowing for analytical access to the purification transition at large $N$ for higher-order R\'enyi entropies $n > 2$. Furthermore, this setup can be extended to a combination of different $q$-body measurements, without the unitary part, allowing for exploration of measurement-only dynamics within the hybrid Brownian setup. Such measurement-only circuits have recently been shown to harbor symmetry-protected-measurement-only phases~\cite{lavasani2021measurement,sang2020measurement,ippoliti2021entanglement,lang2020entanglement,lavasani2021topological,bao2021symmetry} and phase transitions, which could be investigated in these Brownian setups as well.

Other straightforward generalizations include studying the $(p,q)$ models at higher spin $S$ or for more general degrees of freedom such as $\mathrm{SU}(Q)$ spins or fermions. We expect many of these models to also show measurement-induced transitions governed by boundary conditions and instanton effects similar to the story presented here for the $(2,1)_2$ path integral, although of course the details will differ considerably depending on the specifics of the model. Another exciting direction is to consider chains or lattices of $(p,q)$ models, with nearest-neighbor Brownian spin-spin interactions between individual clusters. This would allow for direct connection to measurement-induced phase transitions in $1+1$d models, including analytical estimates of the spatial critical exponent. We reserve study of these more general models for future work.


\begin{acknowledgements}
The authors thank Shaokai Jian, Christopher Baldwin, and Michael Gullans for helpful discussions. GSB is supported by the DOE GeoFlow program (DE-SC0019380). SS is supported in part by AFOSR under Award FA9550-17-1- 0180. The work of BGS is supported in part by the AFOSR under grant number FA9550-19-1-0360. Krylov ED numerics were run on the HPCC cluster at Brandeis University.
\end{acknowledgements}



\appendix

\section{Physical interpretation of the different averaged observables}

\label{app:disorderavg}

\begin{figure*}
    \centering
    \includegraphics[width=\textwidth]{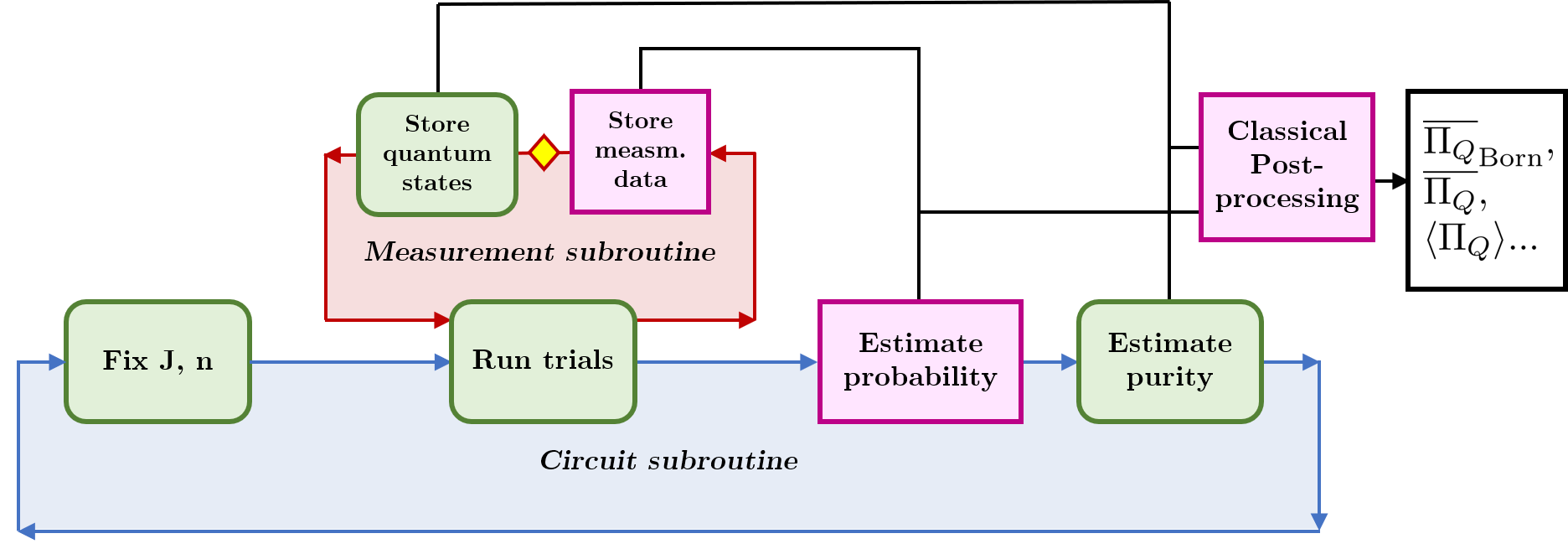}
    \caption{\textbf{Experimental protocol to simulate the various averaged purities.} The protocol is composed of two subroutines - `circuit' (blue) and `measurement' (red). Sampling of circuit realizations is done in the blue subroutine, which calls the sampling of measurement trajectories in the red subroutine. For each run of the measurement subroutine, the measurement data is stored classically, and the resulting quantum state is stored in a quantum memory depending on whether we want to simulate the Born probability or post-selected trajectories (this choice is represented by the diamond in the circuit). For each run of the circuit subroutine, purity can be estimated by doing SWAP tests on identical copies of the stored quantum states and the corresponding probabilities can be estimated by processing the classical data of measurement records. To estimate either of these quantities, the typical number of runs of the measurement subroutine scales exponentially with the number of measurements, i.e. exponentially with the `volume' of the circuit. All quantum processes in the protocol are denoted by `green' rounded boxes and all classical processes are denoted by `pink' boxes.  Finally, once enough statistics is collected, the `classical' data of purity and the probability for each circuit/measurement can be post-processed (as described in the text) to give us $\overline{\left(\Pi_{Q}\right)}_{\text{Born}}$, $\overline{\Pi_{Q}}$, $\langle\Pi_{Q}\rangle$ or any other simple averaged purity-like quantities.}
    \label{fig:protocol}
\end{figure*}

Here we discuss the issue of extracting the averaged purities $\overline{\Pi_Q},\langle \Pi_Q \rangle$ from experiments. A flowchart describing the experimental protocol is provided in Fig. \ref{fig:protocol}. 

To measure the purity $\Pi_Q$, the experimenter first fixes the unitary circuit elements (1), and the measurement bases $\sigma_{\mathrm{aux}}^y$ (3), yielding a single circuit realization with disorder coefficients $\mvec{J},\mvec{n}$. The experimenter then prepares the initial maximally-entangled state $\rho_0 = \ket{\Psi_0} \bra{\Psi_0}$ and applies an alternating sequence of unitary and weak-measurement layers to the system $Q$ as described in Sec. IIA. As discussed in the main text, weak measurements on the system are performed by coupling the system to auxiliary qubits and projectively measuring the auxiliary system. In principle, the experimenter can record the measurement results on a classical memory, and the resulting quantum states can be stored in a quantum memory. At this point the experimentalist keeps all resulting quantum states, even if the measurement results are not all $+1$.

Due to the unpredictability of these measurement outcomes, for each given realization $\mvec{J},\mvec{n}$ of the circuit, there will be a collection of quantum trajectories, which we label $\mathbf{T} = \mathbf{T}(\mvec{J},\mvec{n},\mvec{m})$, where $\mvec{m}$ is the record of measurement outcomes. Each trajectory performs a non-unitary operation on the state, $\ket{\Psi_{QR}} = \mathbf{T}\ket{\Psi_{0}}$ which is an unnormalized pure state, and each trajectory occurs with the Born probability $P\left[\mathbf{T}\right] = \langle\Psi_{QR}\ket{\Psi_{QR}}$. Now, given a sample set of classical data (collection of measurement records), one can estimate the Born probability $P\left[\mathbf{T}\right]$. For a system in an initially mixed state, one would start with a prior of 1/2 for each measurement, which would be updated based on the actual outcomes corresponding to the circuit realization and the measurement randomness.

Using the quantum memory which stores the obtained quantum states, the experimenter can access the purity of the state.  Once enough copies of each state are obtained, the experimenter can perform SWAP tests on the copies of the states to obtain the purity. The purity for each trajectory is given by
\begin{equation}
    \Pi_{Q}\left[\mathbf{T}\right] = \frac{Z_{2}\left[\mathbf{T}\right]}{P^{2}\left[\mathbf{T}\right]},
\end{equation}
where $Z_{2} = \tr{\tilde{\rho}_{Q}^{2}}$, for the unnormalized reduced density matrix on the system, $\tilde{\rho}_{Q} = \trover{R}{\ket{\Psi_{QR}}\bra{\Psi_{QR}}}$. The estimate from experiments improves with the number of copies, but these copies are hard to obtain, requiring a typical number of trials that scales exponentially with the number of measurements as discussed in section IIB of the main text. In the simplified setup we consider for our analytical calculation, we require the state to be stored only for specific measurement records $\mvec{m}$, where the auxiliary qubit measurement only gives the result $+1$.

Now the experimenter can repeat the whole sub-routine, by sampling different circuit realizations, $V = V(\mvec{J},\mvec{n})$, with an underlying probability distribution, $\pi(V(\mvec{J},\mvec{n}))$. For our analytical computation, we considered an analytically-tractable Gaussian probability distributions over the coefficients $\mvec{J},\mvec{n}$ as described in Sec. IIA. While repeating the experiment to collect data, there can be some simplifications due to symmetries in the circuit: for example, applying a weak-measurement layer with the disorder coefficient $\mvec{n}$ and obtaining a $+1$ outcome on the auxiliary qubit is equivalent to using a disorder coefficient $-\mvec{n}$ with an overall negative sign and obtaining the result $-1$.

Armed with the probability distribution of states and the purities, one can estimate a family of observables, related to different kinds of averages of the purity. Firstly, the Born probability averaged purity is given by,
\begin{align}
    &\overline{\left(\Pi_{Q} \right)}_{\text{Born}}  = \sum_{V,\mathbf{T}}\pi(V) P \left[\mathbf{T}\right]\Pi_{Q}\left[\mathbf{T}\right]
\end{align}

In our setup with post-selection, one can avoid averaging with the Born probability $P[\mathbf{T}]$, by averaging the purity $\Pi_Q$ only over post-selected trajectories with the desired measurement record $\mvec{m} = +1$. In this case, there is a single post-selected trajectory for each choice of circuit realization, with a single value of $\Pi_{Q}(V)$, $Z_{2}(V)$ and $P^{2}(V)$. In this case, the circuit-averaged purity for the post-selected trajectories is given by,

\begin{align}
    &\overline{\Pi_{Q}}  = \sum_{V}\pi(V) \Pi_{Q}(V).
\end{align}

Both the Born and post-selected  averaged purity considered above are difficult to access analytically, as one needs to average the ratio of two multi-replica quantities. In principle one could access the disorder-averaged ratio $\overline{Z_2 / P^2}$ by studying the path-integral representation of $\overline{Z_2 P^{2n}}$ analytically continued to $n = -1$. This can be difficult because one would typically need to access the expression over the entire domain of $n$ in order to take the analytic continuation. 

However, by classical post-processing of the probability data, we can access the analytically-tractable re-weighted purity $\langle Z_2 \rangle / \langle P^2 \rangle$ studied in this work. To do so, we first define a re-weighted probability,
\begin{equation}
    \mathcal{N}(V) = \frac{\pi(V)\times P^{2}(V)}{\mathcal{N}_{0}},
\end{equation}
with $\mathcal{N}_{0} = \sum_{V}\pi(V) P^{2}(V)$ to ensure that the probabilities sum to 1. In an experiment, $\mathcal{N}(V)$ can be estimated with just the classical information of the measurement records. Now we consider the purity averaged over this re-weighted probability,

\begin{align}
    \langle\Pi_{Q}\rangle  = \sum_{V}\mathcal{N}(V) \Pi_{Q}(V) = \frac{\sum_{V}\pi(V)Z_{2}(V)}{\sum_{V}\pi(V)P^{2}(V)}  = \frac{\langle Z_{2}\rangle}{\langle P^{2}\rangle},
\end{align}
which being an average of ratios is analytically accessible, and is the quantity we compute in this work. From an experimental point of view, although this estimation requires classical post-processing, it doesn't require any more quantum resources than the other two averages (in fact it requires fewer quantum resources than the Born-averaged quantity due to post-selection and selective storage of quantum states).

\section{Derivation of general $(p,q)$ path integral}
\label{app:deriv_pqpathintegral}

\begin{figure*}
    \centering
    \includegraphics[width=\textwidth]{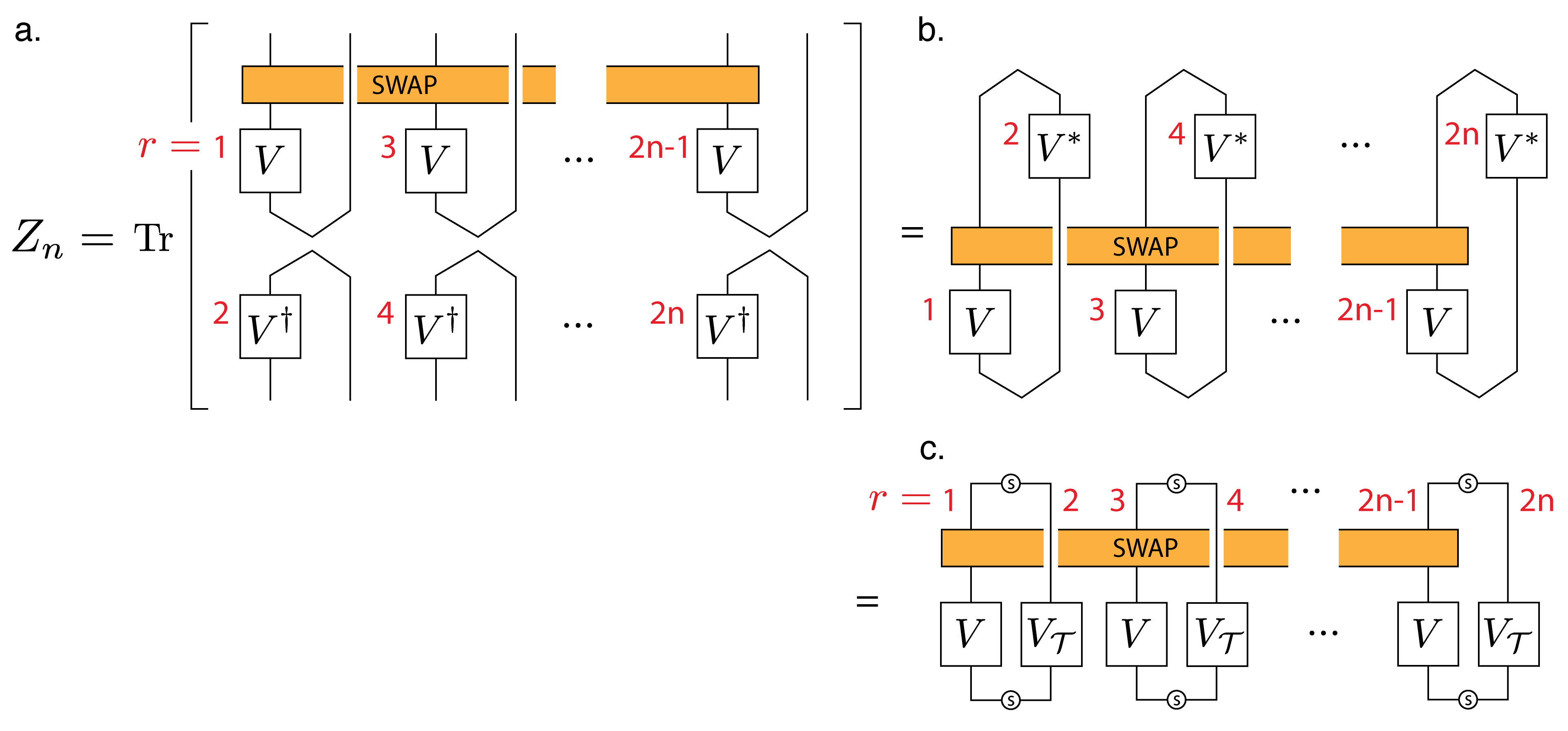}
    \caption{\textbf{Partition function $Z_n$ for the $n$th moment of the density matrix.} The $n$th-order R\'enyi entropy $S_Q^{(n)} = -\ln \tr{\rho_Q^n}$ is defined in terms of the $n$th moment of the system density matrix $\rho_Q$. The associated circuit for computing the $n$th moment $\tr{\unrho_Q^n}$ of the unnormalized density matrix $\unrho_Q$ can be transformed into pure-state dynamics on $2n$ replicas $r = 1,2,\ldots,2n$ with nontrivial boundary conditions at times $t = 0,T$ coming from the generalized $n$-system $\mathrm{SWAP}$ operator (orange).} 
    \label{fig:miptcircuit}
\end{figure*}

Here we derive a path integral representation for general $(p,q)$ hybrid Brownian circuits probed by the $n$th moment of the unnormalized density matrix $Z_n = \langle \tr{\unrho_Q^n} \rangle$. To compute this object, we introduce $n$ copies of the system $Q$ and reference $R$, and calculate the expectation value of the generalized $n$-system $\mathrm{SWAP}$ operator as shown in Fig. \ref{fig:miptcircuit}a. Using the circuit identities in Fig. \ref{fig:tensornetwork} and introducing factors of $iY$ as in section IIC of the main text, we can bring this circuit to the form shown in Fig. \ref{fig:miptcircuit}c describing pure-state dynamics on $2n$ replicas $r = 1,2,\ldots,2n$ with evolution operator
\begin{equation}
    \mathbb{V} = V \otimes V_{\mathcal{T}} \otimes \cdots \otimes V \otimes V_{\mathcal{T}}.
\end{equation}
and generalized SWAP boundary condition at $t = T$ which cyclically permutes the $n$ odd replicas $r = 1,3,\ldots,2n-1$. The associated probability $P^n = \langle \tr{\unrho_Q}^n \rangle$ is described by the same circuit but with trivial boundary condition at $t = T$.

The operators $V$ can be constructed by stacking any sequence of unitary Brownian layers or weak-measurement Brownian layers to give a variety of interaction and damping terms in the final action. Regardless of the particular choice of unitary and weak measurement dynamics, we always choose Brownian coefficients $J_{ij}^{\alpha \beta}(t),n_i^{\alpha}(t),\ldots$ for these layers that are statistically uncorrelated in time. As a result, the disorder average factorizes over the different timesteps $\Delta t$ and we may compute the disorder average over each layer independently as shown in Fig. \ref{fig:tensornetwork}c. We therefore first consider contributions to the effective action from unitary $p$-body Brownian circuit layers in Sec. \ref{app:pbodybrown}. We then consider contributions from non-unitary $q$-body Brownian circuit layers in Sec. \ref{app:qbodybrown}. Finally, we combine these results in Sec. \ref{app:coherentstatepathint} to give a path integral expression in Eq. \eqref{eq:fullgeneralpathintegral} for $Z_n$ for general $(p,q)$ hybrid Brownian models.

\subsection{$p$-body Brownian interactions} 
\label{app:pbodybrown}
Unitary Brownian dynamics are generated by $p$-body spin interactions $U(t) = \exp[ -i H(t) \Delta t/2]$ with time-dependent Hamiltonian
\begin{equation}
    \label{eq:pbodyint}
    H(t) = \sum_{\substack{i_1 < \ldots < i_p \\ \alpha_1 \ldots \alpha_p}} J_{i_1 \ldots i_p}^{\alpha_1 \ldots \alpha_p}(t) \ S_{i_1}^{\alpha_1} S_{i_2}^{\alpha_2} \cdots S_{i_p}^{\alpha_p}
\end{equation}
where the $S_i^{\alpha}$ are $\su{2}$ spin-$S$ degrees of freedom on sites $i = 1, \ldots, N$ and the Brownian coefficients $J_{i_1 \ldots i_p}^{\alpha_1 \ldots \alpha_p}(t)$ are white-noise-correlated Gaussian random variables
\begin{align}
    &\left\langle J_{i_1 \ldots i_p}^{\alpha_1 \ldots \alpha_p}(t) J_{i_1' \ldots i_p'}^{\alpha_1' \ldots \alpha_p'}(t') \right \rangle_{\mvec{J}} \nonumber \\
    &\quad \quad = \frac{J}{N^{p-1} (S+1)^{2p}} \delta(t-t') \delta_{i_1 i_1'} \cdots \delta^{\alpha_1 \alpha_1'} \cdots
\end{align}
where the normalization $1/N^{p-1} (S+1)^{2p}$ ensures that the Hamiltonian $H(t)$ is extensive and independent of spin size. We regulate the delta function via the replacement $\delta(t-t') \approx \delta_{tt'} (\Delta t/2)^{-1}$ and consider the limit $\Delta t \to 0$. Note that under the time-reversal operation $\mathcal{T}$ the Hamiltonian and unitary operator transform as
\begin{align}
    H_{\mathcal{T}}(t) &= (-1)^p H(t) \nonumber \\
    U_{\mathcal{T}}(t) &= \exp \left[i (-1)^p H(t) \Delta t/2 \right].
\end{align}
Expanding this layer to second order in $\Delta t$ and performing the disorder average over the Brownian coefficients $J_{i_1 \ldots i_p}^{\alpha_1 \ldots \alpha_p}(t)$ we find
\begin{widetext}
\begin{align}
    \left \langle U \otimes U_{\mathcal{T}} \otimes \cdots \otimes U \otimes U_{\mathcal{T}} \right \rangle_{\mvec{J}} &\approx \left \langle \left( 1 - i H \frac{\Delta t}{2} - \frac{1}{2} H^2 \frac{\Delta t^2}{4}\right) \otimes \left( 1 + i H_{\mathcal{T}} \frac{\Delta t}{2} - \frac{1}{2} H^2_{\mathcal{T}} \frac{\Delta t^2}{4}\right) \otimes \cdots \right \rangle_{\mvec{J}} \nonumber \\
    & = \left \langle \left( 1 - i H \frac{\Delta t}{2} - \frac{1}{2} H^2 \frac{\Delta t^2}{4} \right) \otimes \left( 1 + i (-1)^p H \frac{\Delta t}{2} - \frac{1}{2} H^2 \frac{\Delta t^2}{4}\right) \otimes \cdots \right \rangle_{\mvec{J}} \nonumber \\
    & = 1 - \frac{\Delta t^2}{4} \sum_{r < s} \mu_{rs}^p \left \langle H^r H^s \right \rangle_{\mvec{J}} - \frac{1}{2} \frac{\Delta t^2}{4} \sum_r \left \langle \left( H^r \right)^2 \right \rangle_{\mvec{J}}
\end{align}
\end{widetext}
where $H^{r,s}$ denote copies of the Hamiltonian \eqref{eq:pbodyint} acting on replicas $r,s = 1, \ldots, 2n$, and we have defined
\begin{align}
    \label{eq:mucoeff}
    \mu_{rs}^p \equiv \begin{cases} 
      (-1)^{r+s} &p \ \mathrm{even} \\
      1 &p \ \mathrm{odd}.
   \end{cases}
\end{align}
Evaluating the disorder average, we find
\begin{align}
    \left \langle H^r H^s \right \rangle_{\mvec{J}} \frac{\Delta t^2}{4} &= \frac{J \Delta t}{N^{p-1} (S+1)^{2p}} \frac{1}{2 p!} \nonumber \\
    &\quad \times \sum_{\substack{i_1 \ldots i_p \\ \alpha_1 \ldots \alpha_p}} \left( S_{i_1}^{\alpha_1,r} S_{i_1}^{\alpha_1,s} \right) \cdots \left( S_{i_p}^{\alpha_p,r} S_{i_p}^{\alpha_p,s} \right) \nonumber \\
    &= \frac{J \Delta t}{N^{p-1} (S+1)^{2p}} \frac{N^p}{2 p!} \left( \frac{1}{N} \sum_i \mvec{S}_i^r \cdot \mvec{S}_i^s \right)^p
\end{align}
as an operator equation, where the additional factor of $1/p!$ comes from converting the ordered sum in Eq. \eqref{eq:pbodyint} to an unordered sum. The disorder-averaged Brownian circuit layer can therefore be written as a propagator:
\begin{align}
    \label{eq:pbodyscrambleprop}
    &\left \langle U \otimes U_{\mathcal{T}} \otimes \cdots \otimes U \otimes U_{\mathcal{T}} \right \rangle_{\mvec{J}} \approx e^{- N I_p(t) \Delta t} \nonumber \\
    I_p(t) &\equiv n \frac{J S^{p}}{2 p! (S+1)^{p}} \nonumber \\
    &\quad + \sum_{r<s} \mu_{rs}^{p} \frac{J}{ 2 p! (S+1)^{2p}}\left(\frac{1}{N}\sum_{i} \mvec{S}_{i}^{r}\cdot \mvec{S}_{i}^{s}\right)^{p}
\end{align}
to lowest order in $\Delta t$, which holds as an operator equation.

\subsection{$q$-body Brownian measurements}
\label{app:qbodybrown}
\begin{figure}
    \centering
    \includegraphics[width=0.5\textwidth]{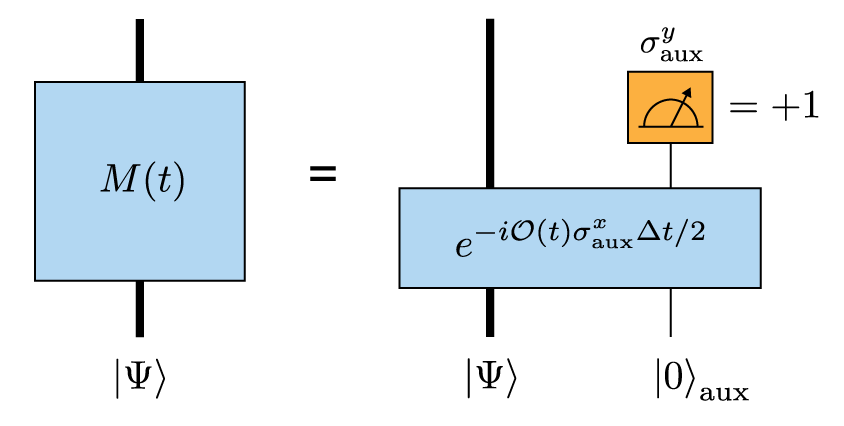}
    \caption{\textbf{Brownian weak measurement protocol.} The operator $M(t)$ weakly measures the Brownian operator $\op{}(t)$ by coupling it to an auxiliary qubit $\ket{\psi}_{\mathrm{aux}}$ for a time $\Delta t / 2$ (blue) and projectively measuring the auxiliary qubit in the $\sigma^y_{\mathrm{aux}}$ basis, post-selecting for $+1$ results (orange). Due to the coupling between the system $\ket{\Psi}$ and auxiliary qubit $\ket{\psi}_{\mathrm{aux}}$ this projective measurement alters the many-body state $\ket{\Psi} \to M(t) \ket{\Psi}$.}
    \label{fig:meascircuit}
\end{figure}

Consider making a weak measurement of a Hermitian $q$-body operator 
\begin{equation}
    \op{}(t) = \sum_{\substack{i_1 < \ldots < i_q \\ \alpha_1 \ldots \alpha_q}} \op{i_1 \ldots i_q}^{\alpha_1 \ldots \alpha_q}(t) \ S_{i_1}^{\alpha_1} \cdots S_{i_q}^{\alpha_q}
\end{equation}
at some time $t$ during the circuit evolution. For the moment we leave the coefficients $\op{i_1 \ldots i_q}^{\alpha_1 \ldots \alpha_q}(t)$ of this operator unspecified.
This $q$-body operator is the analogue of the 1-body spin operators $\op{}(t) = \sum_{i,\alpha} n_i^{\alpha}(t) S_i^{\alpha}$ weakly measured in the $(2,1)$ model described in section II of the main text. 
To measure this operator we introduce an auxiliary qubit initialized in $\ket{\psi}_{\mathrm{aux}} = \ket{0}_{\mathrm{aux}}$ and couple it to $\op{}$ via a unitary interaction
\begin{equation}
    \exp{\left[- i \op{}(t) \sigma^x_{\mathrm{aux}} \Delta t/2 \right]} \ket{\Psi}\ket{0}_{\mathrm{aux}}
\end{equation}
for a short time $\Delta t / 2$, where $\ket{\Psi}$ is the state of the system prior to the weak measurement and $\sigma^x_{\mathrm{aux}}$ is the Pauli-$x$ operator acting on the auxiliary qubit.
By projectively measuring the auxiliary qubit in the eigenbasis of $\sigma^y_{\mathrm{aux}}$ and post-selecting only for $+1$ results, the original state is transformed to
\begin{equation}
    \ket{\Psi} \to M(t) \ket{\Psi} \equiv \left( 1 - \op{} \frac{\Delta t}{2} - \frac{1}{2} \op{}^2 \frac{\Delta t^2}{4} + \cdots \right) \ket{\Psi}
\end{equation}
to lowest order in $\Delta t$. The circuit diagram for this measurement protocol is shown in Fig. \ref{fig:meascircuit}. The strength of the measurement is controlled by the magnitude of the operator $\magn{\op{}(t)}$ in units of the timestep $\Delta t/2$.
Under the time-reversal operation $\mathcal{T}$ the operators $\op{}(t)$ and $M(t)$ transform as
\begin{align}
    \op{\mathcal{T}}(t) &= (-1)^q \op{}(t) \nonumber \\
    M_{\mathcal{T}}(t) &= \left( 1 - (-1)^q \op{} \frac{\Delta t}{2} - \frac{1}{2} \op{}^2 \frac{\Delta t^2}{4} + \cdots \right).
\end{align}

We now apply weak-measurement operators $M(t)$ in the circuit at each odd timestep $t = (2m+1) \Delta t / 2$ and take the operator coefficients to be white-noise-correlated Gaussian random variables
\begin{align}
    &\left\langle \op{i_1 \ldots i_q}^{\alpha_1 \ldots \alpha_q}(t) \op{i_1' \ldots i_q'}^{\alpha_1' \ldots \alpha_q'}(t') \right \rangle_{\mvec{O}} \nonumber \\
    &\quad \quad= \frac{\gamma}{N^{q-1} (S+1)^{2q}} \delta(t-t') \delta_{i_1 i_1'} \cdots \delta^{\alpha_1 \alpha_1'} \cdots
\end{align}
whose strength is controlled by the parameter $\gamma$.
As before, the delta function can be regulated by the replacement $\delta(t-t') \approx \delta_{tt'} (\Delta t/2)^{-1}$. Performing the disorder average over the coefficients $\op{i_1 \ldots i_p}^{\alpha_1 \ldots \alpha_p}(t)$ in the $2n$-replica system we find
\begin{widetext}
\begin{align}
    \left \langle M \otimes M_{\mathcal{T}} \otimes \cdots \otimes M \otimes M_{\mathcal{T}} \right \rangle_{\mvec{O}} &\approx \left \langle \left( 1 - \op{} \frac{\Delta t}{2} - \frac{1}{2} \op{}^2 \frac{\Delta t^2}{4} \right) \otimes \left( 1 - \op{\mathcal{T}} \frac{\Delta t}{2} - \frac{1}{2} \op{\mathcal{T}}^2 \frac{\Delta t^2}{4} \right) \otimes \cdots \right \rangle_{\mvec{O}} \nonumber \\
    & = \left \langle \left( 1 - \op{} \frac{\Delta t}{2} - \frac{1}{2} \op{}^2 \frac{\Delta t^2}{4} \right) \otimes \left( 1 - (-1)^q \op{} \frac{\Delta t}{2} - \frac{1}{2} \op{}^2 \frac{\Delta t^2}{4} \right) \otimes \cdots \right \rangle_{\mvec{O}} \nonumber \\
    & = 1 + \frac{\Delta t^2}{4} \sum_{r < s} \chi_{rs}^q \left \langle \op{}^r \op{}^s \right \rangle_{\mvec{O}} - \frac{1}{2} \frac{\Delta t^2}{4} \sum_r \left \langle \left( \op{}^r \right)^2 \right \rangle_{\mvec{O}}
\end{align}
where we have defined
\begin{align}
    \label{eq:chicoeff}
    \chi_{rs}^q \equiv \mu_{rs}^{q+1} = \begin{cases} 
      1 &q \ \mathrm{even} \\
      (-1)^{r+s} &q \ \mathrm{odd}
   \end{cases}
\end{align}
similar to Eq. \eqref{eq:mucoeff}.
Evaluating the disorder average, we find
\begin{align}
    \left \langle \op{}^r \op{}^s \right \rangle_{\mvec{O}} \frac{\Delta t^2}{4} &= \frac{\gamma \Delta t}{N^{q-1} (S+1)^{2q}} \frac{1}{2q!} \sum_{\substack{i_1 \ldots i_q \\ \alpha_1 \ldots \alpha_q}} \left( S_{i_1}^{\alpha_1,r} S_{i_1}^{\alpha_1,s} \right) \cdots \left( S_{i_q}^{\alpha_q,r} S_{i_q}^{\alpha_q,s} \right) \nonumber \\
    &= \frac{\gamma \Delta t}{N^{q-1} (S+1)^{2q}} \frac{N^q}{2 q!} \left( \frac{1}{N} \sum_i \mvec{S}_i^r \cdot \mvec{S}_i^s \right)^q
\end{align}
as an operator equation. The disorder-averaged Brownian-measurement circuit layer can therefore be written as a propagator:
\begin{align}
    \label{eq:qbodymeasprop}
    &\left \langle M \otimes M_{\mathcal{T}} \otimes \cdots \otimes M \otimes M_{\mathcal{T}} \right \rangle_{\mvec{O}} \approx e^{- N I_q(t) \Delta t} \nonumber \\
    I_q(t) &\equiv n \frac{\gamma S^{q}}{2 q! (S+1)^{q}} - \sum_{r<s} \chi_{rs}^{q} \frac{\gamma}{2 q! (S+1)^{2q}}\left(\frac{1}{N}\sum_{i} \mvec{S}_{i}^{r}\cdot \mvec{S}_{i}^{s}\right)^{q}
\end{align}
\end{widetext}
to lowest order in $\Delta t$, which holds as an operator equation.
Comparing Eqs. \eqref{eq:pbodyscrambleprop} and \eqref{eq:qbodymeasprop}, we conclude that the $q$-body Brownian measurement propagator is nearly identical to the unitary $p$-body propagator -- the only differences are in the numerical coefficients $\mu_{rs}^p,\chi_{rs}^q$ and in the overall sign of the interaction term.



\subsection{Coherent spin state path integral}
\label{app:coherentstatepathint}

We now stack a repeating sequence of $p$-body Brownian interactions and $q$-body Brownian measurements, insert resolutions of the identity between each layer, and take the limit $\Delta t \rightarrow 0$ with $T$ fixed to express the dynamics as a path integral over $2nN$ unit-norm $\mathrm{SO}(3)$ spins $\mvec{S}_i^r$, using spin coherent states as the basis. The completeness relation for the coherent states for a single spin is given by,
\begin{equation}
    \id = \int \frac{2S+1}{4\pi}d\mvec{\Omega}_{i}\ket{\mvec{\Omega}_{i}}\bra{\mvec{\Omega}_{i}}.
\end{equation}
To turn the spins into coherent states, we use the upper symbols for single spin-$S$ Pauli operators \cite{Klauder1985},
\begin{align}
    S_{i}^{\alpha} &= \int \frac{2S+1}{4\pi}d\mvec{\Omega}_{i} \ket{\mvec{\Omega}_{i}}\bra{\mvec{\Omega}_{i}}(S+1) \Omega_{i}^{\alpha}\\
    \left(S_{i}^{\alpha}\right)^{2} &= \int \frac{2S+1}{4\pi}d\mvec{\Omega}_{i} \ket{\mvec{\Omega}_{i}}\bra{\mvec{\Omega}_{i}}\left[(S+1)\left(S+\frac{3}{2}\right)\left(\Omega_{i}^{\alpha}\right)^{2} \right. \nonumber \\
    & \left. \hspace{1.7in} -\frac{S+1}{2}\right].
\end{align}
We introduce a measure for the coherent spin states in the path integral,
\begin{equation}
    \mathcal{D}\Omega_i^r = \prod_{t_{n}}\frac{2S+1}{4\pi}d\mvec{\Omega}^r_{i,t_{n}}\langle\mvec{\Omega}^r_{i,t_{n+1}}|\mvec{\Omega}^r_{i,t_{n}}\rangle,
\end{equation}
such that it includes the overlap of spin coherent states at discrete times $t_{n}$ and $t_{n+1}$. (Explicit evaluation of these overlap terms leads to `kinetic energy' or Berry-phase terms $\sim \Omega \partial_t \Omega$ in the path integral \cite{fradkin2013field}; the above choice of integration measure allows us to keep these time-dependent terms implicit, but the reader should keep in mind that these terms are always present.)

In terms of the coherent states, the action for $2n$ copies, combining both scrambling and measurement can be defined as follows,
\begin{widetext}
\begin{align}
    \left \langle V \otimes V_{\mathcal{T}} \otimes \cdots \otimes V \otimes V_{\mathcal{T}} \right \rangle_{\mvec{J},\mvec{O}} &= e^{- N I[\mvec{\Omega}](t) \Delta t} \nonumber \\
    I[\mvec{\Omega}] &= \int_{0}^{T}dt \left[  \frac{n J S^p}{2 p! (S+1)^p} +  \frac{n \gamma  S^q}{2 q! (S+1)^q}+\frac{J}{2p!}\sum_{r<s}\mu_{rs}^{p}\left(\frac{1}{N}\sum_{i}\mvec{\Omega}_{i}^{r}\cdot \mvec{\Omega}_{i}^{s}\right)^{p}\right. \nonumber\\
    &\left. \quad \quad -\frac{\gamma}{2q!}\sum_{r<s}\chi_{rs}^{q}\left(\frac{1}{N}\sum_{i}\mvec{\Omega}_{i}^{r}\cdot \mvec{\Omega}_{i}^{s}\right)^{q}\right].
\end{align}
To deal with the non-linear interactions in $\mvec{\Omega}_{i}^{r}\cdot\mvec{\Omega}_{i}^{s}$, we introduce decoupling fields $F_{rs}(t)$ and $G_{rs}(t)$ with the following operator identity,
\begin{equation}
    \id = \int \mathcal{D}F_{rs}\mathcal{D}G_{rs}\exp\left[ i N \int_0^T dt \ F_{rs}\left(G_{rs}-\frac{1}{N} \sum_{i}\mvec{\Omega}_{i}^{r}\cdot\mvec{\Omega}_{i}^{s} \right) \right]
\end{equation}


We can now treat $F_{rs}$ and $G_{rs}$ as the dynamical fields for the problem which couple different replicas $r,s$, and integrate out the spins, which gives us a propagator for spin problem. Note, we actually have a $N$-spin propagator when we evaluate the $\mvec{\Omega}$ path integral. However, since all the sites $i$ are identical and have been decoupled by the disorder average, we can rewrite the $2nN$-spin problem as the $N$-th power of a $2n$-spin problem,
\begin{equation}
    \int \prod_{i,r}\mathcal{D}\Omega^{r}_i  \exp\left(-\sum_{r<s}iF_{rs}\sum_{i}\mvec{\Omega}_{i}^{r}\cdot\mvec{\Omega}_{i}^{s}\right) = \int \prod_{r} \left( \mathcal{D}\Omega^{r} \right)^N \exp\left(-N\sum_{r<s}iF_{rs}\mvec{\Omega}^{r}\cdot\mvec{\Omega}^{s}\right).
\end{equation}
where the new single-site integration measure is
\begin{equation}
    \mathcal{D} \Omega^r = \prod_{t_n} \frac{2S+1}{4 \pi} d \mvec{\Omega}^r_{t_n} \bracket{\mvec{\Omega}^r_{t_{n+1}}}{\mvec{\Omega}^r_{t_n}}
\end{equation}

Putting it all together, along with the boundary conditions for the spin propagator, we get the action density
\begin{align}
    \label{eq:fullgeneralpathintegral}
    I[F_{rs},G_{rs}] &= I_0 + I_1 - \ln K(\psi_0,\psi_T,T) \ \quad \quad \text{where} \nonumber\\
    \nonumber \\
    I_0 &\equiv  J T \frac{n S^p}{2 p! (S+1)^p} + \gamma T \frac{n S^q}{2 q! (S+1)^q} \nonumber\\
    I_1 &\equiv \int_0^T dt\left[  \frac{ J}{2 p!}\sum_{r<s}\mu_{rs}^{p}G_{rs}^{p} - \sum \frac{\gamma}{2 q!}\sum_{r<s}\chi_{rs}^{q}G_{rs}^{q} -i\sum_{r<s}F_{rs}G_{rs}\right] \nonumber \\
    K(\psi_0,\psi_T,T)  &\equiv \bra{\psi_T} \exp\left[-\int_0^T dt \sum_{r<s} \frac{iF_{rs}}{(S+1)^{2}}\mvec{S}^{r}\cdot \mvec{S}^{s} \right] \ket{\psi_0} 
 \end{align}
\end{widetext}

Since the propagator $K$ is composed of $\su{2}$-symmetric Heisenberg couplings, the dynamics of the propagator are highly constrained. For $S = 1/2$ and $n = 2$ we showed in the main text that these constraints reduce the dynamics to a two-dimensional subspace; we expect similar constraints to simplify the problem for more general cases, but this remains a problem for future work.

\section{Replica symmetry}
\label{app:replicasymm}

As discussed in the main text, the microscopic bulk dynamics $\mathbb{V} = V^{(1)}\otimes V^{(2)}_{\mathcal{T}}\otimes V^{(3)}\otimes V^{(4)}_{\mathcal{T}}$ on replicas $r = 1,2,3,4$ for $n = 2$ is manifestly invariant under the replica symmetry group
\begin{align}
    \label{eq:replicasymmgroupsm}
    G = (S_2\times S_2)\rtimes \mathbb{Z}_{2}
\end{align}
where the inner $S_2 \cong \mathbb{Z}_2$ groups denote the permutation groups on replicas $1,3$ and $2,4$ with generators $\sigma = (13)$, $\sigma' = (24)$, respectively, where we use standard cycle notation in this section to represent permutations of replicas. The outer $\mathbb{Z}_2$ in the semidirect product is generated by $\tau = \mathcal{T} (12) (34)$, where the operation $\cal{T}$ represents time-reversal $V \leftrightarrow V_{\cal{T}}$ on all four replicas. Under the semidirect product, the generator $\tau$ simply exchanges the generators $\sigma,\sigma'$:
\begin{eqnarray}
    \sigma' = \tau \sigma \tau.
\end{eqnarray}
Explicitly, these generators act on the bulk dynamics as
\begin{align}
    \sigma \left( V^{(1)}\otimes V^{(2)}_{\mathcal{T}}\otimes V^{(3)}\otimes V^{(4)}_{\mathcal{T}} \right) &= V^{(3)}\otimes V^{(2)}_{\mathcal{T}}\otimes V^{(1)}\otimes V^{(4)}_{\mathcal{T}} \nonumber \\
    \sigma' \left( V^{(1)}\otimes V^{(2)}_{\mathcal{T}}\otimes V^{(3)}\otimes V^{(4)}_{\mathcal{T}} \right) &= V^{(1)}\otimes V^{(4)}_{\mathcal{T}}\otimes V^{(3)}\otimes V^{(2)}_{\mathcal{T}} \nonumber \\
    \tau \left( V^{(1)}\otimes V^{(2)}_{\mathcal{T}}\otimes V^{(3)}\otimes V^{(4)}_{\mathcal{T}} \right) &= V^{(2)}\otimes V^{(1)}_{\mathcal{T}}\otimes V^{(4)}\otimes V^{(3)}_{\mathcal{T}}
\end{align}
where superscripts denote replica indices $r = 1,2,3,4$. The replica symmetry group $G$ for $n = 2$ is isomorphic to the dihedral group $D_4 = \mathrm{Dih}_4$ (the group of symmetries of the geometrical square) via the representation
\begin{equation}
    G = \left \langle a,b | a^4 = b^2 = 1, bab = a^{-1} \right \rangle
\end{equation}
with the identification $a = \tau \sigma$ and $b = \sigma$.

While the bulk dynamics $\mathbb{V}$ are invariant under the full group $G$, the boundary conditions at $t = 0,T$ break this down to a subgroup $H \subset G$ generated by the mutually-commuting operators $\tau,c$, where
\begin{equation}
    c = \sigma \tau \sigma = b a = \mathcal{T}(14)(23)
\end{equation}
corresponds to a `reflection' $1234 \leftrightarrow 4321$ in replica space followed by time-reversal $\mathcal{T}$ on all replicas. This subgroup is isomorphic to the Klein four-group $H \cong \mathbb{Z}_2 \times \mathbb{Z}_2$ composed of the four elements $\{ e, \tau, c, \tau c \}$ with $e$ the identity element. Left-multiplication by $\sigma = b$ yields the left coset $\sigma H = \{\sigma, \sigma \tau, \sigma c, \sigma \tau c\}$, and together the left cosets $H,\sigma H$ generate the full group $G$. In this sense, the generator $\sigma = b$ represents the $\mathbb{Z}_2$ symmetry that is explicitly broken by the boundary conditions and spontaneously broken in the bulk.

For $n > 2$ the replica symmetry group is
\begin{equation}
    G = (S_n \times S_n') \rtimes \mathbb{Z}_2
\end{equation}
where $S_n = S_{135 \ldots}, S_n' = S_{246 \ldots}$ are the order-$n!$ symmetric groups on replicas $r = 1,3,5,\ldots$ and $r = 2,4,6,\ldots$ with generators $\sigma,\sigma'$, respectively. Similar to above, the outer $\mathbb{Z}_2$ is generated by an element $\tau = \mathcal{T} (12) (34) \cdots (2n-1 \ 2n)$ that exchanges the generators $\sigma' = \tau \sigma \tau$. The boundary conditions at $t = 0,T$ break this bulk symmetry down to a subgroup $H \subset G$ that depends on the details of the boundary states.

Because they contain the time-reversal operation $\mathcal{T}$, which itself contains the complex conjugation operation $*$, the operators $\tau,c$ are \emph{antilinear} operators on the Hilbert space of quantum states, in contrast to the generators $\sigma,\sigma'$ which are conventional linear operators \cite{haake2010quantum}. Whereas conventional linear operators (as their name suggests) are linear in their arguments:
\begin{equation}
    \sigma\left(\alpha \ket{\psi} + \beta \ket{\psi'}\right) = \alpha \sigma \ket{\psi} + \beta \sigma \ket{\psi'}
\end{equation}
antilinear operators are antilinear in their arguments:
\begin{equation}
    \tau\left(\alpha \ket{\psi} + \beta \ket{\psi'}\right) = \alpha^* \tau \ket{\psi} + \beta^* \tau \ket{\psi'}
\end{equation}
where complex conjugation of the $c$-numbers $\alpha,\beta$ arises due to the complex conjugation in the definition of time-reversal $\mathcal{T}$. The antilinearity of $\tau,c$ leads to restrictions on the spectrum of the operator $\mathbb{V}$ via the same mechanism that guarantees the reality of eigenvalues of certain non-Hermitian Hamiltonians in PT-symmetric quantum mechanics \cite{bender1998real,bender2005introduction}.

Here we show that an operator $\mathbb{V}$ with \emph{unbroken PT symmetry} has a real spectrum, while an operator with \emph{broken PT symmetry} has a spectrum consisting of complex-conjugate pairs. Assume that there is an antilinear operator $\tau$ that commutes with the operator $\mathbb{V}$:
\begin{equation}
    \label{eq:tauvcomm}
    [\tau,\mathbb{V}] = 0
\end{equation}
and suppose $\ket{\Psi}$ is an eigenstate of $\mathbb{V}$ with eigenvalue $V$:
\begin{equation}
    \label{eq:veig1}
    \mathbb{V} \ket{\Psi} = V \ket{\Psi}.
\end{equation}
Multiplying this eigenvalue equation on the left by $\tau$ and using the fact that $\tau V = V^* \tau$ by antilinearity of $\tau$, along with the commutativity of $\tau,\mathbb{V}$, we find that $\ket{\Psi'} = \tau \ket{\Psi}$ is also an eigenstate of $\mathbb{V}$ with eigenvalue $V^*$:
\begin{equation}
    \label{eq:veig2}
    \mathbb{V} \ket{\Psi'} = V^* \ket{\Psi'}.
\end{equation}
Hence any eigenstate $\ket{\Psi}$ of $\mathbb{V}$ with eigenvalue $V$ is always accompanied by a second eigenstate $\tau \ket{\Psi}$ with eigenvalue $V^*$.

This shows that the spectrum of any PT-symmetric operator $\mathbb{V}$ always consists of complex-conjugate pairs $V,V^*$, but there is no guarantee that these eigenvalues lie on the real line. To guarantee reality of the spectrum one requires the additional assumption that the eigenstate $\ket{\Psi}$ is simultaneously also an eigenstate of $\tau$:
\begin{equation}
    \label{eq:taueig}
    \tau \ket{\Psi} = \lambda \ket{\Psi}.
\end{equation}
If $\tau$ were a conventional linear operator, this would follow immediately from the commutativity of $\tau,\mathbb{V}$ \eqref{eq:tauvcomm}; but when $\tau$ is antilinear this is an additional independent assumption. Because $\tau$ is antilinear the eigenvalue $\lambda$ can be any pure phase $\lambda = e^{i \phi}$ but we may always appropriately redefine the eigenstate $\ket{\Psi}$ such that $\lambda = 1$ \cite{bender1998real,bender2005introduction,haake2010quantum}. In this case we have $\ket{\Psi'} = \tau \ket{\Psi} = \ket{\Psi}$ and therefore by combining Eqs. \eqref{eq:veig1} and \eqref{eq:veig2} we immediately obtain $V = V^*$. Thus, if $\ket{\Psi}$ is a simultaneous eigenstate of both $\tau$ and $\mathbb{V}$, then its eigenvalue $V$ is real.

If all of the eigenstates of $\mathbb{V}$ are also eigenstates of the antilinear operator $\tau$ then the spectrum is guaranteed to be real by the above arguments and we say that the PT symmetry of $\mathbb{V}$ is \emph{unbroken}. Conversely, if there are eigenstates of $\mathbb{V}$ that are not eigenstates of $\tau$, then the spectrum consists of complex-conjugate pairs and we say that the PT symmetry of $\mathbb{V}$ is \emph{broken}. 



\section{Simplification of the path integral at saddle point}
\label{app:pathintegraltechnical}

We introduce symmetric and anti-symmetric fields defined as

\begin{align*}
    F_{a}^{\pm} &= \frac{2}{9}(iF_{12}\pm iF_{34}),\  G_{a}^{\pm} = \frac{9}{2}(G_{12}\pm G_{34}) \\
    F_{b}^{\pm} &= \frac{2}{9}(iF_{14}\pm iF_{23}),\  G_{b}^{\pm} = \frac{9}{2}(G_{14}\pm G_{23})\\
    F_{c}^{\pm} &= \frac{2}{9}(iF_{13}\pm iF_{24}),\  G_{c}^{\pm} = \frac{9}{2}(G_{13}\pm G_{24}).
\end{align*}

This re-definition simplifies the 4-replica propagator, as the only fields appearing in the propagator are the symmetric combinations, $F_{a,b,c}^{+}$. This implies that the saddle-point equations of motion for the anti-symmetric fields $F_{a,b,c}^{-}$ set $G_{a,b,c}^{-}=0$, which reduces the number of fields to consider down from 12 to 6. Since all the anti-symmetric fields are thus integrated away, we drop the $\pm$ superscript and define $G_{a,b,c}=G_{a,b,c}^{+}$ (and similarly for the F fields).

In terms of these fields, the action density can be rewritten as, 
\begin{align}
    I &= \int_{0}^{T} dt \left[\frac{J}{162}\left(-G_{a}^{2}-G_{b}^{2}+G_{c}^{2}\right)-\frac{\gamma}{9}\left(-G_{a}-G_{b}+G_{c}\right) \right. \nonumber \\
    & \quad \quad \left. -F_{a}G_{a}-F_{b}G_{b}-F_{c}G_{c}\right]-\ln K.
\end{align}

The saddle-point equations of motion corresponding to this action are given by,
\begin{align}
    -\frac{J}{81}G_{a,b}+\frac{\gamma}{9} &= F_{a,b} \nonumber \\ \frac{J}{81}G_{c}-\frac{\gamma}{9} &= F_{c}\label{eq:saddleeom1}, \ \ G_{a,b,c} = -\frac{d \ln K}{d F_{a,b,c}}
\end{align}
The G fields can thus be integrated out by replacing them with the saddle-point solutions. We can also rewrite the action in terms of the magnetic field variables $\vec{B}$. The $B_{0}$ field can also be integrated out since it appears in a quadratic form. 

Finally we have a path integral over just two fields, with the action,
\begin{align}
     I &= \int_{0}^{T} dt \left[\frac{27 B_{x}^{2}}{4 J}-\frac{81 B_{z}^{2}}{4 J}+B_{z}(1+18\gamma)-\frac{J}{72} \right. \nonumber \\ 
     & \quad \quad \left. -\frac{4\gamma^{2}}{J}-\frac{\gamma}{2}\right]-\ln K\nonumber\\
     K & = \bra{\psi_{T}}\exp\left[\frac{1}{2}\int_{0}^{T} dt \left(B_x \sigma_{x}+ B_z \sigma_{z}\right)\right]\ket{\psi_{0}}.
\end{align} 

We need to determine the integral contour such that the integral is converged. This implies that $B_x$ is to be integrated from $-\infty \to \infty$, while $B_z$ is to be integrated along the imaginary axis, $-i\infty \to i\infty$. 


Note that until now we have not made any assumptions about the time dependence of the fields. This simplified expression for the path integral over just 2 fields follows naturally from the symmetry of the 4-spin Hamiltonian and the fact that the boundary states belong to a particular spin sector.

\section{Numerical gradient descent}
\label{app:gradientdesc}

We can estimate the time dependent solutions to Eq. \eqref{eq:2fieldPathIntegralmain} by performing numerical gradient descent on discretized field configurations of $B_x$ and $B_z$. For this section in order to perform gradient descent over real valued $B_x$ and $B_z$ fields we change the definition of $B_z$ to an imaginary `magnetic field', $iB_z$. The action is given by,

\begin{align} \label{eq:2fieldPathIntegral}
     I &= \int_{0}^{T} dt \left[\frac{27 B_{x}^{2}}{4 J}+\frac{81 B_{z}^{2}}{4 J}+i B_{z}(1+18\gamma)-\frac{J}{72}\right. \nonumber \\
     & \quad \quad \left. -\frac{4\gamma^{2}}{J}-\frac{\gamma}{2}\right]-\ln K\nonumber\\
     K & = \bra{\psi_{T}}\exp\left[\frac{1}{2}\int_{0}^{T} dt \left(B_x \sigma_{x}+i B_z \sigma_{z}\right)\right]\ket{\psi_{0}}.
\end{align} 

For the gradient descent, we consider units where $Jdt = 0.05$. In Figs. 6 and 7 we consider total times of $2430 Jt$ and $3240 Jt$ respectively. Starting from the bulk saddle-point configurations we perform the gradient descent with the action given in Eq. \eqref{eq:2fieldPathIntegral} until the difference in action is below a threshold of $\delta I \sim 10^{-7}$. Each of the configurations in Fig. \ref{fig:numgraddesc} require $\sim 10000$ iterations of the gradient descent to reach the required threshold. For Fig. \ref{fig:crit_exp}a, we initialize the configuration of the fields to correspond to the instanton configuration in Eq. \eqref{eq:exact_instanton} and find that the action is already below the threshold for gradient descent.

\begin{figure*}
    \centering
    \includegraphics[width=0.8\textwidth]{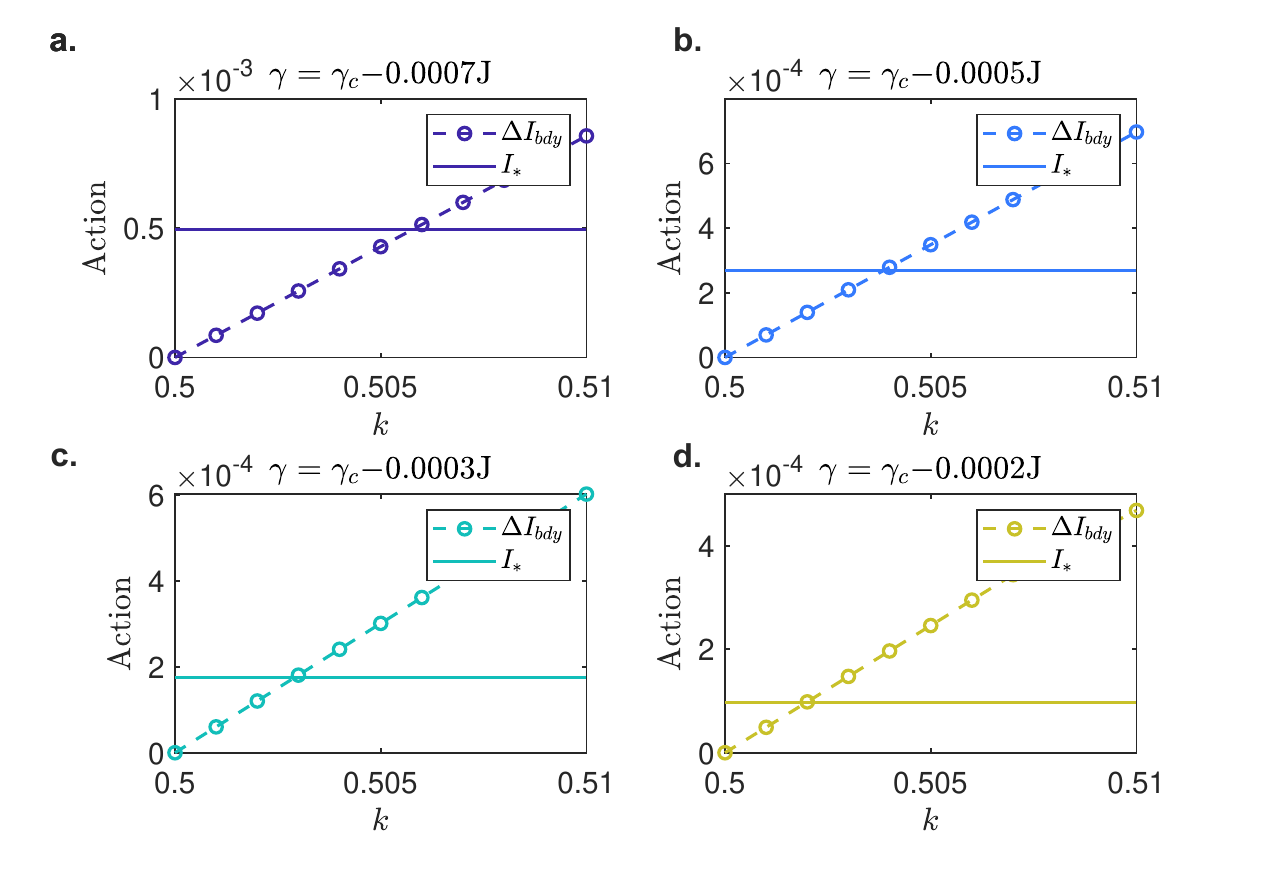}
    \caption{\textbf{Subsystem purity critical exponent $\cek$ from gradient descent numerics.} The critical subsystem fraction $k_c(\gamma)$ is identified for measurement rates $\gamma < \gamma_c$ just below the critical point by finding points in the $k,\gamma$ plane where the boundary action $\Delta I_{\mathrm{bdy}}$ is equal to the single-instanton action $I_*$.}
    \label{fig:k_crit_appendix}
\end{figure*}

To explore the subsystem purity phase diagram Fig. \ref{fig:kphasediag}, we perform numerical gradient with the $k$ dependent action with the propagator given by Eq. \eqref{eq:pathintegral_k}. For $\gamma<\gamma_c$ but close to criticality, we consider the two configurations in Fig. \ref{fig:kphasediag}b(iii-iv) and perform gradient descent to find local minima near these solutions, for a range of $k\in {0.50,0.501,..,0.51}$ as shown in Fig. \ref{fig:k_crit_appendix}. We then interpolate to find the value of $k$ for which the two configurations exchange in total action, which is the numerical estimation of $k_c$, used to plot Fig. \ref{fig:k_crit_exp}. The error bars are the errors due to the resolution of the $k$ values considered for the numerics.



\section{Exact diagonalization}
\label{app:exactdiag}

Numerical simulations of the hybrid dynamics for small system sizes $N$ confirm the presence of a long-lived mixed phase as shown in Fig. \ref{fig:timedependencepurity} of the main text. These data were obtained by numerically simulating the $(2,1)$ hybrid Brownian model on $N = \magn{Q} = 6$ qubits maximally entangled with $\magn{R} = 6$ reference qubits using the Krylov subspace method \cite{lanczos1950iteration,park1986unitary,liesen2013krylov}.

\begin{figure*}
    \centering
    \includegraphics[width=\textwidth]{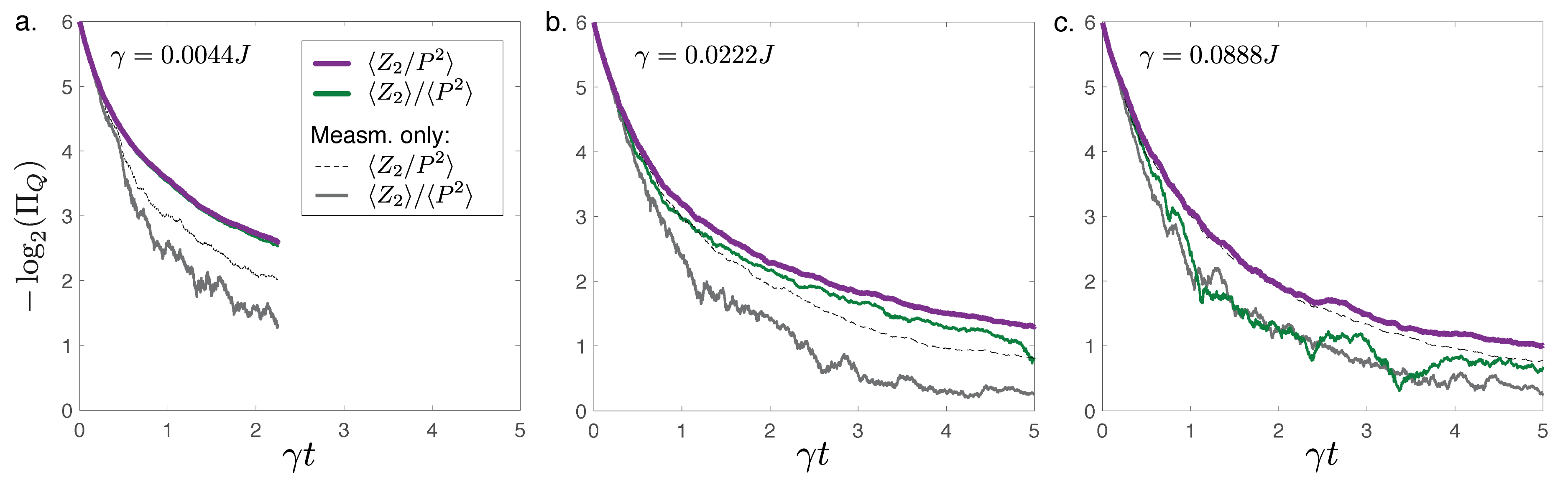}
    \caption{\textbf{Disorder averaging in exact diagonalization numerics.} (a) Far below the critical point, the two averaging protocols (purple, green) yield nearly identical results in numerical simulations with $N = 6$ qubits averaged over 50 circuit realizations. (b) Closer to the critical point the two estimates begin to diverge, while above the critical point (c) they disagree sharply. Nevertheless, both disorder averages appear to faithfully diagnose the transition when compared to the respective disorder averages performed in circuits featuring measurements only (dotted grey, solid grey).}
    \label{fig:edresults}
\end{figure*}

We plot the results of these numerical simulations in Fig. \ref{fig:edresults}, which shows the entropy $-\ln \Pi_Q = -\ln Z_2/P^2$ as a function of time, averaged over $50$ circuit realizations. We perform the disorder average in two different ways: the `physical' disorder average $\left \langle Z_2 / P^2 \right \rangle$ that one obtains from the Born rule (purple); and the `tractable' disorder average $\left \langle Z_2 \right \rangle / \left \langle P^2 \right \rangle$ studied in this work (green). At low measurement rates $\gamma$ we find that these two ways of doing the disorder average give nearly identical answers (Fig. \ref{fig:edresults}a), while they disagree for higher values of $\gamma$ (Figs. \ref{fig:edresults}b-c). Both disorder averages, however, deviate substantially from the corresponding curves computed in measurement-only circuits (dotted black, solid black), indicating that both disorder averages are sensitive to the purification transition.

Our exact diagonalization calculations are limited to short simulation times and values of $\gamma / J$ that are not too small. At small $\gamma$ one must distinguish the phase from the initial exponential decay, necessitating simulation times longer than $t > 1/\gamma$; accessing these long timescales is challenging for exact diagonalization due to the propagation of successive errors in the Krylov approximation. Specifically, numerical accuracy of the Krylov algorithm requires $\epsilon = \sqrt{J \delta t} \ll 1$ (the square root comes from the fact that $J$ controls the variance of the couplings in Eq. \eqref{eq:jrandvar}, not the standard deviation). Together with the requirement $t > 1/\gamma$ this gives a lower bound
\begin{equation}
    \frac{t}{\delta t} > \frac{1}{\epsilon^2} \frac{J}{\gamma}
\end{equation}
on the number of timesteps $t/\delta t$ required to access the mixed phase given fixed numerical precision $\epsilon$ and phase parameter $\gamma / J$. Our simulation for $\gamma = 0.0044J$ above, for example, has $\epsilon \approx 0.16$ and $J/\gamma = 225$, requiring on the order of $t/\delta t \approx 10^4$ timesteps or more to access the mixed phase.






\section{Continuous Monitoring of Disordered Spin Observables with Optical Cavities}
\label{app:cavitymonitor}
While the projective qubit model introduced in Appendix \ref{app:qbodybrown} is conceptually useful for deriving the effective weak measurement operator $M(t)$, high-fidelity single-site projective measurements are challenging to implement experimentally. Fortunately, such high-fidelity single-qubit projective measurements are not strictly necessary for our scheme to work. Instead, one can generate equivalent weak measurement dynamics by continuously monitoring the collective spin operator $\op{} = \sum_{i,\alpha} n_i^{\alpha} S_i^{\alpha}$ directly. Such collective spin variables can be monitored naturally in state-of-the-art cavity quantum electrodynamics setups by coupling a quasi-one-dimensional cold atomic ensemble to the optical mode of an all-to-all optical cavity \cite{gleyzes2007quantum,schleier2010squeezing,leroux2010implementation,leroux2012unitary,barontini2015deterministic,davis2018painting}.

In such a cavity setup, each spin $\mvec{S}_i$ is encoded into the electronic states of the $i$th atom, which resides at position $z_i$ along the longitudinal cavity axis. The atoms act like a spin-dependent refractive index for the cavity mode, which causes the cavity resonance to shift by an amount proportional to the total magnetization $S^z_{\mathrm{Tot}} = \sum_i S_i^z$ \cite{schleier2010squeezing,leroux2010implementation,leroux2012unitary}. By probing the cavity with near-resonant light we can therefore continuously monitor the total spin projection $S^z_{\mathrm{Tot}}$. If the atoms are coupled unequally to the cavity mode, the shift in cavity resonance is instead proportional to the disordered magnetization $\tilde{S}^z_{\mathrm{Tot}} = \sum_i n_i S_i^z$, where the weights $n_i$ are determined by the coupling between the $i$th atom and the cavity mode. These couplings can be modified by shifting the physical locations of the atoms relative to the cavity mode, or by applying nonuniform local ac Stark shifts to the ensemble.
Further, one can couple different spin components $S_i^{\alpha}$ to the cavity mode by applying additional non-uniform magnetic fields or optical drive fields to rotate the local coordinate frame at each atomic site $z_i$. The combination of these tools in principle allows for continuous monitoring of disordered spin-linear operators of the form $\op{} = \sum_{i,\alpha} n_i^{\alpha} S_i^{\alpha}$ without requiring single-site projective measurements of single qubits.

\bibliography{References}

\end{document}